\documentclass[acmsmall,screen]{acmart}

\usepackage{listings}
\usepackage{algorithm}
\usepackage{algpseudocode}
\usepackage{xcolor}
\usepackage{needspace}

\lstdefinestyle{qqpy}{
  language=Python,
  basicstyle=\ttfamily\footnotesize,
  keywordstyle=\bfseries,
  commentstyle=\color{gray},
  stringstyle=\color{gray},
  numbers=none,
  frame=none,
  breaklines=true,
  columns=fullflexible,
  aboveskip=8pt,
  belowskip=8pt,
  captionpos=t,
}

\acmJournal{TQC}
\acmYear{2026}
\acmVolume{0}
\acmNumber{0}
\acmArticle{0}
\acmMonth{0}
\acmDOI{XX.XXXX/XXXXXXX}
\setcopyright{acmlicensed}
\copyrightyear{2026}



\begin{document}

\title[QQLink: Declarative Multi-QPU Quantum Orchestration]{QQLink: A Declarative Framework for Multi-QPU Quantum Workflow Orchestration}
\titlenote{This manuscript has been authored by UT-Battelle, LLC, under
Contract No.\ DE-AC05-00OR22725 with the U.S. Department of Energy. The
United States Government retains and the publisher, by accepting the article
for publication, acknowledges that the United States Government retains a
non-exclusive, paid-up, irrevocable, worldwide license to publish or
reproduce the published form of this manuscript, or allow others to do so,
for United States Government purposes. The Department of Energy will provide
public access to these results of federally sponsored research in accordance
with the DOE Public Access Plan.}

\author{Vicente Leyton-Ortega}
\email{leytonorteva@ornl.gov}
\affiliation{%
  \institution{Oak Ridge National Laboratory}
  \department{Quantum Information Science Section}
  \city{Oak Ridge}
  \state{TN}
  \country{USA}
}

\author{Narasinga Rao Miniskar}
\email{miniskarnr@ornl.gov}
\affiliation{%
  \institution{Oak Ridge National Laboratory}
  \department{Architectures and Performance}
  \city{Oak Ridge}
  \state{TN}
  \country{USA}
}

\author{Thomas Naughton}
\email{naughtont@ornl.gov}
\affiliation{%
  \institution{Oak Ridge National Laboratory}
  \department{Intelligent Systems and Facilities}
  \city{Oak Ridge}
  \state{TN}
  \country{USA}
}

\author{Travis S. Humble}
\email{humblets@ornl.gov}
\affiliation{%
  \institution{Oak Ridge National Laboratory}
  \department{Quantum Science Center}
  \city{Oak Ridge}
  \state{TN}
  \country{USA}
}

\renewcommand{\shortauthors}{Leyton-Ortega et al.}

\begin{abstract}
Distributed quantum protocols require measurement outcomes produced at one
node to control operations at another. We present QQLink, a programming and
orchestration layer that expresses these dependencies in a single program,
using workflow nodes, deferred measurement references, and declared
entanglement, and derives the cross-node coordination automatically. Its
planner constructs a provably acyclic dependency graph in time linear in the
total size of the instructions' qubit footprints. The same programs execute
through simulator backends or are lowered to dynamic circuits on partitioned
quantum hardware. We evaluate seven protocol workflows. Exhaustive
feedforward-branch tests on the simulator runtime path, over tomographically
complete input sets, yield output-state infidelities of at most
$5\times10^{-12}$ and no incorrect classical outputs, and replaying every
plan in random topological orders leaves each final state unchanged.
Separate executor processes run the plans with delayed messages, with the
joint quantum state held by a centralized simulation service. Every run with
the full plan is correct, and removing required cross-node dependencies
causes protocol failures. The lowered circuits match hand-written
implementations of the same construction in two-qubit gate and
conditional-block counts. Runs on two IBM Heron processors, each partitioned
into virtual QPUs, demonstrate same-chip execution in two sessions a week
apart. The runs also show large variation between
sessions and losses beyond what calibration-derived noise models predict.
Coherence-sensitive checks do not establish coherent Toffoli operation or
preservation of the intended repetition-code superposition. These results
support QQLink's workflow semantics on the tested execution paths and
identify limits of the current hardware realization; execution across
physically networked QPUs remains future work.
\end{abstract}

\maketitle

\section{Introduction}
\label{sec:intro}

A quantum program that spans several quantum processing units (QPUs) is not
merely a collection of circuits compiled separately. Teleportation, superdense
coding, and non-local gate constructions all rely on coordination across
program boundaries. A measurement produced in one part of the program must be
referenced in another, and a conditional correction must be applied at the
right execution point. The same program should also be portable between a
simulator, where it is developed, and hardware, where it is eventually run.
Distributed quantum computing has long been studied as a route past the size
of a single processor~\cite{cuomo2020,caleffi2022}, and modular architectures
have been proposed for several hardware platforms~\cite{monroe2014,jnane2022}.
A recent experiment executed a distributed quantum algorithm across an
optical network link between two processors~\cite{main2025}. This progress widens the gap between what
hardware can do and what existing software can readily express. A recent
survey of production quantum--high-performance computing (HPC) software stacks~\cite{shehata2026} names
as an open problem the extension of their communication semantics from
one-way circuit submission to bidirectional interaction.

Quantum software development kits (SDKs) such as Qiskit~\cite{qiskit2024},
Cirq~\cite{cirq2024}, and PennyLane~\cite{bergholm2022} provide mature support
for circuit construction, transpilation, and, in some cases, dynamic control
flow within one circuit on one backend~\cite{corcoles2021,cross2022}. They do
not provide a reusable orchestration layer in which application intent is
expressed separately from backend placement and execution decisions. Adjacent
problems have been addressed at other levels of the stack. Circuit
cutting~\cite{peng2020,tang2021} replaces quantum communication with classical
post-processing and cannot express real-time feedforward. Compiler-level
partitioning~\cite{andres2019,ferrari2023,baker2020,wu2022} decides where
gates run but not how mid-circuit outcomes are routed at runtime. Network
simulators~\cite{coopmans2021,wu2021,diadamo2021} model link-layer physics
below the application. Message-passing proposals~\cite{haner2021} define
communication primitives, and quantum-network node software, from the
NetQASM instruction set and SDK~\cite{dahlberg2022} to the QNodeOS operating
system~\cite{delledonne2025} and the Qoala execution
environment~\cite{vandervecht2025qoala}, executes per-node programs whose
cross-node coordination the programmer writes; none derives that
coordination from one program over all nodes. Section~\ref{sec:related}
positions each of these lines of work in detail.

QQLink addresses this gap. It is a programming and orchestration layer that
separates \emph{what} a workflow computes from \emph{where} and \emph{how} it
executes. Applications declare workflow nodes, deferred measurement
references, conditional feedforward, and cross-node interactions through a
backend-neutral interface. The platform layer maps those abstractions onto
simulators or same-chip hardware, including partitioned virtual-QPU layouts on
IBM backends. QQLink is not a replacement for backend-native SDKs but a layer
above them that captures multi-QPU workflow semantics in a portable form.

\begin{figure}[t]
\centering
\includegraphics[width=0.8\linewidth]{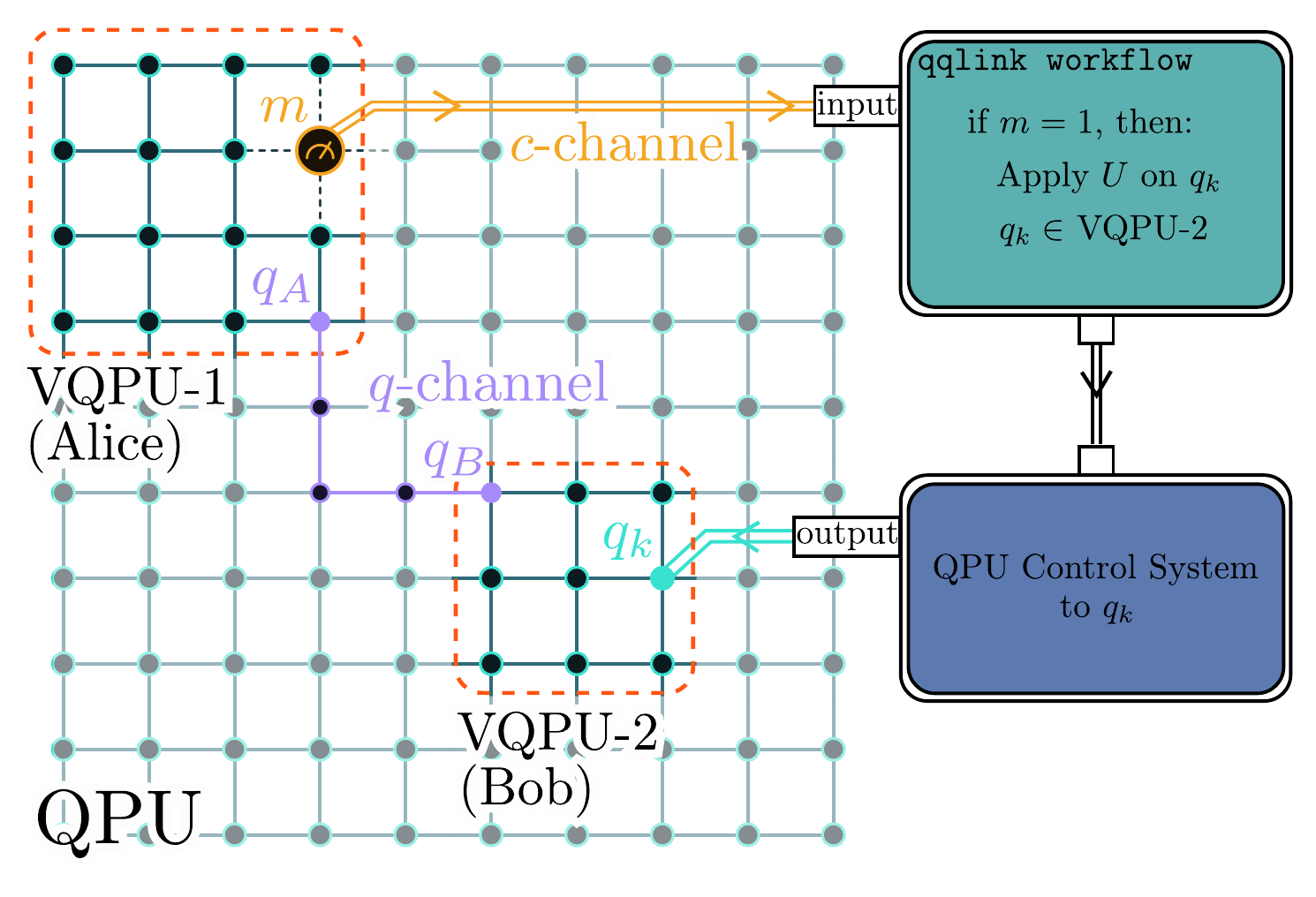}
\Description{Schematic of a qubit lattice divided into two dashed sectors labelled VQPU-1 (Alice) and VQPU-2 (Bob). A purple quantum channel joins qubit q_A in Alice's sector to qubit q_B in Bob's sector through three routing qubits that belong to neither sector, and an orange classical channel carries a measurement outcome m from Alice's sector to a gate U on qubit q_k in Bob's sector.}
\caption{Same-chip execution of a cross-node conditional. One QPU (qubit
lattice) hosts two virtual QPUs, VQPU-1 (Alice) and VQPU-2 (Bob), as disjoint
sectors (dashed). Two kinds of link cross the sector boundary. The
\emph{q-channel} (purple) is the quantum path between $q_A$ and $q_B$ that
realizes a Bell pair or a cross-virtual CNOT; on a partitioned chip it is a
native two-qubit gate routed through intermediate qubits, which is the
co-located path of the channel-selection policy. Dark qubits are allocated;
the three between $q_A$ and $q_B$ are claimed by routing and belong to
neither sector. The \emph{c-channel}
(orange) is classical feedforward, in which the outcome $m$ of a measurement on Alice's
sector conditions a gate $U$ on $q_k$ in Bob's sector. On the IBM path the
whole workflow is lowered before submission into one dynamic circuit, and
the condition is evaluated by the QPU control system during the shot; no
host-side QQLink code runs between the measurement and $U$. Neither the workflow nor the application names a
physical qubit; the sector mapping is decided at planning time.}
\label{fig:vqpu}
\end{figure}

Two terms recur throughout the paper. A \emph{workflow node} is a
program-level participant in the application, such as Alice or Bob in a
communication protocol; each node declares the qubits it owns. A
\emph{virtual QPU} (vQPU) is an execution-level partition of a shared backend
used to realize a workflow node at runtime. The first belongs to the
application layer, the second to the platform layer, and the mapping between
them is decided at planning time. Our vQPUs are \emph{entangled} partitions of
one backend that exchange Bell pairs and feedforward; the term is also used for
independent, non-interacting partitions that execute unrelated circuits in
parallel~\cite{claudino2024}, which is a different construction.
Fig.~\ref{fig:vqpu} shows the two on a partitioned chip. Two sectors of one
QPU act as Alice and Bob, a quantum path between sectors carries the
entanglement, and a classical path through the control system carries a
measurement outcome back as a conditional gate.

The question the paper answers is whether the coordination semantics of
multi-QPU protocols can be captured in a backend-agnostic workflow abstraction
such that (a)~one program runs unchanged on different simulators and on
partitioned hardware, (b)~planning is deterministic, acyclic, and linear in
the size of the program, (c)~the abstraction adds no circuit-level cost relative to
hand-written dynamic circuits of the same construction, and (d)~the plan suffices to run a program on
separate executors, one per node, that share no controller. Condition (d)
separates QQLink from a single dynamic circuit with named registers, which
satisfies (a) and (c) on one backend but fixes one global order and one
controller. The test of (d), one controller process per node with delayed
messages and no dispatcher (Section~\ref{sec:separate}), is therefore the
central evidence for what the abstraction adds; the evaluation reports it
directly after the correctness checks. This paper makes the following
contributions.
\begin{enumerate}
\item A programming model for multi-QPU-style workflows based on workflow
  nodes, typed deferred measurement references, declarative entanglement
  links, cross-node conditional gates, and an explicit ordering primitive
  (Section~\ref{sec:model}).
\item A planning algorithm that derives dependencies from qubit footprints
  in authoring order and from measurement references, with proved guarantees
  (deterministic output, an edge for every local and cross-node
  measurement reference, acyclicity at the granularity of execution units,
  and independence of the result from the topological order) and measured cost of
  8 to 17 microseconds per instruction, linear over three orders of
  magnitude of program size (Section~\ref{sec:planning}).
\item Execution mechanisms that realize the plan on heterogeneous targets,
  namely virtual-QPU partitioning, a channel-selection policy, per-instruction
  conditional dispatch, retry and timeout controls, and a hardware path
  (Section~\ref{sec:mechanisms}).
\item Seven protocol workflows authored entirely at the application
  layer, namely superdense coding, register teleportation, entanglement swapping, a
  Grover search and a Bernstein--Vazirani algorithm built from teleported
  gates, a three-node Toffoli gate, and a three-node repetition code; and a
  single-node accept/reject control loop for the task layer
  (Section~\ref{sec:demos}).
\item A systems evaluation answering (a)--(d), with exhaustive forced-branch
  verification of every feedforward branch on the simulator runtime path, on
  tomographically complete inputs, and replay of every plan in random topological orders;
  execution on one controller process per node with delayed messages, and
  separate ablations of the measurement and footprint edges between
  nodes; a cost-of-abstraction
  comparison against hand-written dynamic circuits over transpiler settings
  and against an independent distribution compiler; planning overhead; and
  same-chip runs on two IBM Heron processors in two sessions, with controls
  that break the correlation between outcomes and corrections,
  coherence-sensitive checks, and a study of the excess loss of the deeper
  workflows
  (Section~\ref{sec:eval}).
\end{enumerate}

We are deliberate about scope. The hardware results reported here are
same-chip, virtual-QPU demonstrations, each on a single partitioned device. They
are not physically distributed, multi-chip experiments across networked QPUs;
that regime is the goal the abstractions are designed for, but realizing it
on hardware remains future work (Section~\ref{sec:limitations}). The
separate-executor experiment distributes the control plane across processes
but keeps the quantum state in one simulator. Throughout, ``distributed''
describes the class of protocols QQLink expresses, not the physical execution
substrate demonstrated.

\section{Background and Requirements}
\label{sec:background}

A multi-QPU-style quantum workflow consists of several circuit regions
related by entanglement and by feedforward, in which measurements in one
region drive conditional operations in another. This pattern arises in quantum teleportation~\cite{bennett1993},
superdense coding~\cite{bennett1992}, non-local gates derived from gate
teleportation~\cite{gottesman1999,eisert2000}, and distributed search and
factoring algorithms~\cite{main2025,yimsiriwattana2004}. In each case the
protocol is a workflow whose correctness depends on the ordering of local
operations, the availability of shared entanglement, and the use of
measurement outcomes to drive later computation. These dependencies must
remain well defined independently of the backend on which the workflow runs.

These constraints yield four requirements for a software layer aimed at
multi-QPU workflow semantics.
\begin{description}
\item[R1: Backend-agnostic authoring.] A workflow is written in terms of
  workflow nodes and protocol structure, without embedding backend placement.
\item[R2: Deferred measurement and feedforward.] Measurement outcomes are
  referenceable handles that can drive conditional operations later in the
  workflow, including across qubit-allocation boundaries.
\item[R3: Virtual-QPU execution support.] The same workflow model maps onto
  simulator targets and shared-backend virtual-QPU configurations without
  rewriting the application.
\item[R4: Dependency-aware orchestration.] Execution respects workflow
  dependencies and provides the runtime support to schedule, dispatch, and
  recover multi-stage tasks.
\end{description}
QQLink satisfies these requirements by treating a multi-QPU protocol as a
workflow with explicit application-level structure rather than as a one-off
composition of backend calls. Sections~\ref{sec:model}
to~\ref{sec:mechanisms} describe the model, the planner, and the execution
mechanisms in turn; Section~\ref{sec:demos} returns to R1--R4 and shows how
the eight workflows exercise each of them.

\section{Programming Model}
\label{sec:model}

\begin{figure}[t]
\centering
\includegraphics[width=0.92\linewidth]{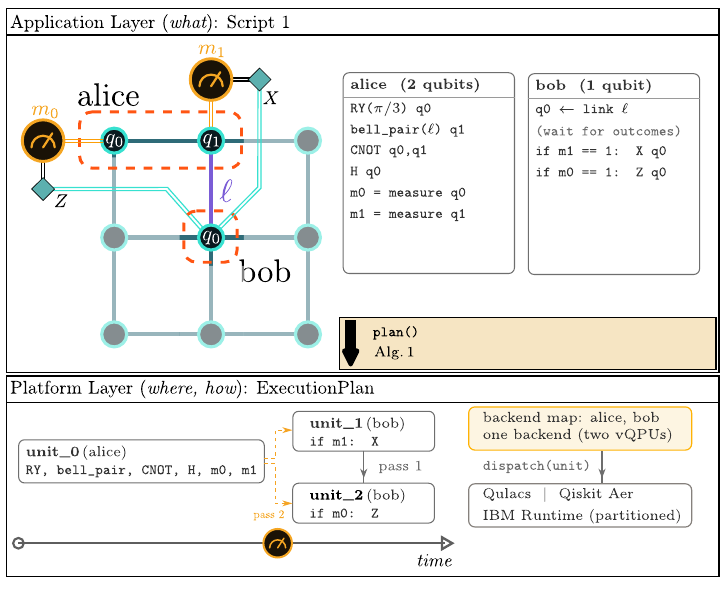}
\Description{Two-layer diagram. The top layer shows Alice's two-qubit and Bob's one-qubit workflow circuits with their instruction lists, an entanglement link between them, and orange arrows from Alice's two measurements to Bob's X and Z corrections. The bottom layer shows the execution plan as three units on a time axis, Alice's unit first, then Bob's two units chained to each other and each linked to Alice's unit by a dashed orange edge.}
\caption{The two-layer model on the running example (Script~\ref{lst:teleport1}).
Orange carries classical information, cyan carries quantum
information, and purple marks an entanglement link. The top panel shows the
application layer with the two \emph{WorkflowCircuits}, Alice ($q_0$, $q_1$) and Bob ($q_0$), drawn
on the qubits they will occupy (grey qubits are unallocated) and listed as the
backend-free instruction streams the developer wrote. The
\emph{EntanglementLink} $\ell$ ties Alice's $q_1$ to Bob's $q_0$. Alice's
measurements return \emph{MeasureRef} handles $m_0$, $m_1$ (meters); Bob's
two \texttt{if\_then} conditionals reference them across the circuit boundary
and are drawn as the $X$ and $Z$ corrections applied to Bob's $q_0$, the
c-channel of Fig.~\ref{fig:vqpu}. No backend name or physical
qubit address appears. In the bottom panel, the platform layer's \texttt{plan()}
(Algorithm~\ref{alg:plan}) turns the program into an \emph{ExecutionPlan} of
three units. Alice's stream is one unit; Bob is split at each conditional into
\texttt{unit\_1} and \texttt{unit\_2}, chained by the intra-circuit edge of
pass~1 (grey) and tied to Alice's measurements by the two cross-circuit edges
of pass~2 (orange, dashed), which place both of Bob's units after Alice's
measurements on the time axis. The backend map co-locates the two circuits as
virtual QPUs on one backend, and the executor dispatches units to a simulator
or to a partitioned chip.}
\label{fig:arch}
\end{figure}

QQLink is organized into two layers, application and platform, connected by
a planning interface (Fig.~\ref{fig:arch}). The application layer captures
\emph{what} a workflow computes in terms of workflow nodes and the
dependencies between them; the platform layer decides \emph{where and how}
it executes. No component in the application layer references a physical
backend.

\subsection{Workflow programs and circuits}

A developer constructs a \emph{WorkflowProgram} containing one or more named
\emph{WorkflowCircuits}, one per workflow node. Each circuit records a
sequence of instructions (gates, measurements, conditional blocks, Bell-pair
requests, and cross-circuit CNOT operations) without referencing any physical
backend. A circuit declares $n$ qubits; the platform layer maps them to
physical indices later. The application layer is purely descriptive. A
circuit does not allocate qubits, open backend connections, or evaluate
conditionals. It appends instructions to a list. The same circuit object can
therefore be inspected, serialized, or transformed by the orchestrator without
side effects, and several planning strategies can be applied to one workflow
without re-authoring it.

\subsection{Deferred measurement references}

The abstraction that makes cross-node protocols expressible is the
\emph{MeasureRef}. When a circuit measures a qubit, the call returns a
lightweight typed handle carrying the circuit's identity and a string label.
Comparing the handle with a classical value (\texttt{m0 == 1}) produces a
condition, which is passed to \texttt{if\_then(condition, gates)} on
\emph{any} circuit in the program, including a different one. Because the
condition encodes its source circuit, a single \texttt{if\_then} call on Bob's
circuit expresses ``apply $X$ to my qubit~0 if Alice's measurement
\texttt{m0} returned~1.'' This is a cross-circuit conditional feedforward that
requires no imperative branching and no explicit messaging. Script~\ref{lst:teleport1}
shows the complete single-qubit teleportation program, which we use as the
running example in Section~\ref{sec:planning}.

A second, gate-free use of the same handle is \texttt{wait\_for(ref)}. It
records that the circuit's subsequent instructions must be ordered after the
referenced measurement without applying any gate. The planner derives
ordering from the qubits each instruction acts on
(Section~\ref{sec:planning}), so \texttt{wait\_for} is needed only for
constraints that are not visible in the instruction streams, for example a
decision that must follow an outcome without touching a qubit. The planner
treats it as a dependency edge; executors and the hardware lowering treat it
as a no-op.

\begin{lstlisting}[style=qqpy,float=tp,caption={Single-qubit teleportation: the running example. Bob's corrections reference Alice's outcomes by handle; no backend appears.},label={lst:teleport1}]
def build_teleportation():
    wp = WorkflowProgram("teleportation")
    alice = wp.create_circuit("alice", num_qubits=2)
    bob = wp.create_circuit("bob", num_qubits=1)
    bell = wp.declare_entanglement("alice", 1, "bob", 0)
    alice.gate("RY", [0], params=[pi / 3])   # state to send
    alice.request_bell_pair(1, bell)
    alice.gate("CNOT", [0, 1])               # Bell measurement
    alice.gate("H", [0])
    m0 = alice.measure(0, label="m0")
    m1 = alice.measure(1, label="m1")
    bob.if_then(m1 == 1, [("X", [0])])       # cross-circuit
    bob.if_then(m0 == 1, [("Z", [0])])       # feedforward
    return wp
\end{lstlisting}

\subsection{Declarative entanglement}

Entanglement follows the same philosophy. The program registers Bell-pair
requirements between circuit--qubit pairs via
\texttt{declare\_entanglement()}, producing \emph{EntanglementLink} objects
that carry no placement information. Each link names two circuit--qubit pairs
and is assigned a unique identifier that circuits reference when they call
\texttt{request\_bell\_pair()}. This call emits a \texttt{bell\_pair}
instruction into the circuit's stream, recording the link identifier and the
local qubit index, but performs no quantum operation. Similarly,
\texttt{cross\_cnot()} records a non-local CNOT request that names the target
circuit and qubit by identifier rather than by physical address. In both cases
the instruction is self-describing, and the platform layer determines how and
where each operation is realized at dispatch time.

\subsection{Ordering semantics}
\label{sec:ordering}

``Declarative'' in QQLink refers to entanglement and feedforward. The program
states which Bell pairs exist and which outcome drives which correction, and
the planner derives the messages and the order. Operation order is not
declarative. Every instruction carries its position in a counter shared by
all circuits of a program, and the planner orders two operations on the same
qubit, including the two endpoints of a Bell-pair or cross-CNOT request, the
way they were written. The model is therefore closer to a single-threaded
tracing interface with namespaced registers than to a set of independently
authored node programs. The program is written in one thread, and swapping
two calls on different circuits that touch a common qubit changes the plan.
This is a deliberate choice. It makes the plan a deterministic function of
the program text (P1 in Section~\ref{sec:planning}), and it lets the planner
recover cross-circuit order from qubit footprints without asking the
developer for explicit synchronization. Authoring node programs concurrently,
by separate parties or processes, would need an explicit merge of their
instruction streams, or a partial order stated through \texttt{wait\_for}
alone, and is outside the present model (Section~\ref{sec:limitations}).
Execution, by contrast, need not follow the authoring thread. Units whose
footprints are disjoint have no edge between them and may run concurrently on
different nodes, which Section~\ref{sec:separate} exercises with one
controller process per node.

\section{Planning}
\label{sec:planning}

The platform layer translates the declarative description into concrete
execution. Its central component, the \emph{Orchestrator}, consumes a
\texttt{WorkflowProgram} and produces an \emph{ExecutionPlan}, an ordered
list of \emph{ExecutionUnits} with dependency edges, backend assignments, and
planning metadata. Algorithm~\ref{alg:plan} states the procedure.

\begin{algorithm}[t]
\caption{Two-pass plan construction}
\label{alg:plan}
\begin{algorithmic}[1]
\Require WorkflowProgram $P$ with circuits $c_1,\dots,c_m$; backend catalog $B$
\Ensure ExecutionPlan $(U, E, \beta)$: units, dependency edges, backend map
\State $U \gets \emptyset$;\; $E \gets \emptyset$;\; $\Lambda \gets \emptyset$ \Comment{$\Lambda$: map from (circuit, label) to unit}
\ForAll{circuits $C_i$} \Comment{Pass 1: segmentation}
  \If{every qubit of $C_i$ is measured, no operation acts on a qubit after its measurement, and $C_i$ has no conditional}
    \State one segment for the whole stream (whole-circuit mode)
  \Else
    \State split the stream at every measurement, conditional, Bell-pair, and cross-CNOT boundary into segments $s_{i,1},\dots,s_{i,k_i}$
  \EndIf
\EndFor
\State $\mathit{last} \gets \emptyset$ \Comment{Pass 1b: units and footprint edges in authoring order}
\ForAll{instructions $x$ of $P$ in authoring order, in segment $s$}
  \State $u \gets$ the open unit of $s$, or a new one if none is open
  \If{some $(c,q) \in \mathrm{footprint}(x)$ has $\mathit{last}[(c,q)] \notin \{\bot, u\}$ written after the first instruction of $u$}
    \State close $u$;\; $u \gets$ new open unit of $s$ \Comment{refinement for (P3)}
  \EndIf
  \State add $x$ to $u$;\; $U \gets U \cup \{u\}$
  \If{$x$ measures with label $\ell$} $\Lambda[(\mathrm{circuit}(x), \ell)] \gets u$ \Comment{a repeated pair is rejected} \EndIf
  \ForAll{$(c, q) \in \mathrm{footprint}(x)$} \Comment{both endpoints of Bell-pair and cross-CNOT requests}
    \If{$\mathit{last}[(c,q)] \notin \{\bot, u\}$} $E \gets E \cup \{(\mathit{last}[(c,q)], u)\}$ \EndIf
    \State $\mathit{last}[(c,q)] \gets u$
  \EndFor
\EndFor
\ForAll{units $u \in U$} \Comment{Pass 2: measurement-reference edges}
  \ForAll{conditionals and \texttt{wait\_for} in $u$, referencing $(c', \ell)$ on any circuit $c'$, local or remote}
    \If{$\Lambda[(c', \ell)] \neq u$} $E \gets E \cup \{(\Lambda[(c', \ell)], u)\}$ \EndIf
  \EndFor
\EndFor
\State group circuits transitively by shared entanglement links and cross-circuit references; for each group choose $\beta \in B$ with sufficient capacity
\State \Return $(U, E, \beta)$
\end{algorithmic}
\end{algorithm}

\paragraph{Pass 1: segmentation and qubit footprints.}
The orchestrator walks each circuit's instruction stream. A circuit whose
qubits are all measured, which applies no operation to a qubit after
measuring it, and which contains no conditional is emitted as a single whole-circuit unit;
otherwise the stream is segmented at measurement, conditional, Bell-pair,
and cross-CNOT boundaries, and each segment becomes one execution unit
unless the refinement described below splits it. A
conditional both starts and ends a segment, so it is always a unit of its
own. The gates authored before it are not held back until its outcome is
known, and the gates after it cannot run before it. Dependencies between
units are then derived from qubit footprints; we call these footprint edges. Every instruction carries the position at which it was
authored, from a counter shared by all circuits of a program, and the planner
replays the instructions in that order, recording for each (circuit, qubit)
pair the last unit that acted on it; a unit depends on the last unit that
acted on any qubit in its footprint. Bell-pair and cross-CNOT requests act on
both endpoints, so a request on Alice's side whose other end is Bob's qubit
is ordered after Bob's last operation on that qubit, and Bob's next operation
on it is ordered after the request. Segments of one circuit that act on
disjoint qubits receive no edge between them. A \texttt{wait\_for}
instruction acts as a barrier, and its footprint is every qubit of its circuit. Because a request also acts on the other circuit's qubit, the
instructions of a segment need not be contiguous in authoring order with
respect to the qubits it touches, because Bob may act on Alice's qubit between two of
Alice's gates on it. The same replay therefore refines segments into units.
An instruction opens a new unit of its segment when one of its qubits was
last touched by another unit after the first instruction of the current
one. Without this rule, two circuits that issue requests on each other's
qubits can produce a dependency cycle. As units are formed, the
measurement-label map $L$ associates every labeled measurement, keyed by
its circuit and label, with the unit that produces it; two circuits may use
the same label, and a label repeated within one circuit is rejected.

\begin{figure}[t]
\centering
\includegraphics[width=\linewidth]{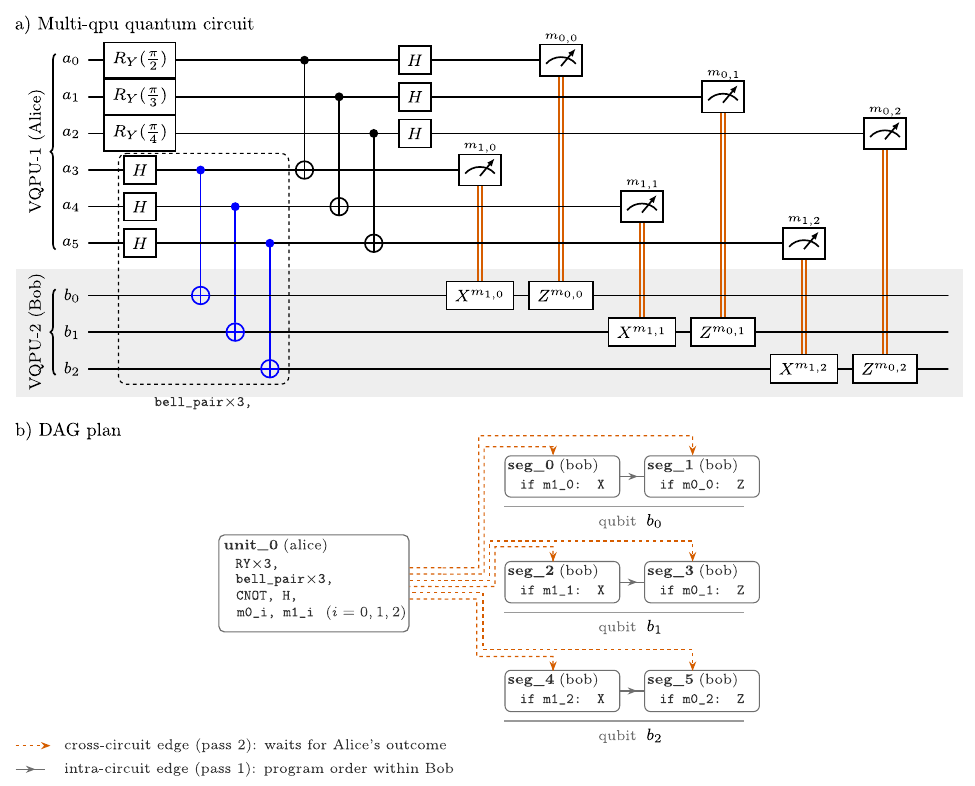}
\Description{Panel (a) is a nine-qubit circuit diagram for three-qubit teleportation with Alice's six qubits above Bob's three, three Bell-pair CNOTs crossing between them, and orange classical wires from Alice's six measurements to Bob's corrections. Panel (b) is a graph with one unit for Alice fanning out by dashed orange edges to six units for Bob, arranged as three independent two-unit chains, one per qubit.}
\caption{Three-qubit register teleportation (Script~\ref{lst:teleport}) as
authored and as planned. (a)~The multi-QPU circuit on the nine allocated
qubits, with Alice's data qubits $a_0$--$a_2$ and Bell halves $a_3$--$a_5$
(VQPU-1), Bob's $b_0$--$b_2$ (VQPU-2). The three Bell pairs (blue CNOTs) are
the only gates that cross the sector boundary; each of Alice's measurement
outcomes drives one correction on Bob through a classical wire (orange). The
corrections are drawn in the order the planner emits them. (b)~The execution
plan as a DAG produced by Algorithm~\ref{alg:plan}.
Alice's single unit produces six labeled outcomes; each of Bob's six units
depends on it through a cross-circuit edge (orange, dashed) from pass~2. The
footprint analysis of pass~1 chains only units that act on the same qubit
(grey). The three per-qubit chains are independent, so the critical path is
three units for any register size, although the work in Alice's unit grows
with the register.}
\label{fig:dag}
\end{figure}

\paragraph{Pass 2: measurement-reference edges.}
The second pass scans all units for conditional and \texttt{wait\_for}
instructions. For each one it looks up the unit that produces the referenced
measurement in the label map and inserts a measurement-reference edge
(measurement edge for short) from that unit, whether the measurement was
taken on the same circuit or on another one. The
footprint edges of pass~1 do not cover a local reference. In a circuit that
measures qubit 0 as $m$ and then applies $X$ to qubit 1 if $m = 1$, the
conditional's footprint is disjoint from the measured qubit, and only the
measurement-reference edge orders the two. A single pass would fail whenever a
conditional on one circuit references a measurement from a circuit that has
not yet been processed. Deferring the resolution of references to a pass that runs
after all units have been formed handles arbitrary reference patterns
regardless of circuit processing order.

\paragraph{Backend selection.}
Backend selection does not depend on the two passes; the implementation
runs it before pass~1. The orchestrator groups circuits that are connected
by entanglement links or cross-circuit references, so that every Bell pair
can be realized through a co-located cross-virtual CNOT and every
cross-circuit reference is resolved on one backend, and then selects a
backend with sufficient qubit capacity for each group. The grouping is transitive. If Alice shares a
link with Bob and Bob shares a link with Carol, all three are placed on the
same backend.

\paragraph{Running example.}
For Script~\ref{lst:teleport1}, Alice's circuit becomes one unit (state
preparation, Bell-pair request, CNOT, Hadamard, two measurements) and Bob's
becomes two conditional units. The second pass links each of Bob's units to
Alice's measurement unit, yielding a directed acyclic graph (DAG) of units
that dispatches the corrections only after the outcomes are available.
Fig.~\ref{fig:dag} shows the circuit and the graph for the three-qubit
register case. Alice's single unit fans out to three independent two-unit
chains on Bob's side, one per teleported qubit, because corrections on
different qubits have disjoint footprints.

\paragraph{Guarantees and cost.}
Three properties of the plan follow from the construction and are used in the
evaluation.
\begin{enumerate}
\item[(P1)] \emph{Determinism.} Both passes depend only on the instruction
  streams and the backend catalog, not on runtime state. Repeated planning of
  the same program yields the same units, edges, and backend assignments.
\item[(P2)] \emph{Dependency completeness.} Every conditional and
  \texttt{wait\_for}, on the producing circuit or on another one, references
  a labeled measurement (a measurement without a label receives one
  automatically); pass~1 records every labeled measurement before
  pass~2 begins, and a reference whose measurement no unit produces is
  rejected at planning; hence every reference receives an edge from its
  producing unit, independent of the order in which circuits were authored or
  processed. Every pair of units that act on a common (circuit, qubit) pair
  is connected by a chain of footprint edges, since the replay links each
  access to the previous one.
\item[(P3)] \emph{Acyclicity.} Let $f(u)$ be the authoring position of the
  first instruction of unit $u$. Every edge $(u, v)$ satisfies
  $f(u) < f(v)$. A footprint edge arises from consecutive accesses $x \in u$
  and $y \in v$ to one (circuit, qubit) pair, with $x$ before $y$. If $x$
  came after $f(v)$, the refinement would have opened a new unit at $y$, so
  $x$ precedes $f(v)$ and $f(u) \le x < f(v)$. A measurement edge points from
  the unit producing a label to the unit of a conditional or
  \texttt{wait\_for} that consumes it; a \texttt{wait\_for} is recorded as a
  conditional without gates, so like every conditional it starts its unit, and
  it can only be recorded after the measurement whose handle it consumes, so
  again $f(u) < f(v)$. Hence $f$ increases strictly along every edge and
  the graph is acyclic, so a dispatcher that receives every completion never
  waits on a cycle. Acyclicity does not exclude a stall caused by a lost
  message, an unavailable resource, or a failed service. The co-located
  dispatcher can bound each unit with a timeout and report the failure, while
  the separate executors have no timeout (Section~\ref{sec:reliability}). The dispatcher's
  topological sort also checks acyclicity at run time and rejects a plan
  with a cycle.
\end{enumerate}
We measure the size of a program by $S = \sum_{x \in I} |\mathrm{footprint}(x)|$,
the total length of its instructions' footprints, which counts every qubit of
every gate in a conditional body, both endpoints of a Bell-pair or
cross-CNOT request, every qubit of a multi-qubit custom gate, and every qubit
of the circuit for a \texttt{wait\_for} or another instruction that acts as
a barrier. Both passes touch each footprint entry a constant number of times,
and the label map and the refinement give constant-time hashed lookups, so
planning takes $O(S + |E|)$ time. In every workflow and synthetic program of
this paper, no footprint has more than two entries and no instruction is a
barrier, so $S \le 2|I|$. Each footprint entry adds at most one footprint edge and
each measurement reference at most one measurement edge, so
$|E| \le S + |I|$ and planning is linear in the number of instructions. Dispatch requires a
topological order of the units; we compute it with Kahn's algorithm, using a
heap that releases ready units in authoring order, in
$O((|U| + |E|)\log|U|)$, so that execution follows the program as written
wherever the dependencies allow. Any topological order yields the same
result as the authoring order. Two units that no path connects act on
disjoint (circuit, qubit) pairs (P2), and neither reads a measurement the
other produces (P2), so their operations, measurements included, commute,
and every linear extension of the plan composes the same quantum
instruments. For each record of measurement outcomes, the post-measurement
state and the probability of the record are therefore independent of the
order. Section~\ref{sec:correctness} checks this statement by replaying
plans in random topological orders.

\section{Execution Mechanisms}
\label{sec:mechanisms}

The plan is dispatched by walking the dependency graph in topological order
and invoking a user-supplied callback for each unit. QQLink provides a default
\emph{WorkflowExecutor} that allocates qubits sequentially across vQPUs on a
shared physical backend and translates each instruction at dispatch time.
Standard gates and measurements map directly to QPU operations, and Bell-pair
and cross-CNOT instructions are realized through a cross-virtual CNOT on the
shared state. This section describes the mechanisms that operate across the
application--platform boundary.

\subsection{Per-instruction conditional dispatch}

Classical feedforward is fully declarative. The application writes
\texttt{bob.if\_then(m0 == 1, [("X", [0])])} where \texttt{m0} is a
\texttt{MeasureRef} from Alice's circuit; no imperative branch,
message-passing call, or backend name appears. Because the mechanism is driven
by the measurement label rather than by the consuming circuit, it generalizes
to chains, in which a measurement on $c_1$ can drive corrections on $c_2$, whose own
measurement can drive corrections on $c_3$, with the orchestrator resolving
the full dependency graph.

Conditions are evaluated per instruction at dispatch time. Every
conditional is its own unit (Section~\ref{sec:planning}), and the executor
reads the recorded outcome and applies the conditional gates only if it
matches; the unit is dispatched either way, so nothing is silently dropped
and the plan's bookkeeping (which unit produced which outcome, how many
attempts it took) stays complete. The gate-teleportation workflows of
Section~\ref{sec:demos} depend on this behavior, since their ancilla resets
are conditional gates interleaved with unconditional ones.

For this to be correct, a backend adapter must honor one contract. A
mid-circuit measurement must collapse the simulated state exactly once, and
subsequent operations must act on the branch that was observed. An adapter
that re-executes the accumulated circuit on every measurement and returns a
freshly sampled bit violates this contract; the corrections it applies are
then uncorrelated with the branch that later measurements sample, and every
feedforward protocol degrades to the statistics of uncorrected branches
(Section~\ref{sec:lessons}).

\subsection{Entanglement channels and virtual-QPU partitioning}

A \texttt{request\_bell\_pair()} instruction is recorded in the stream but not
realized until dispatch time, when the executor applies a Hadamard and a
cross-virtual CNOT to the partner qubit. A channel-selection policy chooses,
in order of precedence, among three paths, namely a \emph{co-located} path that
executes the CNOT directly on the shared state when both vQPUs share a
backend; a built-in \emph{simulated} channel when they do not; and a
\emph{provider-native} channel that delegates to a hardware-specific executor
(for example an ion-trap shuttle or a photonic link) when a physical quantum
network is available. The simulated channel is a control-flow model of a
link, not a physics model. It has a success probability, an optional Pauli
error injected on both endpoints, and a latency, and it reports every attempt
in the operation record. It cannot create entanglement between two
independent state vectors, so protocol outcomes obtained through it are not
meaningful; its purpose is to exercise placement, retry, and latency
accounting for workflows whose nodes live on different backends
(Section~\ref{sec:robustness}). Only the co-located path yields
protocol-level results in this paper.

The co-located path relies on virtual-QPU partitioning, which maps several
workflow nodes onto disjoint qubit ranges of one backend, for example $c_1$ on
$[0,k)$ and $c_2$ on $[k,n)$. Bell pairs and cross-virtual CNOTs between
partitions then act on the same state, modeling entanglement correlations
without a physical network or multiple simulator instances.
Fig.~\ref{fig:vqpu} (Introduction) shows both kinds of link on a partitioned
chip. The quantum path between sectors is the co-located path, and the
classical loop through the control system is the feedforward.

\subsection{Reliability controls and failure contract}
\label{sec:reliability}

The orchestrator dispatches a unit only after all its dependencies, on the
same circuit or another one, have completed. The runtime makes four
commitments about failure.
(i)~\emph{Retry eligibility.} A unit whose dispatch fails before any of its
instructions has completed is retried up to a configurable number of times
with optional backoff. A unit that fails after one or more instructions have
acted on the state is not retried, because re-dispatching it would apply its
first operations twice; the failure is propagated at once, together with the
results of the units that did complete. Recovery from such a failure needs a
restart of the workflow or the error-detection structure of the protocol.
(ii)~\emph{Missing references.} In the co-located executor, a conditional
whose measurement has not been recorded is an error, not a skipped
correction. With a complete plan (P2) this cannot happen, so the error
signals an executor that violates the plan. The separate executors evaluate a
missing outcome as no correction, a case that only the edge ablations of
Section~\ref{sec:separate} exercise.
(iii)~\emph{Ownership.} A Bell-pair request or cross-circuit CNOT is issued by
the unit of the requesting circuit and orders the other circuit's qubit
through its footprint; the remote circuit performs no matching operation.
(iv)~\emph{Delivery.} The co-located executors share one process and need no
messages, and their dispatcher can bound each unit with a timeout. The separate executors of Section~\ref{sec:separate} assume that
every completion message is delivered exactly once, with arbitrary delay;
they have no timeout, so a lost message would leave its consumers waiting,
and loss, duplication, and process failure are outside what we test.
Channel operations carry their own retry budget and timeout, so a failed link
attempt is repeated without re-dispatching the unit. Program, task, and queue
state can be serialized to JSON and restored, which enables post-hoc trace
analysis and restart of the classical orchestration state; the quantum state
of a simulator or device is not checkpointed. The accept/reject workflow of
Section~\ref{sec:demos} instead retries at the application level, in a loop
that plans and dispatches each attempt through the orchestrator.

\subsection{Hardware path}
\label{sec:hwpath}

The hardware path submits one dynamic circuit per workflow. A plan is
lowered by emitting its units in topological order on a global register in
which each circuit occupies a contiguous block; labeled measurements receive
one classical bit each and conditionals become \texttt{if\_test} blocks on
that bit. The same lowering is used in
Section~\ref{sec:cost} to compare against hand-written circuits and in
Section~\ref{sec:robustness} for noise sweeps, so the circuit that reaches
the device is the circuit those experiments analyze. The hardware runner
then chooses the vQPU
\emph{sector}, the physical qubits that host the lowered register, either by
letting the Qiskit preset pass manager score layouts with the backend's
calibration data or by an exhaustive search for the connected line of qubits
with the lowest sum of two-qubit-gate and readout errors. Along that line,
the qubits are either placed in Cuthill--McKee order of their interaction
graph, which can interleave nodes, or, as in the main runs reported here,
node by node, so that each virtual QPU occupies a contiguous stretch and the sectors
are disjoint. Readout-calibration circuits on the same physical qubits are
bundled into the same Sampler job, so that a tensored readout-error
inversion of the application-level check bits is available as a
post-processing step; feedforward bits are consumed on the device and cannot
be mitigated. The numbers reported in Section~\ref{sec:hardware} are raw. Every run record
stores the UTC submission and completion times, the backend name and
processor family, the timestamp of the calibration the device reported, the
$T_1$, $T_2$, readout and single-qubit gate errors of every physical qubit
used and the two-qubit gate error of every coupler between them, the
physical qubits assigned to each circuit, the transpiled depth and two-qubit
count, the job identifier and its timing metrics, the raw counts, and the
versions of the software stack including the QQLink commit. None of this is
visible at the application layer. The same program object is planned once
and executed on a simulator, on a fake backend with the device's noise
model, or on the device.

The fake-backend dry run uses Qiskit Aer's simulator built from the
backend's calibration (\texttt{AerSimulator.from\_backend}), either from the
snapshot shipped with Qiskit's fake \texttt{ibm\_kingston} or from the
calibration that the device reports at the time of the run. Its noise model
contains, for every gate, a depolarizing error together with thermal
relaxation over the gate's duration; for every qubit, the reported readout
error; and an error on reset. A measurement carries only its readout error,
with no relaxation during its duration ($2.3\,\mu$s on the fake backend), and the circuits are not
scheduled, so a qubit that waits while another qubit is measured or while a
conditional is resolved accrues no error. Explicit \texttt{delay}
instructions do receive thermal relaxation from $T_1$ and $T_2$. The dry run
therefore models gate and readout infidelity but not the time structure that
feedforward introduces (idle intervals and conditional execution), which the
probes of Section~\ref{sec:hardware} examine directly.

\section{Implementation}
\label{sec:impl}

\begin{figure}[t]
\centering
\includegraphics[width=\linewidth]{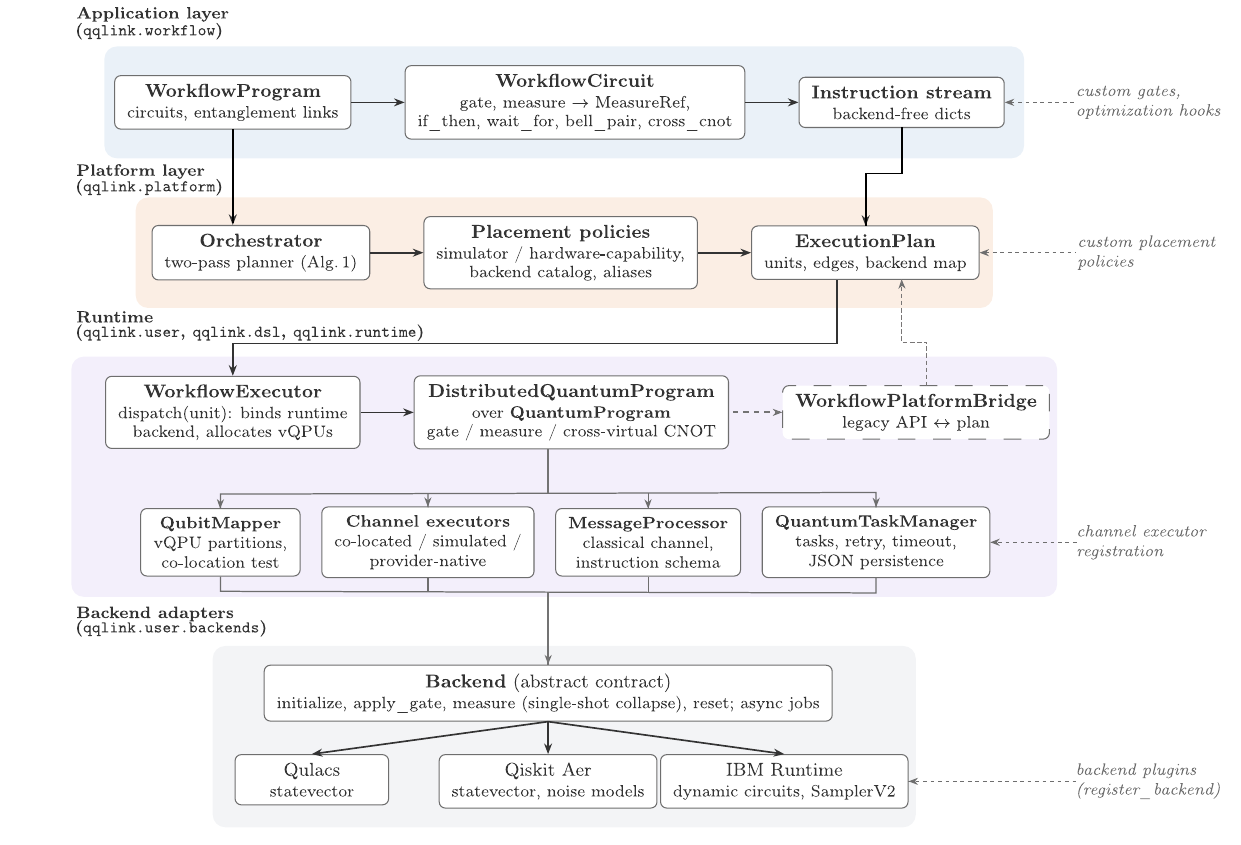}
\Description{Layered block diagram of the QQLink Python packages, namely an application layer, a platform layer with the planner and execution plan, the workflow executor, the message-driven runtime with four subsystems on one bus, and a backend contract leading to simulators and hardware. Arrows trace a program from authoring to execution, and dashed leaders mark extension APIs.}
\caption{Software architecture of QQLink. Bands are Python packages. The
application layer records backend-free instruction streams; the platform
layer plans them into an \texttt{ExecutionPlan}; the \texttt{WorkflowExecutor}
binds each unit to a runtime backend and drives the pre-existing
message-driven runtime and its four subsystems; all execution reaches
hardware or simulators through one \texttt{Backend} contract. Black arrows
are the path a program takes from authoring to execution; grey arrows are
delegation inside the runtime (the four subsystems join one bus into the
\texttt{Backend} contract). Dashed arrows mark the compatibility bridge
between the legacy API and the workflow path; dashed leaders on the right
mark the registration APIs through which the framework is extended.}
\label{fig:software}
\end{figure}

Fig.~\ref{fig:software} maps the concepts of Sections~\ref{sec:model}
to~\ref{sec:mechanisms} onto the software. QQLink is a Python package
organized in four layers that correspond to the four bands of the figure.

\paragraph{Layering.}
\begin{sloppypar}
The application layer (\texttt{qqlink.workflow}) holds the
\texttt{WorkflowProgram}, \texttt{WorkflowCircuit}, \texttt{MeasureRef}, and
\texttt{EntanglementLink} classes. A circuit's methods append
backend-free instruction dictionaries to a stream; nothing in this layer
imports a simulator or a provider SDK. The platform layer
(\texttt{qqlink.platform}) contains the \texttt{Orchestrator}, the
\texttt{ExecutionPlan} and its units, and the placement policies, which
implement one \texttt{PlacementPolicy} protocol (a simulator policy and a
hardware-capability policy are provided). The runtime
(\texttt{qqlink.user}, \texttt{qqlink.dsl}, \texttt{qqlink.\allowbreak runtime}) predates
the workflow layer. It is the message-driven core of the framework, in which
\texttt{DistributedQuantumProgram} wraps \texttt{QuantumProgram} and delegates
to four subsystems, the \texttt{QubitMapper} that owns virtual-QPU
partitions and answers the co-location test, the channel executors, the
\texttt{MessageProcessor} that validates and applies classical instruction
payloads (not exercised by the experiments reported here), and the \texttt{QuantumTaskManager} that provides task
dependencies, retries, timeouts, and JSON persistence. The backend layer
(\texttt{qqlink.user.\allowbreak backends}) implements one abstract \texttt{Backend}
contract for Qulacs, Qiskit Aer, and IBM Runtime.
\end{sloppypar}

\paragraph{From plan to backend.}
The \texttt{WorkflowExecutor} is the only component that spans the platform
and runtime layers. It reads the runtime backend name resolved during
planning, registers one shared backend, allocates each circuit a contiguous
virtual-QPU block on it, and exposes a \texttt{dispatch(unit)} callback that
translates each instruction of a unit into runtime calls. Gates and
measurements become \texttt{apply\_gate} and \texttt{measure} on the
circuit's vQPU, Bell-pair and cross-CNOT instructions become
\texttt{execute\_cross\_virtual\_cnot}, and conditionals are evaluated
against the measurements the executor has recorded. Because the executor is
a callback, a different executor can be substituted without touching the
planner, for example the two-backend executor of Section~\ref{sec:robustness}, the
dynamic-circuit lowering of the hardware path, and the process-per-node
executor of Section~\ref{sec:separate}, which does not use the dispatcher at
all and gives each node only its own units. The workflow layer was added on
top of the existing runtime rather than replacing it; a compatibility bridge
to the earlier API (dashed in Fig.~\ref{fig:software}) exists for parity
tests and is not used in this paper.

\paragraph{Adapter contract.}
Every backend implements \texttt{initialize\_state}, \texttt{apply\_gate},
\texttt{measure}, and \texttt{reset}, plus an asynchronous job interface
for hardware, and must honor the single-shot measurement contract of
Section~\ref{sec:mechanisms}; the contract is tested by running the
feedforward workflows on every adapter.

\paragraph{Extension points.}
Four registration APIs extend the framework without modifying it. They add custom
gates and optimization hooks on the program, placement policies on the
orchestrator, channel executors (simulated, provider-native, or
user-supplied) on the runtime, and backend plugins through
\texttt{register\_backend}. The provider-native channel executor is the
intended attachment point for physical inter-chip links; it exists today as
a scaffold with deterministic executor selection and no hardware behind it.

\section{Application Workflows}
\label{sec:demos}

We validate QQLink with seven protocol workflows of increasing complexity
and one control loop. Superdense
coding, multi-qubit teleportation, and a two-node Grover search exercise
entanglement declaration, cross-circuit feedforward, and multi-round gate
teleportation with ancilla reuse. Entanglement swapping is the
quantum-repeater primitive and tests feedforward across a chain of three
nodes joined by two links; a distributed Bernstein--Vazirani algorithm splits an oracle across
two nodes; a three-node Toffoli gate tests tripartite orchestration; a
three-node repetition code extracts syndromes across nodes and corrects an
injected error by feedforward. A single-node accept/reject loop, with no
cross-node edge, tests the task layer. All eight are authored
entirely at the application layer, with no backend name in any circuit
definition; the seven protocol workflows are the ones run on hardware.

\subsection{Multi-channel superdense coding}

Superdense coding~\cite{bennett1992} transmits two classical bits through one
qubit and a shared Bell pair. We generalize the protocol to $k$ parallel
channels, in which Alice and Bob each hold $k$ qubits and share $k$ independent Bell
pairs, so $2k$ classical bits are communicated in one round. For each channel,
Alice prepares her half of the Bell pair via \texttt{request\_bell\_pair},
encodes two bits by applying one of $\{I, X, Z, Y\}$, and decodes with a
\texttt{cross\_cnot} onto Bob's qubit followed by a Hadamard and a
measurement of her own; Bob measures his qubit. We sweep
$k \in \{1,2,4\}$, from 2 to 8 physical qubits. Because the protocol involves
no conditional feedforward, it isolates the entanglement declaration and
channel realization mechanisms. Script~\ref{lst:dense} in Appendix~\ref{app:scripts} shows the construction;
the loop body is identical for every channel. This protocol is also the
cleanest entry point for same-chip hardware experiments because it requires
only terminal measurement.

\subsection{Multi-qubit teleportation}

Quantum teleportation~\cite{bennett1993} transfers an unknown state using a
shared Bell pair and two classical bits of feedforward. We generalize to an
$n$-qubit register, in which Alice holds $2n$ qubits ($n$ data and $n$ Bell halves),
Bob holds $n$ qubits, and each qubit is teleported through its own link, Bell
measurement, and correction round. Data qubit $i$ is prepared as
$R_Y(\pi/(i{+}2))\,|0\rangle$ so that every qubit carries a distinct known
state. After Alice measures her data and Bell qubits, her two outcomes per
qubit drive Pauli corrections on Bob's side through \texttt{if\_then} calls
whose conditions reference \texttt{MeasureRef}s from Alice's circuit. The
orchestrator resolves these references during planning, inserting $2n$
cross-circuit edges without any coordination code from the developer
(Script~\ref{lst:teleport}).

We sweep $n \in \{1,3,5,7\}$, from 3 to 21 physical qubits. The sweep
separates two costs. The planning-level resources (units and cross-circuit
edges) grow linearly with $n$, whereas the statevector simulation cost grows
exponentially with the total qubit count $3n$. This provides a baseline
against which hardware experiments can distinguish orchestration overhead
from noise-induced degradation.

\begin{lstlisting}[style=qqpy,caption={Multi-qubit teleportation with cross-circuit feedforward.},label={lst:teleport}]
def build_multi_teleportation(n):
    wp = WorkflowProgram("multi_teleportation")
    alice = wp.create_circuit("alice", num_qubits=2*n)
    bob = wp.create_circuit("bob", num_qubits=n)
    for i in range(n):
        alice.gate("RY", [i], params=[pi / (i + 2)])
    for i in range(n):
        a_data, a_bell, b_bell = i, n + i, i
        link = wp.declare_entanglement(
            "alice", a_bell, "bob", b_bell,
            link_id=f"bell_{i}")
        alice.request_bell_pair(a_bell, link)
        alice.gate("CNOT", [a_data, a_bell])
        alice.gate("H", [a_data])
        m0 = alice.measure(a_data, label=f"m0_{i}")
        m1 = alice.measure(a_bell, label=f"m1_{i}")
        bob.if_then(m1 == 1, [("X", [b_bell])])
        bob.if_then(m0 == 1, [("Z", [b_bell])])
    return wp
\end{lstlisting}

\subsection{Entanglement swapping}
\label{sec:demo-swap}

Entanglement swapping is the elementary step of a quantum
repeater~\cite{coopmans2021,dahlberg2022}. Two nodes that never interact end
up sharing a Bell pair because a third node measures its halves of two
independent pairs. Alice--Charlie and Charlie--Bob share Bell pairs; Charlie
performs a Bell measurement on his two halves, and Bob applies
$X^{m_1} Z^{m_0}$ (Script~\ref{lst:swap} in Appendix~\ref{app:scripts}). Alice's and Bob's qubits are then
in $|\Phi^+\rangle$. The workflow adds a third node and a feedforward chain
in which both outcomes of one Bell measurement drive one remote qubit. The check
measures Alice's and Bob's qubits in the same basis, $Z$ or $X$, where a
Bell state yields equal outcomes with certainty.

\subsection{Two-node Grover search via gate teleportation}
\label{sec:demo-grover}

This workflow combines entanglement, feedforward, and ancilla management.
We implement a two-qubit Grover search~\cite{grover1996,main2025} in which
each search qubit resides on a separate node, each accompanied by a network
ancilla. The oracle and the diffusion operator both require a non-local
controlled-$Z$ (CZ) between the nodes. Each CZ is decomposed as
$\mathrm{CZ} = (I \otimes H)\,\mathrm{CNOT}\,(I \otimes H)$, where the CNOT is
realized through the Gottesman--Chuang gate teleportation
protocol~\cite{gottesman1999}. The two nodes consume a shared Bell pair on
their ancillas, perform two sequential mid-circuit measurements with
cross-circuit corrections, and reset the ancillas to $|0\rangle$ for reuse in
the next teleported gate.

Script~\ref{lst:gtcnot} shows the teleported CNOT as a reusable building
block. Each invocation declares a fresh entanglement link, requests a Bell
pair, and emits four \texttt{if\_then} conditionals, two Pauli corrections and
two ancilla resets. Two of the four reference a measurement on the other
node, so each teleported CNOT contributes two cross-circuit edges to the plan.
Script~\ref{lst:grover} in Appendix~\ref{app:scripts} composes two teleported CZ gates (oracle and
diffusion) into a complete Grover iteration. For two qubits a single
iteration amplifies the marked state to unit probability; we verify all four
marked states. Per search the workflow consumes two Bell pairs, generates
eight conditional instructions, and requires that ancillas be correctly reset
between teleported gates. It therefore exercises per-instruction conditional
evaluation, multi-round dependency resolution, and mid-circuit qubit reuse at
once.

\begin{lstlisting}[style=qqpy,caption={Teleported CNOT via Gottesman--Chuang gate teleportation. \texttt{ctrl} and \texttt{tgt} are circuits; \texttt{cname} and \texttt{tname} their names.},label={lst:gtcnot}]
def add_teleported_cnot(wp, ctrl, cname, tgt, tname,
                        prefix, lc):
    """CNOT(ctrl[0] -> tgt[0]) via gate teleportation.
    Uses ancilla qubits ctrl[1], tgt[1]."""
    link = wp.declare_entanglement(
        cname, 1, tname, 1, link_id=f"epr_{prefix}_{lc}")
    ctrl.request_bell_pair(1, link)
    lc += 1
    ctrl.gate("CNOT", [0, 1])
    m1 = ctrl.measure(1, label=f"{prefix}_m1")
    tgt.if_then(m1 == 1, [("X", [1])])
    tgt.gate("CNOT", [1, 0])
    tgt.gate("H", [1])
    m2 = tgt.measure(1, label=f"{prefix}_m2")
    ctrl.if_then(m2 == 1, [("Z", [0])])
    ctrl.if_then(m1 == 1, [("X", [1])])   # reset ancillas
    tgt.if_then(m2 == 1, [("X", [1])])
    return lc, m1, m2
\end{lstlisting}

\subsection{Distributed Bernstein--Vazirani}
\label{sec:demo-bv}

The Bernstein--Vazirani algorithm recovers a hidden string $s$ from one
query to the oracle $f(x) = s \cdot x$. Distributed variants have been
proposed~\cite{zhou2023bv}, and a distributed Deutsch--Jozsa and phase
estimation over teleported CNOTs were demonstrated between nodes 7\,km
apart~\cite{liu2024nonlocal}. We split the input register across two nodes.
Alice holds two input qubits and a network ancilla, Bob holds two input
qubits, the oracle target prepared in $|-\rangle$, and a network ancilla
(Script~\ref{lst:bv} in Appendix~\ref{app:scripts}). The oracle is a CNOT from each input with $s_i = 1$
into the target, local on Bob's side and teleported for Alice's bits through
the primitive of Script~\ref{lst:gtcnot} with explicit qubit indices. After a
Hadamard on every input, measuring the inputs returns $s$. The workflow is
the paper's second algorithm, and its oracle, unlike the Grover oracle, is
data-dependent, because the number of teleported gates equals the number of ones in
Alice's part of $s$.

\subsection{Three-node Toffoli gate}
\label{sec:demo-toffoli}

A Toffoli (CCNOT) gate spread over
three nodes tests whether the same primitives compose when entanglement links
and feedforward chains span every pair of a set of nodes. We use the standard
decomposition of the Toffoli gate into six CNOTs and single-qubit $T$ and
$T^\dagger$ rotations~\cite{nielsen2010}. Nodes A, B, and C each hold one
circuit qubit and one network ancilla; the six CNOTs span the pairs B--C,
A--C, and A--B, and each is realized by the teleported CNOT of
Script~\ref{lst:gtcnot}. Script~\ref{lst:toffoli} in Appendix~\ref{app:scripts} shows the construction.

This construction tests ancilla reuse across interactions with two
different remote nodes. Node C's ancilla is reused by four consecutive teleported gates
whose control sides alternate between A and B. The second gate's first
instruction on A, a Bell-pair request whose other end is C's ancilla, carries
no measurement reference to the first gate's final reset of that ancilla, so
a planner that ordered only by measurement references would be free to
schedule it first. The footprint analysis of pass~1 orders it correctly, because the
request acts on C's ancilla, so it depends on C's last operation on that
qubit, the reset. No synchronization instruction is needed. By the accounting
of the primitive, the workflow consumes six Bell pairs, performs twelve
mid-circuit measurements plus three final ones, and emits 24 conditional
gates, twelve of which reference a measurement on another node. Because node C participates in four of the six
teleported gates, its instruction stream interleaves conditionals sourced from
both A and B; pass~2 of Algorithm~\ref{alg:plan} resolves these regardless of
the order in which the three circuits are processed (P2). We verify the full
eight-row truth table, in which qubit C flips exactly when $A = B = 1$.

\subsection{Distributed repetition code}
\label{sec:demo-repetition}

Error correction across modules is one motivation for a distributed
programming model, and its smallest instance is a bit-flip
repetition code whose three physical qubits sit on three nodes
(Script~\ref{lst:repetition} in Appendix~\ref{app:scripts}). Node A prepares the logical input
($\lvert 0\rangle$, $\lvert 1\rangle$, or $\lvert +\rangle$) and encodes it
as $\alpha\lvert 000\rangle + \beta\lvert 111\rangle$ with two teleported
CNOTs into B and C. A bit flip is then injected on one node, or on none.
Two syndrome ancillas on node B collect $s_1 = Z_A Z_B$ (teleported CNOT
from A, local CNOT from B) and $s_2 = Z_B Z_C$ (local CNOT from B,
teleported CNOT from C). Decoding needs one classical AND, since the error is on B
exactly when $s_1 = s_2 = 1$. A QQLink conditional tests a single
measurement bit, so the AND is computed on a third ancilla by a CNOT from
the $s_2$ ancilla that runs only when $s_1 = 1$, and measured as $t_B$. The
corrections are then the single-bit conditionals $X_B^{t_B}$ on the syndrome
node and, on the other two nodes, $X_A^{s_1 \oplus t_B}$ and
$X_C^{s_2 \oplus t_B}$, each written as two conditionals on the same qubit.
The workflow consumes four Bell pairs and fourteen measurements and emits
22 conditional instructions, of which twelve reference a measurement on
another node, including the four decoder corrections on A and C, which are
sourced from syndromes measured on B. After correction the three data qubits agree
in the computational basis and equal the logical input when it is
$\lvert 0\rangle$ or $\lvert 1\rangle$; we verify all twelve combinations of
three inputs and four error locations, and the syndrome triple
$(s_1, s_2, t_B)$ identifies the error location in every case.

\subsection{Accept/reject control loop}
\label{sec:demo-magic}

The final workflow exercises the task layer rather than a quantum
primitive. It borrows the verify-and-discard pattern of resource-state
factories~\cite{bravyi2005} without performing any distillation. A
factory node prepares a $T$-state with a per-round rotation error
$\epsilon$ drawn uniformly from $[-a, a]$, applies a check, and accepts or
rejects the attempt on the check outcome; a rejected attempt is retried up to a
bound (Script~\ref{lst:magic} in Appendix~\ref{app:scripts}). The program declares a consumer node, but no
operation reaches it. The workflow has no cross-node edge (Table~\ref{tab:planning}
in Appendix~\ref{app:tables})
and is a single-node control loop. We evaluate two check circuits. The \emph{parity} check entangles the data qubit with an ancilla and measures
the ancilla; because the prepared state is an equal superposition in the
computational basis, this check rejects with probability $1/2$ independently
of $\epsilon$. The \emph{verification} check applies $T^\dagger$ and $H$ to
the data qubit and measures it, which rejects with probability
$\sin^2(\epsilon/2)$ and therefore responds to the injected error. With $R$
retries and a per-attempt acceptance probability $q$, the per-round
acceptance probability is $1-(1-q)^{R+1}$, with
$q = 1/2$ for the parity check and $q = 1/2 + \sin(a)/(2a)$ for the
verification check. The verification
check measures, and so consumes, the state it verifies, and the loop is in
effect the sampling of a biased coin with retries. The purpose of the
workflow is narrower. It shows that accept/reject decisions, bounded retries,
and per-round provenance are expressible with the same orchestrator and task
primitives used by the protocol workflows, and that the observed statistics
match the model. It does not exercise cross-node features, is not run on
hardware, and is counted separately from the seven protocol workflows above.

\subsection{Coverage of the requirements}

Taken together, the eight workflows cover R1--R4. Every protocol is authored
without backend names (R1). Teleportation, entanglement swapping, Grover,
Bernstein--Vazirani, Toffoli, the repetition code, and the accept/reject loop
rely on \texttt{if\_then} conditionals whose dependencies are resolved by
the orchestrator; all but the last do so across circuits (R2). The orchestrator
co-locates entangled circuit groups, transitively for the three-node
workflows, and provisions Bell pairs through the channel-selection policy of
Section~\ref{sec:mechanisms} (R3). The dependency-safe dispatch that
sequences a measurement before the corrections it drives operates
identically for the three-unit teleportation plan, the eighteen-unit Grover
plan, the 51- to 52-unit three-circuit Toffoli plan, and the 41- to 42-unit repetition-code
plan whose corrections on two nodes are driven by syndromes measured on the
third. An application-level loop over the orchestrator drives the
accept/reject retries (R4).

\section{Evaluation}
\label{sec:eval}

We evaluate QQLink as a software layer for portable multi-QPU workflow
semantics. The evaluation asks (Q1) whether planning is
deterministic and its output and cost scale as the model predicts; (Q2)
whether the workflow model produces correct protocol behavior on independent
simulators; (Q3) whether the plan suffices to run a workflow on separate
executors, one per node, that share no dispatcher and learn of each other's
progress only through delayed messages; (Q4) what the abstraction costs at
the circuit level relative to hand-written dynamic circuits; (Q5) whether
the same-chip IBM path realizes measurement-reference and feedforward
semantics on partitioned virtual-QPU backends; and (Q6) how the runtime
behaves under injected faults, channel failures, and gate noise. We do not
present QQLink as a hardware performance
benchmark or as a study of multi-chip distributed quantum execution. The six
questions map onto the four conditions of the introduction. Q2 and Q5 answer
(a), Q1 answers (b), Q4 answers (c), and Q3 answers (d); Q6 checks the
runtime controls.

\subsection{Methodology}
\label{sec:method}

Simulator experiments use two independent backends, Qulacs
0.6.14~\cite{qulacs2021} (statevector) and Qiskit Aer 0.17.2 with Qiskit
2.5.2, which mirrors the IBM transpilation and execution stack, under Python
3.11 on a Linux workstation (Intel Xeon E5-2680~v4, 28 cores, 62\,GB RAM).
Every number in this section is produced by scripts included in the
review snapshot and in the planned release (Data and Code Availability); each script writes a JSON record from which the tables are
generated. Correctness is established by exhaustive forced-branch execution
on the runtime path (Section~\ref{sec:correctness}). In addition, the
end-to-end runs on both simulators, which are stochastic in their
Bell-measurement outcomes, are repeated with fixed seeds, with 16 trials
per configuration with randomly drawn messages or inputs, and for
teleportation a deterministic per-qubit check in which the inverse
preparation $R_Y(-\pi/(i{+}2))$ is applied to Bob's qubit $i$ after the
corrections and the qubit is measured, which must yield 0. For entanglement
swapping, Alice's and Bob's qubits are measured in the same basis ($Z$ or
$X$, 16 trials each) and must agree; for Bernstein--Vazirani, all 16 hidden
strings are run four times each and the measured string must equal $s$;
for the repetition code, each of the twelve (logical input, error location)
cases is run four times and the three data qubits, measured in the
computational basis, must agree and equal the logical input when it is
$\lvert 0\rangle$ or $\lvert 1\rangle$. Noise sweeps use
Aer with a depolarizing channel of strength $p$ on two-qubit gates, $p/10$ on
single-qubit gates, and a readout error that flips each measured bit with
probability $p$, the same for 0 and 1, at 4000 shots. Cost
comparisons transpile on a Qiskit \texttt{GenericBackendV2} with linear
coupling and control-flow support at optimization level~1 with a fixed
transpiler seed. Hardware runs execute on one IBM backend at a time, partitioned
into vQPUs through the path of Section~\ref{sec:hwpath}, with 4096 shots per
circuit and three independent repeats per configuration, in two sessions a
week apart on each of \texttt{ibm\_kingston} and \texttt{ibm\_fez}; the per-run
records (calibration snapshot, physical qubits, job identifiers, counts) are
included in the review snapshot and in the planned release.

\subsection{Planning determinism, structure, and cost (Q1)}
\label{sec:overhead}

\begin{figure}[t]
\centering
\includegraphics[width=0.6\linewidth]{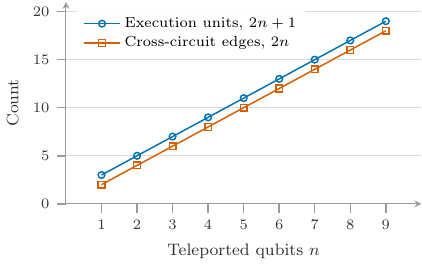}
\Description{Line plot against the number of teleported qubits n. Execution units rise as 2n+1 and cross-circuit edges as 2n, both straight lines, while the critical path stays flat at three units.}
\caption{Planning resources for $n$-qubit teleportation. Execution units grow
as $2n{+}1$ and cross-circuit edges as $2n$; both are exact over the sweep,
so plan size is linear in the number of teleported qubits, while the critical
path is three units throughout.}
\label{fig:resources}
\end{figure}

Appendix~\ref{app:tables} (Table~\ref{tab:planning}) lists the plans
produced by the orchestrator for each workflow across its sweep. Planning each program five times from
fresh orchestrator instances yields byte-identical serialized plans in every
case, as (P1) predicts. Dense coding plans to $k{+}1$ units, namely Alice's
whole-circuit unit and one measurement unit per channel on Bob's side, each
joined to Alice's by a footprint edge. Bob's side splits into one unit per
channel because Alice's cross-circuit CNOT on channel $i$ acts on Bob's qubit
$i$ before Bob measures it, and her requests on later channels act on Bob's
qubits in between; the refinement of Section~\ref{sec:planning} separates
them. Teleportation plans to $2n{+}1$ units with $2n$
cross-circuit edges, one per correction, $n$ intra-circuit edges (the $X$
then $Z$ on each of Bob's qubits), and a critical path of three units for
every $n$, because the per-qubit chains have disjoint footprints
(Fig.~\ref{fig:dag}). Entanglement swapping plans to five units across
three circuits with a critical path of five, because the plan is a single chain
from the Bell-pair requests through Charlie's measurement and Bob's
corrections to the check.
Grover plans to 18 units with a critical path of 11, Bernstein--Vazirani
(with two teleported terms) to 19 units with a critical path of 11, and the
Toffoli workflow to 51 or 52 units with 23 cross-circuit edges and a
critical path of 30; the critical paths of these three are dominated by the
sequential reuse of the network ancillas. The Toffoli plan has 52 units when
node A's input is 1, because the refinement then separates A's input
preparation from its first Bell-pair request. The repetition code plans to 41 or 42 units, depending on
whether an error is injected and where, with 18 cross-circuit edges and a
critical path of 22 or 23; its four teleported CNOTs all reuse node B's
network ancilla, which again sets the critical path. Fig.~\ref{fig:resources} plots the teleportation planning
resources over the extended sweep $n = 1,\dots,9$; units and edges are
exactly linear while the critical path stays constant.

\begin{figure}[t]
\centering
\includegraphics[width=\linewidth]{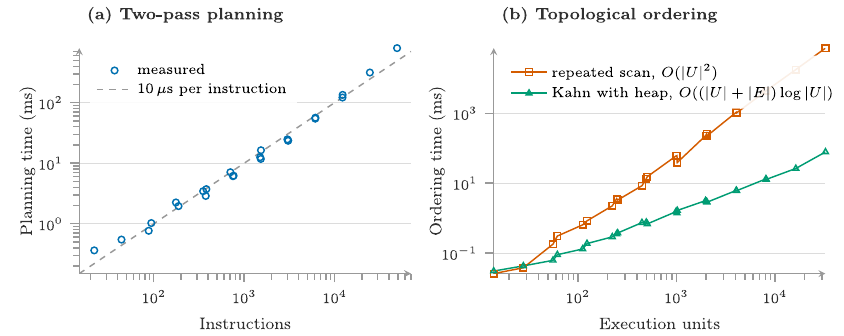}
\Description{Two plots. (a) Planning time against the number of instructions follows a straight line close to a 10 microseconds per instruction reference. (b) Ordering time against the number of units grows quadratically to 73 seconds for the original implementation and stays below 80 milliseconds for Kahn's algorithm.}
\caption{Orchestration cost on synthetic workflows (2--64 circuits,
22--49{,}088 instructions). (a)~Two-pass planning time is linear in the
number of instructions at 8--17\,$\mu$s per instruction (dashed reference:
10\,$\mu$s). (b)~Topological
ordering time versus number of units for the original repeated-scan
implementation and for Kahn's algorithm; the former is quadratic and reaches
73\,s at 32{,}704 units, the latter stays below 80\,ms.}
\label{fig:overhead}
\end{figure}

To measure planning cost beyond the protocol workflows, we generate synthetic
programs with $C \in \{2,\dots,64\}$ circuits and $R \in \{4,\dots,256\}$
rounds of (gate, labeled measurement, conditional), where half of the
conditionals reference a measurement on another circuit, up to
$49{,}088$ instructions, $32{,}704$ units, and $48{,}896$ edges.
Fig.~\ref{fig:overhead}(a) shows that the two-pass planner runs in
8--17\,$\mu$s per instruction across the whole range, with no super-linear
trend. Fig.~\ref{fig:overhead}(b) shows the topological ordering step before
and after replacing the repeated-scan implementation with Kahn's algorithm.
At the largest size, ordering drops from 73\,s to 78\,ms, and the
orchestration overhead of a no-op dispatch from 2.2\,ms to 8.2\,$\mu$s per
unit. Over the tested range, planning therefore adds microseconds per
instruction, whereas the earlier ordering step added tens of seconds at the
largest size.

\subsection{Correctness on two simulators (Q2)}
\label{sec:correctness}

We verify the protocol workflows exhaustively over their feedforward
branches rather than by sampling. Random trials are a weak test of the
feedforward branches. Sixteen trials of one-qubit teleportation miss one of its four
correction branches with probability about $4\%$, the Toffoli workflow has
$2^{12}$ branches, and the single-shot inverse-preparation check, on the
real-amplitude state $R_Y(\theta)\lvert 0\rangle$, passes an omitted $Z$
correction with probability $\cos^2\theta$, which is $0.85$ for
$\theta = \pi/8$, the angle of the seventh qubit of the largest swept
register. On
the QQLink runtime path (orchestrator, dispatcher, and Qulacs executor),
we force every mid-circuit measurement to each outcome in turn, project and
renormalize the state, and prune a branch when the probability of its forced
outcome is below $10^{-10}$. Each feasible branch is thus run exactly once,
deterministically, rather than sampled. For workflows with a quantum output
(teleportation, entanglement swapping, Toffoli, and the repetition code) the
output measurements are removed, the inputs are loaded from the
tomographically complete set $\{\lvert 0\rangle, \lvert 1\rangle,
\lvert +\rangle, \lvert {+i}\rangle\}$ on every input qubit, complex phases
included, and the fidelity of the output qubits with the ideal output state is
computed in every branch. We also check that each branch probability is the
same for every input. Each branch then acts linearly on the input, and in
exact arithmetic fidelity one on a tomographically complete set proves that
it implements the ideal map. In floating point the check bounds the
infidelity of each tested output state, not a channel norm, and we report
that bound. For workflows with a classical output (dense coding,
Grover, and Bernstein--Vazirani) the output measurements are forced too, and
a wrong output must be infeasible in every branch.

Table~\ref{tab:results} reports the results. In all $2.7 \times 10^5$
branch executions with a quantum output, every tested output state equals
the ideal output to floating-point precision (infidelity at most
$5\times10^{-12}$), and the branch probabilities agree across inputs to the
same precision. This holds
for every branch of three-qubit teleportation, of entanglement swapping, and
of the repetition code for each error location, and for all $2^{12}$
branches of the Toffoli workflow on 64 inputs. For every input of every
workflow, the probabilities of the executed branches sum to one up to
round-off ($3\times10^{-15}$), so the pruned branches carry no measurable
probability. No feasible branch of dense coding, Grover,
or any of the 16 Bernstein--Vazirani strings returns a wrong output. To check
that the procedure discriminates, we remove the $Z$ correction on one qubit
of three-qubit teleportation, and the minimum fidelity drops to zero. The
repetition-code runs also confirm the decoder's syndromes. The triple $(s_1, s_2, t_B)$
equals $(0,0,0)$, $(1,0,0)$, $(1,1,1)$, and $(0,1,0)$ for no error and errors
on A, B, and C. Randomized end-to-end runs on Qulacs and on Qiskit Aer,
16 per configuration, return the expected outcome in every case on both
simulators; they test the Aer path, which the forced-branch procedure does
not reach. The lowered circuit that the hardware path submits is checked
separately. Transpiled with the layout procedure of the hardware runs for
Qiskit's simulated versions of both devices, each of the 38 hardware
configurations returns the expected outcome in every one of 200 shots of a
noiseless simulation. The coverage therefore differs by path. The runtime
path is verified branch by branch on a tomographically complete input set,
while the Aer path, the lowered circuits, and the separate executors of
Section~\ref{sec:separate} are checked by sampling the application-level
outcome, which would not reveal an error that leaves that outcome
unchanged, such as a phase on a qubit measured in the computational basis.

\paragraph{Schedule invariance.}
The dispatcher runs units in authoring order, and the statement of
Section~\ref{sec:planning} that any topological order gives the same result
is checked by replay. For each of the 38 hardware configurations and for 100
random programs of two or three circuits, which contain cross-circuit CNOTs,
local entangling gates, and 132 local and 158 remote measurement references,
we run the plan in authoring order on the runtime path, sampling every
measurement, from the workflow's own preparation and from two random product
states with complex amplitudes on every qubit. We then replay the plan in
random topological orders, a uniformly random ready unit at every step,
forcing the $k$-th measurement of each qubit to the recorded outcome; the
footprint edges fix the order of the measurements on each qubit, so the
forced outcomes define the same branch in every order. All 12{,}096 replays
reach a final state of fidelity one with the authoring-order state to within
$6\times10^{-13}$, with the same branch probability and the same recorded
measurements. As a negative control, we replay the 38 hardware
configurations in random orders that respect the footprint edges but not the
measurement edges. These replays fail in 5{,}109 of 7{,}296 cases and in all
27 configurations whose measurement edges are not implied by footprint edges.
The other configurations are dense coding, which has no
conditional, and teleportation and swapping, where the unit that measures
also issued the Bell-pair request on the corrected qubit, so that a
footprint edge already orders the correction after the measurement.

\paragraph{Accept/reject loop.}
Table~\ref{tab:distill} compares the acceptance fraction over 4000 rounds
with two retries against the analytic model for both check circuits, with
95\% Wilson intervals. The parity check stays at the coin-flip value $0.875$
for every error amplitude, and the verification check follows
$1-(1-q(a))^3$, from certain acceptance at small $a$ to $0.89$ at $a = 3$;
in every cell shown the observed fraction lies within $1.5$ standard errors
of the model. We also ran $a = 0.1$ and $0.3$, which deviate most, by $-1.8$
and $-2.3$ standard errors in the parity check. Over all eight amplitudes,
the parity cells give $\chi^2 = 11.9$ for eight degrees of freedom
($p \approx 0.16$). An earlier run with 400 rounds showed the parity
fraction up to $2.7$ standard errors above the model at $a = 3$; the
tenfold larger sample places that excess within sampling error.

\begin{table}[t]
\centering
\caption{Exhaustive forced-branch verification in simulation, on the QQLink
runtime path with the Qulacs state-vector simulator.
Quantum outputs: minimum output-state fidelity over every feasible branch and
every input from $\{\lvert 0\rangle, \lvert 1\rangle, \lvert +\rangle,
\lvert {+i}\rangle\}$ per input qubit; $F_{\min} = 1$ means
$1 - F_{\min} \le 5\times10^{-12}$. Classical outputs: feasible branches
that return a wrong outcome.}
\label{tab:results}
\begin{tabular}{@{}lcccc@{}}
\toprule
Workflow & Inputs & Branches per input & Branch runs & Result \\
\midrule
Teleportation ($n{=}1$) & 4 & 4 & 16 & $F_{\min} = 1$ \\
Teleportation ($n{=}2$) & 16 & 16 & 256 & $F_{\min} = 1$ \\
Teleportation ($n{=}3$) & 64 & 64 & 4096 & $F_{\min} = 1$ \\
Entanglement swapping & -- & 4 & 4 & $F_{\min} = 1$ \\
Repetition code (per error location) & 4 & 256 & 1024 & $F_{\min} = 1$ \\
Toffoli & 64 & 4096 & 262{,}144 & $F_{\min} = 1$ \\
\midrule
Dense coding ($k{=}1, 2$; six messages) & -- & 1 & 6 & 0 wrong \\
Grover (4 marked states) & -- & 16 & 64 & 0 wrong \\
Bernstein--Vazirani (16 strings) & -- & 1--16 & 100 & 0 wrong \\
\bottomrule
\end{tabular}
\end{table}

\begin{table}[t]
\centering
\caption{Accept/reject loop. Fraction of 4000 rounds accepted within two
retries versus the error amplitude $a$, with 95\% Wilson intervals, for the
parity check ($q = 1/2$) and the verification check
($q = 1/2 + \sin a/(2a)$), against the model value $1-(1-q)^3$.}
\label{tab:distill}
\begin{tabular}{@{}lcccc@{}}
\toprule
$a$ (rad) & Parity: observed & model & Verification: observed & model \\
\midrule
0.0 & 0.879 [0.869, 0.889] & 0.875 & 1.000 [0.999, 1.000] & 1.000 \\
0.6 & 0.868 [0.857, 0.878] & 0.875 & 1.000 [0.999, 1.000] & 1.000 \\
1.0 & 0.876 [0.866, 0.886] & 0.875 & 1.000 [0.999, 1.000] & 1.000 \\
1.5 & 0.875 [0.864, 0.885] & 0.875 & 0.995 [0.993, 0.997] & 0.995 \\
2.0 & 0.873 [0.863, 0.883] & 0.875 & 0.982 [0.977, 0.985] & 0.980 \\
3.0 & 0.879 [0.869, 0.889] & 0.875 & 0.894 [0.884, 0.903] & 0.892 \\
\bottomrule
\end{tabular}
\end{table}

\subsection{Separate executors (Q3)}
\label{sec:separate}

The correctness results above run in one shared state vector, and the
hardware runs of Section~\ref{sec:hardware} run as one dynamic circuit; in
both, the nodes of a workflow collapse into one program whose registers carry
the node names. Qiskit expresses that case directly,
with named registers and \texttt{if\_test}. What QQLink adds is a plan that
does not need the collapse. Each node's units, with the dependencies that
cross node boundaries made explicit, are enough to run the node on its own
controller. Q3 tests exactly that. Each circuit of a workflow runs in its own
operating-system process that receives only that circuit's units and their
dependency lists; there is no dispatcher. A controller runs a unit when the
units it depends on have completed, and learns of completions on other nodes,
with the measurement outcomes they produced, only from messages that arrive in
its inbox after a latency $L + U(0, L)$. Conditionals are evaluated from the
outcomes a node has received. Quantum operations go to an
\emph{entanglement service}, a further process that holds the joint state of
all qubits in one Qulacs state vector and applies requests in arrival order.
The service is a stand-in for the physical layer, the nodes' registers and
the links between them, since a classical simulation of entangled nodes
needs one joint state; it makes no scheduling decision. What is distributed
is the control plane, namely readiness, ordering, feedforward, and messaging.

\begin{figure}[t]
\centering
\includegraphics[width=\linewidth]{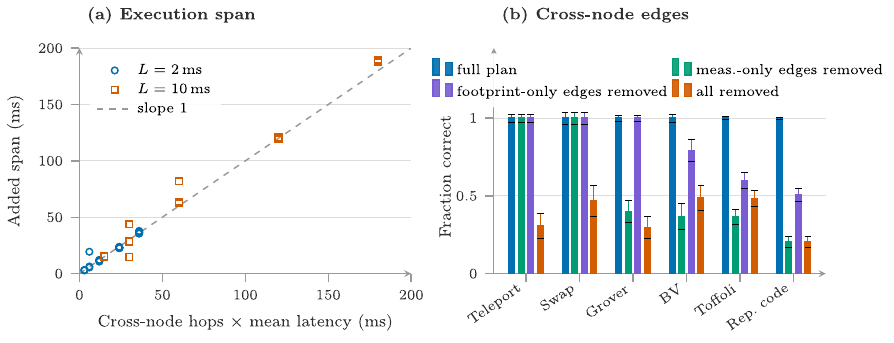}
\Description{Panel (a) is a scatter plot of the increase in execution span against cross-node hops times mean latency for 32 configurations, lying along a dashed line of slope one. Panel (b) is a bar chart of the fraction of correct runs per workflow family for the full plan and three ablations; the full plan is correct in every run, and removing edges lowers correctness for the families that depend on them.}
\caption{Workflows on separate executor processes, one per node, exchanging
completions and outcomes as messages with latency $L + U(0, L)$.
(a)~Execution span above its value at $L = 0$ against the number of
cross-node hops on the plan's critical path times the mean latency, for the
32 workflow configurations; dashed: slope one. (b)~Fraction of correct runs
per workflow family (16 runs per configuration and latency, pooled over
latencies) with the full plan, with the cross-node edges that come only from
measurement references removed, with those that come only from shared qubits
removed, and with all cross-node edges removed; error bars are 95\% Wilson
intervals.}
\label{fig:separate}
\end{figure}

We run the 32 hardware configurations of the six workflows with cross-node
feedforward (dense coding has none), 16 times each, at $L = 0$, 2, and
10\,ms, with the full plan and with three ablations of the edges between
units of different circuits (Fig.~\ref{fig:separate}), namely removing all of them,
removing only the measurement edges, which the planner derived from a
measurement reference alone, and removing only the footprint edges, derived
from a shared qubit alone; an edge with both origins is kept in the two
partial ablations. With the full plan all 1536 runs are
correct. With every cross-node edge removed, 519 are, between 20\% and 49\%
per workflow family pooled over latencies, and the loss occurs at every
latency, including $L = 0$. The partial ablations separate the two kinds of
dependency. Without the measurement edges, 627 of 1536 runs are correct. A
node evaluates a correction before the outcome that drives it has arrived and
skips it, and Grover, Bernstein--Vazirani, Toffoli, and the repetition code
fall to 20--40\%. Without the footprint edges, 1068 are. A node issues a
teleported gate's Bell-pair request or reuses an ancilla before the other
node has finished with the shared qubit, which breaks Bernstein--Vazirani,
Toffoli, and the repetition code (51--79\%) but leaves Grover intact.
Teleportation and swapping survive both partial ablations and fail when all
cross-node edges are removed; in teleportation, the cross-node edges from
Alice's measuring unit to Bob's corrections have both origins.

The execution span, from the first unit's start to the last unit's
end, follows the plan's critical path. After subtracting the span at
$L = 0$, it is proportional to the number of cross-node message hops on the
longest dependency path times the mean latency $1.5L$, with fitted slope $1.04$ ($R^2 = 0.99$), from one hop
for teleportation to twelve for Toffoli. Units on different nodes with no
path between them run concurrently, so the span grows with the depth of the
cross-node dependency chain, not with the number of units or messages.

This experiment answers the alternative of a single dynamic circuit, which
fixes one global order and one controller; the QQLink plan is what
lets the same program run without either, with correctness resting on the
cross-node edges that the planner derives and the developer never writes.
Three limits apply. The quantum state is simulated in one place, so the
experiment says nothing about the physics of remote entanglement or about
entanglement-generation constraints; latencies are emulated on one machine,
not measured over a network; and every message is delivered exactly once,
so loss, duplication, and process failure are not tested
(Section~\ref{sec:reliability}).

\subsection{Instruction-level overhead and cost of abstraction (Q4)}
\label{sec:cost}

\begin{figure*}[t]
\centering
\includegraphics[width=\textwidth]{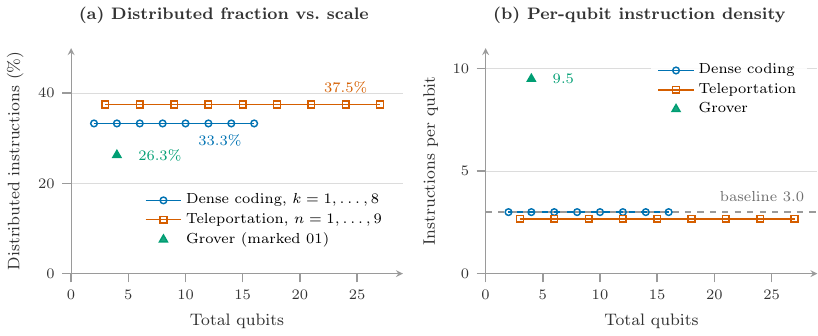}
\Description{Two panels against qubit count. (a) The fraction of distributed instructions is flat at 33.3 percent for dense coding and 37.5 percent for teleportation, with a single Grover point at 26.3 percent. (b) Instructions per qubit are flat at 3.0 for dense coding and 2.67 for teleportation, with a single Grover point at 9.5 above a dashed baseline at 3.0.}
\caption{Scaling of instruction-level overhead across the full protocol
sweeps. (a)~The fraction of distributed instructions (conditionals, Bell-pair
requests, cross-circuit CNOTs) is constant at 33.3\% for dense coding and
37.5\% for teleportation regardless of qubit count. Grover (single point at 4
qubits, marked state $01$) has a lower distributed fraction (26.3\%) because
its instruction count is dominated by local oracle and diffusion gates.
(b)~Per-qubit instruction density is constant for dense coding (3.0) and
teleportation (2.67) across the sweep; Grover's 9.5 instructions per qubit
reflects the intrinsic cost of gate teleportation (dashed line:
direct-entanglement baseline at 3.0).}
\label{fig:scaling}
\end{figure*}

Fig.~\ref{fig:scaling} summarizes the instruction-level resource consumption,
as authored, of dense coding and teleportation, the workflows with a size
parameter, and of Grover; Table~\ref{tab:profile} in
Appendix~\ref{app:tables} lists the full counts. Dense coding and teleportation scale linearly in
every count with $k$ and $n$, and their fraction of distributed instructions
(Bell-pair requests, cross-circuit CNOTs, and conditionals) is constant, at
33.3\% for dense coding and 37.5\% for teleportation at every scale.
Grover carries 9--10 instructions per qubit and
Toffoli 12--13, because each teleported CNOT costs one Bell pair, two
measurements, and four conditionals in addition to the local gates. Apart from
the cost of single-bit conditions in the repetition decoder, discussed below,
this cost is intrinsic to gate teleportation, not to the framework, which
the next comparison makes precise.

To separate protocol cost from abstraction cost, we lower each workflow to a
dynamic circuit with the hardware path of Section~\ref{sec:hwpath} and
compare it against a circuit for the same protocol written directly in
Qiskit, with the same teleported-gate construction, after transpiling both
for the same generic backend (Qiskit's \texttt{GenericBackendV2};
Table~\ref{tab:cost}). On every workflow the lowered
circuit has the same two-qubit gate count and the same number of conditional
blocks as the hand-written one, and the same depth on all but the repetition
code, where the two circuits differ only in the numbering of their classical
bits and routing can give the lowered circuit one more layer of depth. The
equality holds across
transpiler settings. Over ten transpiler seeds at optimization levels 1
and 3, the minimum, median, and maximum two-qubit count and depth of the
lowered and hand-written circuits coincide on every workflow, except for the
repetition code, whose lowered circuit is deeper by one layer at the extremes
of the seed range. This comparison has a
limit that we state plainly. Both circuits are ours and use the same
teleported-gate construction, so the equality shows that the lowering is
faithful to the construction the program describes, and that planning and
lowering add nothing on top of it; it does not show that the construction is
the one an expert or a compiler would choose, nor that the circuit uses
resources optimally. The host-side cost is the planning and dispatch
overhead of Section~\ref{sec:overhead}, microseconds per instruction; on
the hardware path no host code runs during a shot.

One cost of the programming model itself shows only against a baseline that
does not share its restriction. A QQLink conditional tests one measurement
bit, so the repetition-code decoder computes $s_1 \wedge s_2$ on a third
syndrome ancilla with a conditional CNOT and measures it. Written by hand
with the decoder in classical expressions, which OpenQASM~3 and IBM dynamic
circuits accept, the same distributed circuit needs 8 instead of 9 qubits,
13 instead of 14 measurements, and 19 instead of 22 conditional blocks, and
each of its 12 configurations is correct in a noiseless simulation. Most of
the saving appears after routing. On the line-coupled backend of
Table~\ref{tab:cost} at optimization level~1 it needs 39--42 two-qubit gates
over ten seeds (median 42) against 46--79 (median 61) for the lowered
circuit, and transpiled for \texttt{ibm\_kingston}'s coupling map 38 against
78. Conditions over several measurement references would remove this cost
(Section~\ref{sec:limitations}).

The monolithic column, which realizes the Grover, Toffoli, and
repetition-code gates directly without teleportation, quantifies the cost of
distribution instead. A Toffoli over three nodes costs 39 two-qubit gates
against 9, and the repetition code 67 against 25 for a monolithic version
with the same measured syndromes and decoder. To see how much of that cost a
distribution compiler would remove, we distribute the monolithic circuits
with pytket-dqc~\cite{andres2024distributing}, fixing its qubit placement to
the nodes of our workflows and applying its embedding and detached-gate
refinements (Table~\ref{tab:ebits}). For Grover both use two Bell pairs.
For the Toffoli gate pytket-dqc needs three Bell pairs where our workflow,
with one teleported CNOT per nonlocal gate of the textbook decomposition,
uses six, and for the encoding and syndrome extraction of the repetition
code three where ours uses four. QQLink executes the distribution it is
given and does not optimize it; the comparison quantifies what a compiler
placed in front of the planner would save, and is the case for treating
distribution compilers as producers of QQLink programs
(Section~\ref{sec:related}).

Two planner choices matter for the match with hand-written circuits. First,
ordering is plan metadata, not circuit content. An earlier version of the
Toffoli workflow synchronized the reuse of node C's ancilla with trivial conditional gates,
which appeared as ten extra conditional blocks and three extra layers of
depth in the lowered circuit, and the footprint analysis of
Section~\ref{sec:planning} removes them. Second, the lowering emits units in the
planner's topological order, which follows authoring order wherever the
dependencies allow; a topological order that ignored authoring order
produced the same gate counts but two to six extra layers of depth.

\begin{table}[t]
\centering
\caption{Cost of abstraction. Two-qubit gate count, depth, and number of
conditional blocks after transpilation on a \texttt{GenericBackendV2} with
linear coupling (optimization level~1, fixed seed; other seeds and
optimization level~3 in the text). ``Hand-written'' is the same protocol
authored directly in Qiskit with the same teleported-gate construction;
``monolithic'' realizes the non-local gates directly and is the reference
for the cost of distribution (for the repetition code, direct CNOTs with the
same measured syndromes and conditional decoder on six qubits).}
\label{tab:cost}
\begin{tabular}{@{}lccc@{}}
\toprule
 & QQLink lowered & Hand-written & Monolithic \\
Workflow & 2q / depth / cond. & 2q / depth / cond. & 2q / depth \\
\midrule
Dense ($k{=}1$)  & 2 / 11 / 0 & 2 / 11 / 0 & -- \\
Dense ($k{=}4$)  & 8 / 11 / 0 & 8 / 11 / 0 & -- \\
Teleport ($n{=}1$) & 2 / 10 / 2 & 2 / 10 / 2 & -- \\
Teleport ($n{=}7$) & 14 / 10 / 14 & 14 / 10 / 14 & -- \\
Grover ($01$) & 6 / 29 / 8 & 6 / 29 / 8 & 2 / 12 \\
Toffoli ($110$) & 39 / 83 / 24 & 39 / 83 / 24 & 9 / 20 \\
Repetition ($\lvert 1_L\rangle$, flip on B) & 67 / 64 / 22 & 67 / 63 / 22 & 25 / 23 \\
\bottomrule
\end{tabular}
\end{table}

\begin{table}[t]
\centering
\caption{Bell pairs consumed by the QQLink workflows as authored and by the
same monolithic circuits distributed with pytket-dqc over the same qubit
placement, with one Bell pair per nonlocal gate and after its embedding and
detached-gate refinements. The repetition-code row counts the encoding and
syndrome CNOTs.}
\label{tab:ebits}
\begin{tabular}{@{}lccc@{}}
\toprule
 & & \multicolumn{2}{c}{pytket-dqc} \\
Workflow & QQLink & one per gate & refined \\
\midrule
Grover & 2 & 2 & 2 \\
Toffoli & 6 & 6 & 3 \\
Repetition code & 4 & 4 & 3 \\
\bottomrule
\end{tabular}
\end{table}

\subsection{Same-chip hardware runs (Q5)}
\label{sec:hardware}

We ran the seven protocol workflows on \texttt{ibm\_kingston}, a 156-qubit
IBM Heron r2 processor, through the path of Section~\ref{sec:hwpath}, as 38
configurations, one dynamic circuit each, 4096 shots, and three independent
repeats, run between 17:04 and 18:35~UTC on 2026-10-09 against the
calibration the device reported at 15:48~UTC. Each node of a workflow
occupies its own contiguous stretch of the lowest-error line of qubits for
the workflow's size, so the virtual QPUs are disjoint sectors of the chip;
Table~\ref{tab:hardware-summary} lists the physical qubits of every node. On
those qubits, the median $T_1$ was $242\,\mu$s, the median $T_2$
$123\,\mu$s, the readout error at most $0.83\%$, and the CZ error at most
$0.58\%$. All circuits are transpiled at optimization level~1 except dense
coding, which is transpiled without optimization so that, for message 00, the
Bell-pair CNOT and the decoding CNOT are not cancelled and the circuit still
tests entanglement. We report the fraction of
shots that pass the application-level check, pooled over the repeats, with a
95\% Wilson interval. A pass/fail check has a chance level that depends on
its outcome space, and a circuit without working feedforward can pass it
often. Every configuration with feedforward is therefore also run under
controls that break the correlation between outcomes and corrections. In (C1) every
conditional block is removed, and in (C2) every conditional block is applied or
omitted at random, with three fixed random patterns pooled. The ancilla
resets of teleported gates are conditional blocks too, so C1 and C2 also
change what happens to reused ancillas and shorten the circuit. A reset is a
conditional $X$ on the qubit whose outcome it tests; every other conditional
block, including the decoder's conditional CNOT in the repetition code,
counts as a correction. Because C1 and C2 alter the resets, the second
session below adds two controls that keep the physical control structure. In (C1$'$) the corrections are removed and
the resets kept, and in (C3) every correction is kept as a conditional block in its
place but conditioned on an independent fair coin. Each correction has its
own coin, measured on an extra qubit at the start of the same shot, and the
resets stay conditioned on the true outcomes. C3 keeps the conditional blocks
and resets of the nominal circuit, adds the coin measurements before them,
and breaks the outcome--correction correlation in every shot. On
\texttt{ibm\_kingston} every control ran on the physical qubits of the
nominal circuit of its session, C3 with one extra qubit for the coins. C3
then has the same two-qubit gate count and the same number of conditional
blocks as the nominal circuit of the second session, except for the
repetition code (72 instead of 81 two-qubit gates), and a smaller transpiled
depth (126 instead of 174 for Toffoli), so it is close to the nominal circuit but not physically identical.
C1 is run once per session, C1$'$ and C3 once in the second session, each
with 4096 shots, and C2 with three random
patterns of 4096 shots each. Three patterns sample few of the possible
assignments. Their success differs by at most 4.6 points within a
configuration and session, except in the repetition code, where it differs
by up to 27 points. Taking each pattern as a separate control changes no
comparison below, since the same configurations fall under a control level. The
highest control of a session is the \emph{control level} of a check. It is an
empirical reference for these circuits, not a bound on what a classical
strategy could achieve. The \emph{margin} is the success minus the control
level. We give it with a 95\% Newcombe interval for the difference of two
proportions pooled over shots, so the uncertainty of the control enters it;
the interval does not include the variation between repeats.

\begin{table*}[t]
\centering
\caption{Hardware results on \texttt{ibm\_kingston} (Heron r2), 2026-10-09,
4096 shots, three repeats per configuration. Each node occupies a disjoint
stretch of physical qubits (A, B, C: node names; for entanglement swapping,
C is the middle node). ``2q / depth'' is the transpiled two-qubit gate count
and depth. Success is the percentage of shots that pass the
application-level check, pooled over the repeats; the 95\% Wilson interval is
given for single-configuration rows and is at most $\pm 0.9$ points for every
entry. Control is the larger of the success percentages of the same circuit
with every conditional block removed (C1) and applied at random (C2);
Table~\ref{tab:hardware-sessions} in Appendix~\ref{app:tables} gives every
repeat, the second session, and all four controls. Readout mitigation is
not applied.}
\label{tab:hardware-summary}
\scriptsize
\setlength{\tabcolsep}{3pt}
\begin{tabular}{@{}p{1.8cm}p{2.6cm}p{3.2cm}p{1.0cm}p{2.7cm}p{1.5cm}@{}}
\toprule
Workflow & Configuration & Physical qubits per node & 2q / depth & Success (\%) & Control (\%) \\
\midrule
Dense coding & $k{=}1$; message 00 / 01 / 10 / 11 & A: 148; B: 149 & 2 / 15 & 97.5 / 97.4 / 97.6 / 97.2 & -- \\
Dense coding & $k{=}2$; messages (00,01) / (11,10) & A: 109, 118; B: 128, 129 & 7 / 24 & 92.2 / 92.4 & -- \\
Teleportation & $n{=}1$ & A: 109, 118; B: 129 & 2 / 18 & 95.0 [94.6, 95.4] & 51 \\
Teleportation & $n{=}2$ & A: 91, 98, 110, 111; B: 109, 118 & 10 / 32 & 90.0 [89.5, 90.5] & 25 \\
Teleportation & $n{=}3$ & A: 25, 26, 37, 45, 46, 47; B: 38, 48, 49 & 24 / 44 & 82.7 [82.0, 83.4] & 14 \\
Entanglement swapping & $Z$ / $X$ basis & A: 109; C: 118, 129; B: 128 & 3 / 15--18 & 92.1 / 92.2 & 50 / 50 \\
Grover & marked 00 / 01 / 10 / 11 & A: 109, 118; B: 128, 129 & 6 / 43 & 73.6 / 73.1 / 73.5 / 72.4 & 26 / 26 / 25 / 26 \\
Bernstein--Vazirani & $s{=}1001$ / $0110$ (one teleported term) & A: 91, 98, 111; B: 109, 110, 118, 129 & 10 / 35 & 83.1 / 83.6 & 47 / 48 \\
Bernstein--Vazirani & $s{=}1111$ (two teleported terms) & A: 91, 98, 111; B: 109, 110, 118, 129 & 20 / 57 & 44.8 [43.9, 45.7] & 23 \\
Toffoli & input 000 / 001 / 010 / 011 & A: 91, 98; B: 110, 111; C: 109, 118 & 54 / 174 & 73.3 / 63.6 / 58.5 / 51.8 & 32 / 30 / 32 / 29 \\
Toffoli & input 100 / 101 / 110 / 111 & A: 91, 98; B: 110, 111; C: 109, 118 & 54 / 174 & 55.1 / 49.9 / 44.5 / 46.8 & 30 / 28 / 17 / 18 \\
Repetition code & $\lvert 0_L\rangle$; flip none / A / B / C & A: 25, 26; B: 37, 45, 46, 47, 48; C: 38, 49 & 78 / 128--129 & 67.1 / 44.3 / 57.4 / 55.8 & 27 / 10 / 24 / 24 \\
Repetition code & $\lvert 1_L\rangle$; flip none / A / B / C & A: 25, 26; B: 37, 45, 46, 47, 48; C: 38, 49 & 78 / 128--129 & 52.2 / 38.4 / 43.6 / 44.4 & 20 / 9 / 23 / 23 \\
Repetition code & $\lvert +_L\rangle$; flip none / A / B / C & A: 25, 26; B: 37, 45, 46, 47, 48; C: 38, 49 & 78 / 128--129 & 67.4 / 64.9 / 65.9 / 66.1 & 27 / 25 / 25 / 26 \\
\bottomrule
\end{tabular}
\end{table*}

\paragraph{Outcomes.}
On \texttt{ibm\_kingston} every configuration with feedforward succeeds above
its control level, with a margin whose 95\% lower bound is at least 19
points. Within this session, the corrections that the plan routes between
nodes act on the device. Dense coding, entanglement swapping, and one- and
two-qubit teleportation succeed in at least $90\%$ of shots, and three-qubit
teleportation in $83\%$. The workflows built from teleported gates lose more.
Grover succeeds in $72$--$74\%$ of shots for every marked state, Toffoli in
$45$--$73\%$ (highest for input 000, lowest for inputs 110 and 111), and the
repetition code in $38$--$67\%$. The partition into disjoint sectors costs
success only where it adds routing. Compared with a layout in which the
qubits of different nodes interleave along the same line, success agrees
within two points on every workflow except the repetition code, whose
disjoint layout needs 78 instead of 57 two-qubit gates and loses 3 to 11
points.

\paragraph{A second session.}
Seven days later, on 2026-10-16 between 14:40 and 15:01~UTC, we repeated the
suite on \texttt{ibm\_kingston} with three repeats and all four controls,
against the calibrations reported at 14:27~UTC (suite, C1, C2, and the
coherence checks) and 14:46~UTC (C1$'$, C3, and probe (v) below). The sector
search, run on that day's calibration, placed most workflows on other qubits.
Twenty-three of the 38 configurations reproduce within 3 points; dense
coding, one- and three-qubit teleportation, swapping, and Grover change by at
most 2.7 points. Bernstein--Vazirani and Toffoli, which the search placed on
qubits 137 and 147--152, lose 18 to 31 points. Two-qubit teleportation, on
qubits 147--152, loses 6, and the repetition code changes by $-9$ to $+10$
points across error locations. The standard deviation between the three
repeats of a session is at most $1.9$ points on \texttt{ibm\_kingston} and
$3.4$ points on \texttt{ibm\_fez}, so the variation between sessions and
sectors is four to sixteen times larger, and the pooled intervals of
Table~\ref{tab:hardware-summary} describe one session on one set of qubits.
The calibration the device reports does not predict the change. A noise
model built from its calibration of 2026-10-16 (15:21~UTC) and applied to
the qubits of both sessions gives two-term Bernstein--Vazirani $92\%$ and
$91\%$ and Toffoli $84$--$87\%$ on either set of qubits; only three-qubit
teleportation differs between the two sets by more than 3 points ($90\%$
against $93\%$). In the pooled-shot comparison, every configuration with
feedforward still exceeds all four controls of its session, C1$'$ and C3
included, but the smallest lower bound of the margin, for Toffoli input 101,
falls to 0.7 points (Table~\ref{tab:hardware-sessions}).

On \texttt{ibm\_fez} (Tables~\ref{tab:hardware-fez}
and~\ref{tab:hardware-sessions-fez}), two-term Bernstein--Vazirani falls
below its control level on 2026-10-09 ($18.3\%$ against $22\%$). On
2026-10-16 only C1$'$ and C3 were run on \texttt{ibm\_fez}, for
Bernstein--Vazirani and Toffoli, in a separate job whose sector search placed
them on qubits 108--112, 118, 129, and 130 instead of the suite's 137 and
142--147, so that day's comparisons are between sectors. Against these
controls two-term Bernstein--Vazirani exceeds its control level ($30.6\%$,
margin at least 7 points), and Toffoli inputs 100 and 101, which exceed their
controls on 2026-10-09, fall 2.8 points below C1$'$, which keeps the resets,
but still exceed C3. Every other \texttt{ibm\_fez} configuration with
feedforward succeeds above its control level in the sessions in which its
controls were run.

\paragraph{What the checks show.}
The application-level checks differ in what they establish, and four
further checks, summarized in Table~\ref{tab:coherence}, test what a
pass/fail check of the classical output cannot see. Teleportation's
inverse-preparation check tests only the input it prepares. We therefore
teleported each of the six cardinal states, a 2-design, on
\texttt{ibm\_kingston} on 2026-10-16 and checked each by inverse
preparation. The measured quantity is the six-state mean survival
probability, $0.961$ $[0.960, 0.963]$. With ideal state preparation and
readout it would be the average fidelity $F_\mathrm{avg}$ of the
teleportation channel and correspond to a process fidelity of
$(3F_\mathrm{avg} - 1)/2 = 0.942$ $[0.940, 0.944]$. We did not characterize
preparation and readout separately, so both numbers are estimates that
attribute their errors to the channel. For entanglement swapping, the
agreement in the $X$ and $Z$ bases, measured on separate runs of the same
circuit with trusted single-qubit measurements and no readout correction,
bounds the fidelity of the swapped pair with $\lvert\Phi^+\rangle$ by
$F \ge p_{\mathrm{equal},X} + p_{\mathrm{equal},Z} - 1$. The estimate of
the right-hand side is $0.843$ $[0.836, 0.850]$ on 2026-10-09 and $0.864$
$[0.857, 0.870]$ on 2026-10-16, so under these assumptions $F \ge 0.836$
and $F \ge 0.857$ at 95\% confidence, and the swapped pair is entangled.

The Toffoli truth table tests classical behavior only; a channel that
dephased its output would pass it. With one control in $\lvert +\rangle$, the
other in a basis state $b$, and the target in $\lvert -\rangle$, the gate
maps the $\lvert +\rangle$ control to $\lvert -\rangle$ exactly when $b = 1$.
We ran this check for control A in $\lvert +\rangle$ with $b_B \in \{0, 1\}$
and for control B in $\lvert +\rangle$ with $b_A \in \{0, 1\}$. A shot passes
when the $X$-basis readouts of the $\lvert +\rangle$ control and of the
target and the $Z$-basis readout of the other control all take their ideal
values. The check passes in $13$--$24\%$ of shots, against $80$--$82\%$ in
the device-noise simulation. Complete dephasing of the $\lvert +\rangle$
control would halve the success of the matching truth-table inputs, and for
three of the four configurations the check reaches about half of the mean
truth-table success of the two inputs that the control superposes, on the
same day and qubits. The check does not establish coherent Toffoli action on
that sector. The repetition code's computational-basis agreement for
$\lvert +_L\rangle$ would also accept an incoherent mixture of
$\lvert 000\rangle$ and $\lvert 111\rangle$. Read in the $X$ basis, the
logical state has even parity in $48$--$53\%$ of shots for every error
location, against $85$--$86\%$ in simulation. The runs therefore do not
demonstrate that the intended logical superposition is preserved. The
parity alone does not tell dephasing from a coherent phase error, since the state
$(\lvert 000\rangle + i\lvert 111\rangle)/\sqrt{2}$ also gives $50\%$. We
claim no error-correction advantage over an unencoded qubit. The
repetition-code runs test the routing of syndromes to corrections across
nodes, not error correction.

\begin{table*}[t]
\centering
\caption{Coherence-sensitive checks on \texttt{ibm\_kingston}. Shots are
given as configurations $\times$ shots $\times$ repeats. Measured values are
pooled over repeats. Brackets are 95\% confidence intervals, computed as Wilson
intervals for single proportions and as normal-approximation intervals for the
teleportation mean over the six inputs and for the swapping bound, a sum of
two independent proportions. Values joined by a dash are ranges over the
configurations of a row, not intervals; model values are ranges or
point values from 4000-shot simulations. The model is the device-noise simulation with
the calibration of 2026-10-16, 15:21~UTC. Readout mitigation is not
applied.}
\label{tab:coherence}
\scriptsize
\setlength{\tabcolsep}{3pt}
\begin{tabular}{@{}p{2.0cm}p{1.3cm}p{2.4cm}p{1.4cm}p{2.1cm}p{1.1cm}p{2.3cm}@{}}
\toprule
Check & Date & Observable & Shots & Measured & Model & Interpretation \\
\midrule
Teleportation $n{=}1$, six cardinal inputs & 2026-10-16 & mean survival probability of inverse preparation & $6 \times 4096 \times 3$ & 0.961 [0.960, 0.963]; per state 0.949--0.976 & 0.982--0.987 & process fidelity 0.942 if preparation and readout errors are attributed to the channel \\
Entanglement swapping & 2026-10-09 / 2026-10-16 & $p_{\mathrm{equal},X} + p_{\mathrm{equal},Z} - 1$, a lower bound on the fidelity with $\lvert\Phi^+\rangle$ & $2 \times 4096 \times 3$ per date & 0.843 [0.836, 0.850] / 0.864 [0.857, 0.870] & 0.95 / 0.96 & $F \ge 0.836$ / $0.857$ at 95\% confidence; pair entangled \\
Toffoli, one control in $\lvert +\rangle$, target in $\lvert -\rangle$ & 2026-10-16 & all three readouts ideal, $X$ basis on the $\lvert +\rangle$ control and the target & $4 \times 4096 \times 3$ & 13.4--24.3\% & 80--82\% & coherent action not established \\
Repetition code, $\lvert +_L\rangle$, flip none / A / B / C & 2026-10-16 & even logical $X$ parity & $4 \times 4096 \times 3$ & 48.3--53.0\% & 85--86\% & preservation of the intended superposition not established \\
\bottomrule
\end{tabular}
\end{table*}

\paragraph{Where the deeper workflows lose.}
To separate the device's gate and readout errors from the structure of the
circuits, we simulated every configuration, transpiled for the physical
qubits of each session, with the noise model built from the calibration of
2026-10-16 at 15:21~UTC (Section~\ref{sec:hwpath}), the model used above; for
the first session's qubits it describes the device a week after the runs.
Without noise, every circuit transpiled for these qubits passes its check in
every shot. The
gap between device and model divides the workflows in two. On the first
session's qubits, dense coding, teleportation, and entanglement swapping
fall short of the model by at most 7 points, the most for three-qubit
teleportation with its 24 two-qubit gates. The workflows built from
teleported gates fall short by more, and not in proportion to their gate
counts. Grover, with 6 two-qubit gates, falls short by 23--24 points;
Bernstein--Vazirani by 11--12 points with one teleported term and by 47 with
two ($45\%$ on the device against $92\%$ in the model); Toffoli by 13 points
for input 000 up to 40 for input 110; and the repetition code by 18--38
points. On the second session's qubits the gaps of these workflows are 15 to
65 points. This excess loss is not explained by the model's gate and readout
errors, so it lies in what the model omits or misdescribes. Candidates are
qubits that idle while others are measured, corrections executed as
conditional blocks, crosstalk, errors that a measurement induces on
neighboring qubits, and, as the second session suggests, qubit defects that
the reported calibration does not show.

Five probes bear on these candidates; their qubits are those the sector
search selected on the day of each run. The first four ran on \texttt{ibm\_fez},
where two-term Bernstein--Vazirani was weakest on 2026-10-09.
(i)~Idle qubits accrue phase. We ran a Ramsey sequence on an explicit
\texttt{delay} of 0 to $24\,\mu$s, read out in the $X$ and $Y$ bases with
one run of 4096 shots per delay. A linear fit to the phase, with bootstrap
intervals for shot noise only, gives a rotation at $25.7$~kHz
$[25.4, 25.9]$ for one qubit of the Bernstein--Vazirani sector on
2026-10-09, and at $13.5$ $[13.1, 13.8]$ and $1.0$ $[0.6, 1.3]$~kHz for two
qubits of another sector on 2026-10-12. The second qubit of the
Bernstein--Vazirani sector oscillates at $45.1$ $[44.8, 45.4]$~kHz in the
$X$ basis with no $Y$ component, which is consistent with a qubit frequency
that switches between two values on either side of the drive frequency and
not with a fixed detuning. On both qubits of that sector, a single echo at
the midpoint removes the oscillation; the 2026-10-12 run had no echo
variant. On 2026-10-12, two qubits of a third sector, held by a barrier
until eight rounds of measurement and correction on other qubits had
finished, keep their state in $97\%$ of shots, so in that sector waiting
through feedforward is harmless.
(ii)~Conditional execution can disturb the qubits it acts on. In every
round, two qubits each receive a gate twice under the same random condition,
measured on a third qubit, so that the net operation is the identity. On
qubits 142 and 143 of the Bernstein--Vazirani sector on 2026-10-09, both keep
their state after one round in $95$--$96\%$ of shots; after four rounds one
qubit is left in a random state and the other is found inverted in
$88$--$89\%$ of shots, for $X$ as well as $Z$ gates. On qubits 8 and 9 on
2026-10-12, both keep their state in $92$--$93\%$ of shots after one round,
and each keeps it in $77$--$81\%$ after four.

(iii)~Coherent control in place of feedforward. Every measurement stays in
place, so the measurement record is unchanged. Each conditional block on
another qubit becomes the corresponding controlled gate from the measured
qubit, which by the deferred-measurement principle acts as the classically
controlled gate, and the conditional $X$ that resets a reused ancilla
becomes an unconditional reset. No classical feedforward remains. On
\texttt{ibm\_fez} on 2026-10-12, two-term Bernstein--Vazirani then succeeds
in $48\%$ of shots instead of $17\%$, although its two-qubit gate count
doubles from 20 to 39. In the same job, teleportation, Grover, one-term
Bernstein--Vazirani, and Toffoli become worse by 1.7 to 14 points. On
2026-10-09, rewriting the $Z$ correction on the controls as $H$, a
conditional $X$, and $H$ raises two-term Bernstein--Vazirani from $19\%$ to
$31\%$.
(iv)~Echoes on idle intervals. IBM Runtime rejects its own dynamical
decoupling for dynamic circuits, so we wrote one pair of $X$ gates per idle
interval into the lowered circuit, anchored by barriers. This raises
Bernstein--Vazirani from $74\%$ to $83$--$84\%$ with one term and from $17\%$ to
$28\%$ with two, but lowers three-qubit teleportation from $85\%$ to $36\%$
and the repetition code by up to 36 points. For teleportation the barriers
serialize the circuit and lengthen it by about half its depth; for the
repetition code, whose depth does not grow, the cause is not resolved. The
echo pass changes the transpiled circuit beyond the echoes (two-term
Bernstein--Vazirani, for example, goes from 20 to 32 two-qubit gates), so
these comparisons are not between otherwise identical circuits.
(v)~Coherent control on the same qubits. To remove the differences of
device, sector, and day from probe (iii), we ran the feedforward and
coherent-control versions of every teleportation, Grover,
Bernstein--Vazirani, and Toffoli configuration on the same physical qubits
of \texttt{ibm\_kingston} on 2026-10-16, interleaved in one job, two rounds
each. On these qubits, coherent control lowers two-term Bernstein--Vazirani
from $23\%$ to $12\%$ (difference $-11$ $[-12, -10]$ points), Grover by 10
to 12 points, and teleportation by 2 to 5 points. It raises one-term
Bernstein--Vazirani by 6 points and changes Toffoli by $-5$ to $+12$ points,
depending on the input.

These probes establish sensitivity, not a mechanism. The success of the
deeper workflows depends on how the dynamic control is executed (as
feedforward, as coherent control, or with echoes) and on where it runs. The
effect of coherent control on two-term Bernstein--Vazirani changes sign
between the \texttt{ibm\_fez} sector of 2026-10-12 and the
\texttt{ibm\_kingston} sector of 2026-10-16, and the second-session loss on
\texttt{ibm\_kingston} does not appear in the reported calibration. We
therefore do not attribute the excess loss to one process. Characterizing it
would need controlled comparisons across sectors and days, and timing
information that the runs do not give, because Qiskit's scheduler does not accept
conditional blocks, and the runtime reports no per-circuit durations. The
plan identifies every correction and the qubit it acts on, so choosing
coherent control or feedforward per correction is a local decision of the
hardware path.

\paragraph{What the runs establish.}
For Q5, the same program object that runs on two simulators and on separate
executors runs on the devices with no change at the application layer. On
\texttt{ibm\_kingston}, in both sessions, every configuration with
feedforward succeeds above the control level of its check in the pooled-shot
comparison. The lower bound of the smallest margin, 0.7 points in the
second session, comes from an interval that excludes the variation between
repeats, so the comparison does not establish an advantage that holds across
jobs, sectors, or calibrations. On \texttt{ibm\_fez} three configurations
fall below their control level in one of the two sessions. Each number is
tied to a calibration snapshot, a physical sector per node, and a job
identifier. The runs also show what same-chip execution of these protocols
does not yet deliver. The workflows built from teleported gates fall up to
65 points short of a simulation with the device's gate and readout errors,
their success changes by up to 31 points between sessions on qubits that the
calibration does not distinguish, and the coherence-sensitive checks do not
establish coherent Toffoli action or preservation of the intended
repetition-code superposition.

\subsection{Robustness and noise sensitivity (Q6)}
\label{sec:robustness}

\begin{figure}[t]
\centering
\includegraphics[width=\linewidth]{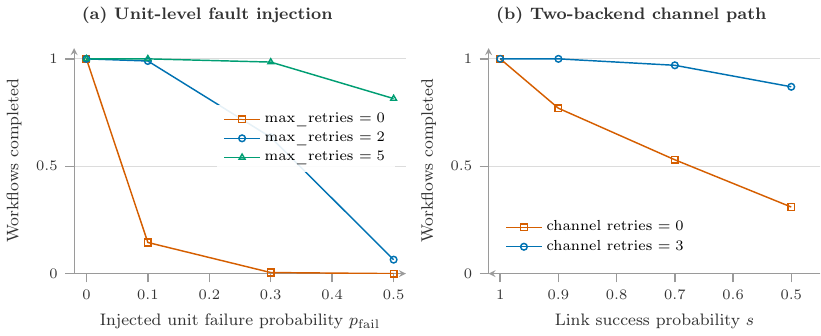}
\Description{Panel (a) plots the fraction of completed Grover workflows against the dispatch failure probability for three retry budgets, with more retries keeping completion higher. Panel (b) plots the fraction of completed teleportation workflows against link success probability, with and without three channel-level retries; retries raise completion.}
\caption{Runtime controls under failure. (a)~Fraction of 200 Grover
workflows (18 units each) that complete when each unit dispatch fails with
probability $p_\mathrm{fail}$, for three retry budgets; every completed run
returned the marked state. (b)~Fraction of 100 two-qubit teleportation
workflows that complete when Alice and Bob run on separate simulator
instances and each Bell pair traverses a simulated link with success
probability $s$, without and with three channel-level retries. The 95\%
binomial interval of each point is at most $\pm0.07$ in (a) and $\pm0.10$
in (b).}
\label{fig:faults}
\end{figure}

\paragraph{Runtime controls.}
This experiment is a functional check of the retry and channel controls, not
a fault-tolerance result. We wrap the executor in a dispatcher that raises a
transient error with probability $p_\mathrm{fail}$ \emph{before} a unit
executes, and run 200 Grover searches (18 units) for each combination of
$p_\mathrm{fail} \in \{0, 0.1, 0.3, 0.5\}$ and retry budget $\{0, 2, 5\}$
(Fig.~\ref{fig:faults}(a)). Completion follows the binomial expectation.
Without retries it is $(1-p_\mathrm{fail})^{18}$, and 29 and 1 of 200 runs
complete at $p_\mathrm{fail} = 0.1$ and $0.3$ (expected 30 and 0.3). With
five retries, 200 and 197 complete (expected 200 and 197), and 163 at $0.5$,
two binomial standard deviations above the expected 151. Every completed run
returned the marked state. Because the fault is injected before the unit
runs, a retry is trivially safe; the cases that matter physically are not
covered. For a fault after some of a unit's
instructions have acted on the state, the runtime applies rule (i) of
Section~\ref{sec:reliability} and reports the failure without retrying it,
which a unit test checks. On the hardware path the workflow is one circuit and
unit-level retry does not apply. Fig.~\ref{fig:faults}(b) checks the
channel-level retries in the same way. In a two-qubit teleportation, each of
the two Bell-pair requests succeeds with probability $s$ on a simulated link
that models availability, not entanglement. Without retries completion is
$s^2$, and three channel-level retries raise it to 100, 97, and 87 of 100 at
$s = 0.9, 0.7, 0.5$. The separate-executor experiment of
Section~\ref{sec:separate} replaces the shared dispatcher with message
passing between processes; it does not model link failures.

\begin{figure}[t]
\centering
\includegraphics[width=0.7\linewidth]{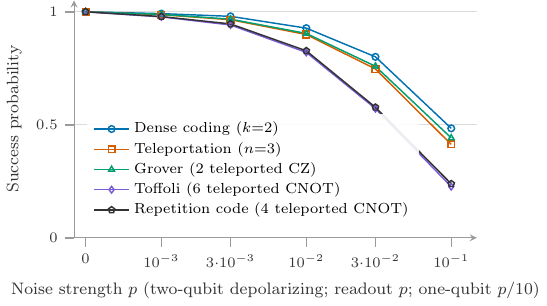}
\Description{Line plot of decoded-outcome success probability against noise strength p for five workflows; all start at one without noise and decline as p grows, the two workflows with the most two-qubit gates declining fastest.}
\caption{Success probability of the decoded outcome versus noise strength
$p$ (depolarizing $p$ on two-qubit gates, $p/10$ on single-qubit gates,
readout error $p$; 4000 shots) for the lowered dynamic circuits of five
workflows. The leftmost point is the noiseless case; the 95\% binomial
interval of each point is at most $\pm0.016$, smaller than the markers.
The two workflows with the most two-qubit gates are the most fragile;
planning adds no gates beyond the authored construction.}
\label{fig:noise}
\end{figure}

\paragraph{Gate and readout noise.}
Fig.~\ref{fig:noise} shows the success probability of the decoded outcome
for the lowered dynamic circuits of five workflows on Aer under the noise
model of Section~\ref{sec:method}, namely dense coding with $k{=}2$, three-qubit
teleportation, Grover with marked state 01, Toffoli with input 110, and the
repetition code on $\lvert 1_L\rangle$ with a flip on B. The orchestration is
unchanged across the sweep; only the noise strength varies. At
$p = 10^{-2}$ dense coding succeeds with probability 0.93,
teleportation 0.90, Grover 0.90, the repetition code 0.83, and Toffoli 0.82;
at $p = 3\times10^{-2}$ the values are 0.80, 0.75, 0.76, 0.58, and 0.57. The
Toffoli and repetition-code workflows, with 39 and 67 two-qubit gates after
transpilation, are the most fragile, which is the protocol cost of
Table~\ref{tab:cost} made visible under noise; the repetition code corrects
the injected flip but, at these physical error rates, not the errors the
protocol itself introduces. The sweep isolates gate and readout noise; it
omits the time structure of the dynamic circuit, which
Section~\ref{sec:hardware} probes on hardware.

\subsection{Lessons learned}
\label{sec:lessons}

Several defects surfaced during this work, each caught by one of the checks
above rather than by inspection. We list them because they show which checks
a framework of this kind needs.
(i)~An earlier Aer adapter re-executed the accumulated circuit on every
mid-circuit measurement and returned the most frequent bit of 1024 shots.
Dense coding, which has no feedforward, passed. Teleportation, Grover, and
Toffoli succeeded only about as often as with uncorrelated corrections
(8/16, 22/64, and 44/128), while Qulacs passed every check. The defect shows
the adapter contract of Section~\ref{sec:mechanisms} in concrete form, and
it was caught only because two simulators were compared.
(ii)~The first topological ordering was quadratic in the number of units
(Section~\ref{sec:overhead}); only the synthetic sweep, not the protocol
workflows, was large enough to show it.
(iii)~The first planner split a segment only at the requesting circuit's own
request, so two circuits issuing requests on each other's qubits could
produce a dependency cycle. None of the eight workflows did, and the
dispatcher's topological sort would have rejected one; the refinement and
proof of Section~\ref{sec:planning} close the gap. Randomly generated
programs with cross-circuit requests now form part of the test suite. Before the change, 211 of
300 produced a cycle, and none did after.
(iv)~Qulacs defines rotation gates as $\exp(+i\theta P/2)$, the opposite sign
to Qiskit, and the Qulacs adapter passed the angle through. Every workflow
in this paper is invariant under the flip, so no result changed, but a
program with an asymmetric rotation would have prepared different states on
the two simulators, contrary to condition (a) of the introduction. A test
now compares single-qubit rotations across backends.
(v)~Qiskit Aer, given one batch that mixes control-flow circuits with
circuits that have fewer classical bits, returned counts for the wrong
circuits in our runs. A batched dry run reported, for example, $0.40$ for
the Bernstein--Vazirani string $0011$, which succeeds with probability
$0.97$ when simulated alone under the same noise model. The dry run now
simulates circuits one at a time; hardware jobs, which return results per
circuit, were not affected.

\section{Related Work}
\label{sec:related}

Several lines of work address parts of the multi-QPU software stack. We group
them by the level at which they operate and state, for each, the delta to
QQLink.

\paragraph{Distributed circuit compilation and partitioning.}
Andr\'es-Mart\'{\i}nez and Heunen~\cite{andres2019} and Ferrari et
al.~\cite{ferrari2023} partition monolithic circuits across QPUs by
minimizing inter-partition communication, through hypergraph partitioning or
integer programming. Baker et al.~\cite{baker2020} partition in time slices
for modular architectures, and AutoComm~\cite{wu2022} aggregates
communication into burst operations at compile time. pytket-dqc~\cite{andres2024distributing}
distributes a monolithic circuit over heterogeneous modular networks and
implements the resulting nonlocal gates with entanglement-assisted
communication, minimizing the number of Bell pairs consumed. These approaches produce
static partitions and leave runtime concerns, such as when a conditional
correction may run or how Bell pairs are provisioned, to the execution layer.
These compilers and QQLink meet at the teleported gate. A compiler such as
pytket-dqc or AutoComm decides where nonlocal gates go and how to batch
their communication, starting from a monolithic circuit; QQLink starts from
per-node programs in which the teleported gates are already written, and
decides when each correction may run and on what target. The two are
complementary, because a distribution produced by such a compiler is an input QQLink
could plan, with QQLink's platform layer choosing placement, resolving
cross-circuit dependencies, and orchestrating execution.
Section~\ref{sec:cost} measures the gap. With the qubit placement of our
workflows fixed, pytket-dqc distributes the Toffoli gate with half the Bell
pairs our hand-authored workflow uses.
At the fault-tolerant end, network-based quantum computing~\cite{naito2026network}
is an architecture and compilation framework for many small nodes holding one
or a few logical qubits each, in which data keeps moving through the network;
it fixes the machine organization that a runtime such as QQLink would then
program against.
Surveys of the field~\cite{caleffi2022,barral2025} catalog these compilers
but do not identify a workflow-level orchestration layer of the kind
proposed here.

\paragraph{Circuit cutting and knitting.}
Circuit cutting~\cite{peng2020,tang2021} decomposes a circuit into
sub-circuits that run independently on separate backends, with classical
post-processing reconstructing the global expectation value. This trades
quantum communication for exponential classical overhead and suits settings
where entanglement distribution is unavailable. In its standard form it cannot express protocols
that require real-time feedforward, such as the Pauli corrections in
teleportation or the conditional resets in gate teleportation, because the
sub-circuits never interact at runtime. QQLink takes the opposite stance. It
assumes that entanglement and classical channels are available and provides
the runtime machinery to consume them.

\paragraph{Quantum network simulators and network stacks.}
NetSquid~\cite{coopmans2021}, SeQUeNCe~\cite{wu2021}, and
QuNetSim~\cite{diadamo2021} model network-layer physics such as photon loss, link
scheduling, purification, and memory decoherence. Interlin-q~\cite{parekh2021}
adds a control-node model for distributed algorithm simulation on top of
QuNetSim. Chen et
al.~\cite{chen2026adaptive} orchestrate Bell-pair generation, routing, and
caching through a hub-and-spoke photonic interconnect, deciding when to
schedule teleportation-based nonlocal gates under loss and decoherence; their
orchestrator allocates entanglement as a physical resource, which is the
service QQLink's entanglement channels consume. These simulators and
orchestrators model or allocate the network; QQLink operates above them and
could target a network simulator as a provider-native channel executor.

\paragraph{Quantum-internet node software.}
Three systems for programmable quantum-network nodes are the closest to
QQLink in their abstractions, and Table~\ref{tab:nodesw} compares them.
NetQASM~\cite{dahlberg2022} is a platform-independent instruction set for
hybrid programs on a network node together with a Python SDK. In the SDK a
host program runs on one node; measurement results are futures that resolve
when the queued operations are flushed to the node controller; entanglement
is generated through EPR (Einstein--Podolsky--Rosen pair) sockets whose \texttt{create} and \texttt{recv}
calls must be paired on the two nodes; and classical information crosses
nodes through explicit socket messages. QNodeOS~\cite{delledonne2025} is an
operating system that executes such per-node programs, with entanglement
request sockets, virtual qubit addresses, and multitasking; it ran
applications on two nitrogen-vacancy (NV) center nodes and was given a driver for a
trapped-ion node, and its architecture is platform independent.
Qoala~\cite{vandervecht2025qoala} extends QNodeOS with a program format of
host-code blocks, local routines, and entanglement-request routines. Its
runtime turns program segments into tasks of a directed acyclic task graph
with precedence and deadline constraints and schedules them jointly and
asynchronously; it is evaluated with a simulator validated against network
hardware. All three partition an application into single-node programs
whose cross-node coordination the programmer writes, namely paired socket calls,
sends and receives, and the order in which they occur. QQLink differs in
where coordination comes from. A program describes all nodes at once;
a correction on one node references the measurement handle of another;
entanglement is declared once as a link; and the planner derives the
cross-node dependencies, both measurement references and the ordering of
shared qubits, together with each node's units (Section~\ref{sec:planning}).
The per-node units with their incoming edges are exactly what the separate
executors of Section~\ref{sec:separate} run, so a QNodeOS or Qoala node could
serve as the executor for one node of a QQLink plan, with the plan's edges
becoming its messages and socket calls.

\begin{table}[t]
\centering
\caption{QQLink and the closest node-level software for quantum networks.}
\label{tab:nodesw}
\small
\begin{tabular}{@{}p{1.75cm}p{2.6cm}p{2.7cm}p{2.5cm}p{2.6cm}@{}}
\toprule
 & Authoring unit & Cross-node coordination & Scheduling & Demonstrated on \\
\midrule
NetQASM SDK & one host program per node & written: paired EPR-socket calls, socket messages; results as futures & program order per node & NetSquid, SimulaQron \\
QNodeOS & per-node programs (NetQASM) & written, as above; entanglement request sockets & node OS, multitasking & two NV-center nodes; trapped-ion driver \\
Qoala & per-node programs (host blocks, routines) & written per node & runtime task DAG with deadlines, joint and asynchronous & simulator validated against hardware \\
QQLink & one program over all nodes & derived: measurement and footprint edges from handles and declared links & topological dispatch, or one controller per node & Qulacs, Aer, separate processes, same-chip IBM Heron \\
\bottomrule
\end{tabular}
\end{table}

\paragraph{Programming models for multi-QPU execution.}
QMPI~\cite{haner2021} proposes a message-passing interface for distributed
quantum programs, transposing the Message Passing Interface (MPI) model to quantum
communication.
NetQMPI, an MPI-inspired model in the same spirit, has been coupled to the
CUNQA emulator, which exposes virtual QPUs on HPC systems, as its execution
backend~\cite{vazquezperez2026emulating}; in both, the programmer writes the
communication explicitly, whereas QQLink derives it from declared
entanglement and measurement references.
OpenQASM~3~\cite{cross2022} and hardware demonstrations of dynamic
circuits~\cite{corcoles2021} support mid-circuit measurement and classical
feedforward within a single circuit on a single backend, but do not extend to
cross-circuit or cross-backend conditionals. Heterogeneous frameworks such as
XACC~\cite{mccaskey2020} and CUDA-Q~\cite{cudaq} target multiple backends
through a common intermediate representation but treat each program as a
single kernel bound to one QPU; XACC's virtual QPUs~\cite{claudino2024}
parallelize independent circuits rather than entangled partitions. QQLink
differs in that the unit of authoring is the workflow node with typed
measurement handles, and the orchestrator, not the programmer, resolves the
resulting dependency graph.

\paragraph{Hybrid workflow and task systems.}
Covalent~\cite{covalent2022}, Qiskit Serverless~\cite{qiskitserverless}, and
Pilot-Quantum~\cite{mantha2025} orchestrate hybrid quantum--classical
workflows as directed task graphs, providing retry logic, distributed
execution, and HPC resource management. From a software-architecture
perspective, Weder et al.~\cite{weder2021hybrid} argue that hybrid
applications need two orchestrations at once, a workflow over quantum and
classical programs and a deployment topology, and Beisel et
al.~\cite{beisel2025operations} collect operations patterns for packaging,
executing, and observing such applications. Qunicorn~\cite{weder2025qunicorn}
gives unified access to heterogeneous quantum cloud offerings by translating
circuit and result formats, and QCOEM~\cite{pham2026qcoem} schedules
independent tasks across noisy, queued backends by multi-objective
optimization. In all of these the quantum program is an opaque task bound to
one backend; none models inter-QPU entanglement, cross-circuit conditional
gates, or entanglement-aware placement. QQLink unifies intra- and inter-circuit feedforward with
entanglement-aware planning and dependency-safe dispatch in one framework,
and its task layer could itself be hosted by such a workflow system.

\paragraph{Quantum--HPC software stacks.}
Production quantum--HPC stacks integrate QPUs into computing centers as
accelerators. The Quantum Framework (QFw)~\cite{chundury2025} orchestrates
hybrid applications over simulators and hardware with HPC process
management, and the Munich Quantum Software Stack~\cite{burgholzer2026}
exposes heterogeneous devices behind a device-management interface. The
openQSE survey~\cite{shehata2026} reviews nine such stacks and proposes a
reference architecture whose Quantum Runtime Interface and
QHPC-Interconnect layers mediate between resource managers and devices; it
names as an open gap that communication semantics in these stacks must
extend beyond one-way circuit submission to bidirectional interaction. QQLink
is not a stack in this sense. It manages neither processes nor devices, its
task layer could be hosted by QFw, and its workflows could be scheduled by a
multi-tenant placement framework such as CloudQC~\cite{zhou2025cloudqc},
which allocates many independent jobs across a QPU cluster. What QQLink
supplies, at the application and workflow layer, is one form of the
bidirectional interaction the survey identifies as missing, with measurement outcomes that
cross between programs, entanglement declared as a resource, and a planner
that turns both into a dependency graph the runtime executes. Three further
works are close in intent.
NetQIR~\cite{cardama2026} extends the Quantum Intermediate Representation (QIR) with distributed primitives and is a
possible lowering target below QQLink's planner. Zhang et
al.~\cite{zhang2025simdisq} simulate distributed circuits end to end with
communication-noise models by inserting teleported gates into a monolithic
circuit, whereas QQLink starts from per-node programs and also executes them
on hardware. Finally, Gupta et al.~\cite{gupta2026} quantify the crossover
between gate teleportation and circuit cutting as a function of interconnect
noise, which indicates when a runtime built around entanglement and classical
channels is preferable to cutting alone.

\paragraph{Experimental distributed quantum computing.}
Main et al.~\cite{main2025} executed a distributed Grover search across an
optical network link between two ion-trap modules using teleported gates. The
Grover workflow of Section~\ref{sec:demo-grover} reproduces the structure of
that experiment at the application layer; the difference is that our execution
substrate is a partitioned single chip rather than two networked modules, a
limitation discussed next.

\section{Limitations and Threats to Validity}
\label{sec:limitations}

\paragraph{Execution substrate.}
QQLink validates multi-QPU workflow semantics in two settings, namely simulator
execution and same-chip virtual-QPU execution on two IBM backends, each
partitioned on its own. These
results support the programming model and execution layer, but they do not
constitute an experimental study of physically distributed multi-chip
execution across networked hardware. Protocol-level results are obtained
through the co-located path, and the separate-executor experiment of
Section~\ref{sec:separate} distributes the control plane but keeps the joint
state in one entanglement-service process, because two independent state
vectors cannot share a Bell pair; its latencies are emulated on one machine.
Extending the channel path to provider-native executors for physical links
across distinct chips is an open engineering task. On a partitioned chip, a
``non-local'' gate is a native two-qubit gate between sectors, so hardware
counts in Table~\ref{tab:hardware-summary} bound only the same-chip
realization and say nothing about link-limited fidelity.

\paragraph{Hardware evaluation depth.}
Section~\ref{sec:hardware} states what the runs establish; four features of
their design limit how far the results generalize. First, the suite comes
from two sessions a week apart on each of two devices, and success changed
between sessions by up to 31 points on \texttt{ibm\_kingston} and 14 on
\texttt{ibm\_fez}; two sessions show this variation but do not
characterize its distribution. Second, the margins over the controls are
pooled-shot intervals that exclude the variation between repeats, and the
\texttt{ibm\_fez} controls of 2026-10-16 ran on other qubits than the
suite. Third, the probes use different sectors on different days, and the
device-noise simulations are unscheduled, assign no error to idle intervals,
and come from a calibration that did not predict the second-session loss.
Fourth, the coherence-sensitive checks ran in one or two sessions on one
sector each, and for the Toffoli gate and the repetition code they fail, so
for these workflows the runs establish classical input--output behavior and
the routing of outcomes to corrections, not coherent operation. The noise
sweep of Section~\ref{sec:robustness} quantifies protocol fragility only
within a depolarizing-plus-readout model, which omits these effects.

\paragraph{Parallelism is exposed but not exploited.}
The plan records dependencies at qubit granularity, so the per-qubit chains
of Fig.~\ref{fig:dag} are independent and the teleportation critical path is
three units for any register size. The default dispatcher still executes
units one at a time in a topological order, so this parallelism is not yet
realized; a concurrent dispatcher is future work, and since the sequential
dispatcher costs 2 to 8 microseconds per unit in all but one configuration
of the synthetic sweep (48 in that one), it would pay off only against
backend latency, not against orchestration cost. The separate-executor
experiment shows the gain in the setting where it matters. There, independent
units on different nodes run concurrently, and the span follows the
cross-node critical path.

\paragraph{Single-bit conditionals.}
A conditional tests one measurement bit against one value. Decisions that
depend on several bits, such as the AND in the repetition-code decoder, are
expressed by computing the function on an ancilla with classically
controlled gates and measuring it, which costs a qubit and a measurement per
Boolean gate, and in the repetition code 19 to 40 two-qubit gates after
routing (Section~\ref{sec:cost}). This is a limitation of QQLink, not of the targets.
OpenQASM~3 compares whole bit registers with integers in its
conditions~\cite{cross2022}, and Qiskit's dynamic circuits branch on
classical expressions over several bits, including the comparisons and
bitwise AND, OR, and XOR that IBM Heron processors accept in real
time~\cite{qiskit2024}. A condition over several measurement
references, lowered to such an expression in \texttt{if\_test} when the
references end up in one circuit and evaluated by the receiving node when
they arrive from several, is a direct extension of the model; it would remove
the decoder ancilla of the repetition code.

\paragraph{Noise-aware planning.}
The sector selector uses backend target data to prefer lower-error
cross-sector entangling pairs and cleaner measurement endpoints, but the
general orchestrator does not incorporate broader device signals such as
$T_1$/$T_2$ coherence times, global routing cost, or multi-hop coupling
constraints into end-to-end placement decisions. The plugin architecture
supports custom placement policies, and integrating richer noise-aware
heuristics is a natural next step.

\paragraph{Scale and construct validity.}
The protocol workflows involve two to three circuits and up to 21 qubits per
configuration; the planner was exercised on synthetic programs up to 64
circuits and $5\times10^4$ instructions, but not on real protocols of that
size. Plan metrics and instruction counts measure the orchestration layer's
bookkeeping; wall-clock overhead is measured with a no-op dispatcher and on
simulators, not against hardware queue and execution times.

\section{Conclusion}
\label{sec:conclusion}

The coordination that multi-QPU protocols need can be derived from one
program over all nodes rather than written by hand per node. In QQLink the
program is a set of workflow nodes with deferred measurement references and
declared entanglement, and the planner derives from it, in linear time, a
provably acyclic dependency graph whose topological orders all give the
same result. The strongest evidence is that the plan alone carries the
coordination. On the simulator runtime path, exhaustive branch tests over
tomographically complete input sets give every tested quantum output state
an infidelity of at most $5\times10^{-12}$ relative to its ideal output, and
when each node runs in its own
process and learns of the others only through delayed messages, every run
with the full plan is correct while removing its cross-node edges breaks the
protocols that need them.

The hardware runs place a limit on same-chip execution that the gate counts
do not show. The workflows built from teleported gates fall up to 65 points
short of a simulation with the device's gate and readout errors, change by
up to 31 points between sessions on qubits that the reported calibration
does not distinguish, and their coherence-sensitive checks do not establish
coherent Toffoli action or preservation of the intended repetition-code
superposition. Their success depends on how the dynamic
control is executed, as feedforward or as coherent control, and the
direction of that effect differs between sectors; we have not identified the
process responsible. The plan marks every correction, so
choosing between feedforward and coherent control is a decision the hardware
path can make per correction. The next steps follow from the same structure. They are a dispatcher that runs
independent units concurrently, a distribution compiler as a producer of
QQLink programs, conditions over several measurement references, and
provider-native channels toward multi-chip execution. The abstractions were
designed so that none of these requires changing existing application
programs.

\section*{Data and Code Availability}

The QQLink source code, the workflow library and example scripts, the
experiment harness that generates every table and figure of
Section~\ref{sec:eval} (\texttt{experiments/}), the resulting JSON records,
and the hardware run records (counts, calibration snapshots, physical-qubit
assignments, and job identifiers) will be released under an open-source
license at \url{https://github.com/ORNL-QCI/qqlink} upon publication, with
the commit corresponding to this manuscript tagged and archived with a DOI.
For review, a snapshot of this commit with the code, scripts, and records is
deposited on Zenodo with restricted access and is available to reviewers at
\url{https://doi.org/10.5281/zenodo.23570273}.

\begin{acks}
This work was supported by the U.S. Department of Energy, Office of Science,
under Contract No.\ DE-AC05-00OR22725, with funding from the Advanced
Scientific Computing Research (ASCR) Accelerated Research in Quantum Computing
(ARQC) Modular and Error-Aware Software Stack for Heterogeneous Quantum
Computing Ecosystems (MACH-Q) project. We acknowledge the use
of resources of the Quantum Computing User Program, which is managed by the
Oak Ridge Leadership Computing Facility, a DOE Office of Science User
Facility supported under Contract DE-AC05-00OR22725.
\end{acks}

\bibliographystyle{ACM-Reference-Format}
\bibliography{references}

@article{caleffi2022,
  author  = {Caleffi, Marcello and Amoretti, Michele and Ferrari, Davide and Illiano, Jessica and Manzalini, Antonio and Cacciapuoti, Angela Sara},
  title   = {Distributed Quantum Computing: A Survey},
  journal = {Computer Networks},
  volume  = {254},
  pages   = {110672},
  year    = {2024},
  doi     = {10.1016/j.comnet.2024.110672},
}

@article{cuomo2020,
  author  = {Cuomo, Daniele and Caleffi, Marcello and Cacciapuoti, Angela Sara},
  title   = {Towards a Distributed Quantum Computing Ecosystem},
  journal = {IET Quantum Communication},
  volume  = {1},
  number  = {1},
  pages   = {3--8},
  year    = {2020},
  doi     = {10.1049/iet-qtc.2020.0002},
}

@article{main2025,
  author  = {Main, D. and Drmota, P. and Nadlinger, D. P. and Ainley, E. M. and Agrawal, A. and Nichol, B. C. and Srinivas, R. and Araneda, G. and Lucas, D. M.},
  title   = {Distributed Quantum Computing across an Optical Network Link},
  journal = {Nature},
  volume  = {638},
  number  = {8050},
  pages   = {383--388},
  year    = {2025},
  doi     = {10.1038/s41586-024-08404-x},
}

@article{gottesman1999,
  author  = {Gottesman, Daniel and Chuang, Isaac L.},
  title   = {Demonstrating the Viability of Universal Quantum Computation Using Teleportation and Single-Qubit Operations},
  journal = {Nature},
  volume  = {402},
  number  = {6760},
  pages   = {390--393},
  year    = {1999},
  doi     = {10.1038/46503},
}

@article{bennett1993,
  author  = {Bennett, Charles H. and Brassard, Gilles and Cr{\'e}peau, Claude and Jozsa, Richard and Peres, Asher and Wootters, William K.},
  title   = {Teleporting an Unknown Quantum State via Dual Classical and {Einstein-Podolsky-Rosen} Channels},
  journal = {Physical Review Letters},
  volume  = {70},
  number  = {13},
  pages   = {1895--1899},
  year    = {1993},
  doi     = {10.1103/PhysRevLett.70.1895},
}

@article{bennett1992,
  author  = {Bennett, Charles H. and Wiesner, Stephen J.},
  title   = {Communication via One- and Two-Particle Operators on {Einstein-Podolsky-Rosen} States},
  journal = {Physical Review Letters},
  volume  = {69},
  number  = {20},
  pages   = {2881--2884},
  year    = {1992},
  doi     = {10.1103/PhysRevLett.69.2881},
}

@inproceedings{grover1996,
  author    = {Grover, Lov K.},
  title     = {A Fast Quantum Mechanical Algorithm for Database Search},
  booktitle = {Proceedings of the 28th Annual ACM Symposium on Theory of Computing (STOC '96)},
  pages     = {212--219},
  year      = {1996},
  publisher = {ACM},
  address   = {New York, NY, USA},
  doi       = {10.1145/237814.237866},
}

@inproceedings{yimsiriwattana2004,
  author    = {Yimsiriwattana, Anocha and Lomonaco Jr., Samuel J.},
  title     = {Distributed Quantum Computing: A Distributed {Shor} Algorithm},
  booktitle = {Quantum Information and Computation II},
  series    = {Proceedings of SPIE},
  volume    = {5436},
  pages     = {360--372},
  year      = {2004},
  publisher = {SPIE},
  doi       = {10.1117/12.546504},
}

@misc{qiskit2024,
  author        = {Javadi-Abhari, Ali and Treinish, Matthew and Krsulich, Kevin and Wood, Christopher J. and Lishman, Jake and Gacon, Julien and Martiel, Simon and Nation, Paul D. and Bishop, Lev S. and Cross, Andrew W. and Johnson, Blake R. and Gambetta, Jay M.},
  title         = {Quantum Computing with {Qiskit}},
  year          = {2024},
  eprint        = {2405.08810},
  archivePrefix = {arXiv},
  primaryClass  = {quant-ph},
  doi           = {10.48550/arXiv.2405.08810},
}

@misc{cirq2024,
  author = {{Cirq Developers}},
  title  = {Cirq: A Python Framework for Creating, Editing, and Invoking Noisy Intermediate Scale Quantum Circuits},
  year   = {2024},
  doi    = {10.5281/zenodo.4062499},
  note   = {\url{https://quantumai.google/cirq}},
}

@misc{bergholm2022,
  author        = {Bergholm, Ville and Izaac, Josh and Schuld, Maria and Gogolin, Christian and others},
  title         = {{PennyLane}: Automatic Differentiation of Hybrid Quantum-Classical Computations},
  year          = {2018},
  eprint        = {1811.04968},
  archivePrefix = {arXiv},
  primaryClass  = {quant-ph},
  doi           = {10.48550/arXiv.1811.04968},
}

@article{peng2020,
  author  = {Peng, Tianyi and Harrow, Aram W. and Ozols, Maris and Wu, Xiaodi},
  title   = {Simulating Large Quantum Circuits on a Small Quantum Computer},
  journal = {Physical Review Letters},
  volume  = {125},
  number  = {15},
  pages   = {150504},
  year    = {2020},
  doi     = {10.1103/PhysRevLett.125.150504},
}

@inproceedings{tang2021,
  author    = {Tang, Wei and Tomesh, Teague and Suchara, Martin and Larson, Jeffrey and Martonosi, Margaret},
  title     = {{CutQC}: Using Small Quantum Computers for Large Quantum Circuit Evaluations},
  booktitle = {Proceedings of the 26th ACM International Conference on Architectural Support for Programming Languages and Operating Systems (ASPLOS '21)},
  pages     = {473--486},
  year      = {2021},
  publisher = {ACM},
  address   = {New York, NY, USA},
  doi       = {10.1145/3445814.3446758},
}

@article{andres2019,
  author  = {Andr{\'e}s-Mart{\'i}nez, Pablo and Heunen, Chris},
  title   = {Automated Distribution of Quantum Circuits via Hypergraph Partitioning},
  journal = {Physical Review A},
  volume  = {100},
  number  = {3},
  pages   = {032308},
  year    = {2019},
  doi     = {10.1103/PhysRevA.100.032308},
}

@article{ferrari2023,
  author  = {Ferrari, Davide and Cacciapuoti, Angela Sara and Amoretti, Michele and Caleffi, Marcello},
  title   = {Compiler Design for Distributed Quantum Computing},
  journal = {IEEE Transactions on Quantum Engineering},
  volume  = {2},
  pages   = {1--20},
  eid     = {4100720},
  year    = {2021},
  doi     = {10.1109/TQE.2021.3053921},
}

@article{coopmans2021,
  author  = {Coopmans, Tim and Knegjens, Robert and Dahlberg, Axel and Maier, David and Nijsten, Loek and de Oliveira Filho, Julio and Papendrecht, Martijn and Rabbie, Julian and Rozp{\k{e}}dek, Filip and Skrzypczyk, Matthew and Wubben, Leon and de Jong, Walter and Podareanu, Damian and Torres-Knoop, Ariana and Elkouss, David and Wehner, Stephanie},
  title   = {{NetSquid}, a {NETwork} Simulator for {QUantum} Information Using Discrete Events},
  journal = {Communications Physics},
  volume  = {4},
  number  = {1},
  pages   = {164},
  year    = {2021},
  doi     = {10.1038/s42005-021-00647-8},
}

@article{wu2021,
  author  = {Wu, Xiaoliang and Kolar, Alexander and Chung, Joaquin and Jin, Dong and Zhong, Tian and Kettimuthu, Rajkumar and Suchara, Martin},
  title   = {{SeQUeNCe}: A Customizable Discrete-Event Simulator of Quantum Networks},
  journal = {Quantum Science and Technology},
  volume  = {6},
  number  = {4},
  pages   = {045027},
  year    = {2021},
  doi     = {10.1088/2058-9565/ac22f6},
}

@article{qulacs2021,
  author  = {Suzuki, Yasunari and Kawase, Yoshiaki and Masumura, Yuya and Hiraga, Yuria and Nakadai, Masahiro and Chen, Jiabao and Nakanishi, Ken M. and Mitarai, Kosuke and Imai, Ryosuke and Tamiya, Shiro and Yamamoto, Takahiro and Yan, Tennin and Kawakubo, Toru and Nakagawa, Yuya O. and Ibe, Yohei and Zhang, Youyuan and Yamashita, Hirotsugu and Yoshimura, Hikaru and Hayashi, Akihiro and Fujii, Keisuke},
  title   = {{Qulacs}: A Fast and Versatile Quantum Circuit Simulator for Research Purpose},
  journal = {Quantum},
  volume  = {5},
  pages   = {559},
  year    = {2021},
  doi     = {10.22331/q-2021-10-06-559},
}

@misc{covalent2022,
  author = {{Agnostiq Inc.}},
  title  = {Covalent: Pythonic Tool for Orchestrating Hybrid Quantum-Classical Workflows},
  year   = {2022},
  note   = {\url{https://github.com/AgnostiqHQ/covalent}},
}

@article{monroe2014,
  author  = {Monroe, C. and Raussendorf, R. and Ruthven, A. and Brown, K. R. and Maunz, P. and Duan, L.-M. and Kim, J.},
  title   = {Large-Scale Modular Quantum-Computer Architecture with Atomic Memory and Photonic Interconnects},
  journal = {Physical Review A},
  volume  = {89},
  number  = {2},
  pages   = {022317},
  year    = {2014},
  doi     = {10.1103/PhysRevA.89.022317},
}

@article{jnane2022,
  author  = {Jnane, Hamza and Undseth, Brennan and Cai, Zhenyu and Benjamin, Simon C. and Koczor, B{\'a}lint},
  title   = {Multicore Quantum Computing},
  journal = {Physical Review Applied},
  volume  = {18},
  number  = {4},
  pages   = {044064},
  year    = {2022},
  doi     = {10.1103/PhysRevApplied.18.044064},
}

@article{corcoles2021,
  author  = {C{\'o}rcoles, A. D. and Takita, Maika and Inoue, Ken and Lekuch, Scott and Minev, Zlatko K. and Chow, Jerry M. and Gambetta, Jay M.},
  title   = {Exploiting Dynamic Quantum Circuits in a Quantum Algorithm with Superconducting Qubits},
  journal = {Physical Review Letters},
  volume  = {127},
  number  = {10},
  pages   = {100501},
  year    = {2021},
  doi     = {10.1103/PhysRevLett.127.100501},
}

@article{cross2022,
  author  = {Cross, Andrew and Javadi-Abhari, Ali and Alexander, Thomas and de Beaudrap, Niel and Bishop, Lev S. and Heidel, Steven and Ryan, Colm A. and Sivarajah, Prasahnt and Smolin, John and Gambetta, Jay M. and Johnson, Blake R.},
  title   = {{OpenQASM} 3: A Broader and Deeper Quantum Assembly Language},
  journal = {ACM Transactions on Quantum Computing},
  volume  = {3},
  number  = {3},
  pages   = {12:1--12:50},
  year    = {2022},
  doi     = {10.1145/3505636},
}

@inproceedings{baker2020,
  author    = {Baker, Jonathan M. and Duckering, Casey and Hoover, Alexander and Chong, Frederic T.},
  title     = {Time-Sliced Quantum Circuit Partitioning for Modular Architectures},
  booktitle = {Proceedings of the 17th ACM International Conference on Computing Frontiers (CF '20)},
  pages     = {98--107},
  year      = {2020},
  publisher = {ACM},
  address   = {New York, NY, USA},
  doi       = {10.1145/3387902.3392617},
}

@inproceedings{wu2022,
  author    = {Wu, Anbang and Zhang, Hezi and Li, Gushu and Shabani, Alireza and Xie, Yuan and Ding, Yufei},
  title     = {{AutoComm}: A Framework for Enabling Efficient Communication in Distributed Quantum Programs},
  booktitle = {55th IEEE/ACM International Symposium on Microarchitecture (MICRO)},
  pages     = {1027--1041},
  year      = {2022},
  publisher = {IEEE},
  doi       = {10.1109/MICRO56248.2022.00074},
}

@article{diadamo2021,
  author  = {DiAdamo, Stephen and N{\"o}tzel, Janis and Zanger, Benjamin and Be{\c{s}}e, Mehmet Mert},
  title   = {{QuNetSim}: A Software Framework for Quantum Networks},
  journal = {IEEE Transactions on Quantum Engineering},
  volume  = {2},
  pages   = {1--12},
  eid     = {2500512},
  year    = {2021},
  doi     = {10.1109/TQE.2021.3092395},
}

@inproceedings{parekh2021,
  author    = {Parekh, Rhea and Ricciardi, Andrea and Darwish, Ahmed and DiAdamo, Stephen},
  title     = {Quantum Algorithms and Simulation for Parallel and Distributed Quantum Computing},
  booktitle = {2021 IEEE/ACM Second International Workshop on Quantum Computing Software (QCS)},
  pages     = {9--19},
  year      = {2021},
  publisher = {IEEE},
  doi       = {10.1109/QCS54837.2021.00005},
}

@article{dahlberg2022,
  author  = {Dahlberg, Axel and van der Vecht, Bart and Delle Donne, Carlo and Skrzypczyk, Matthew and te Raa, Ingmar and Kozlowski, Wojciech and Wehner, Stephanie},
  title   = {{NetQASM}---A Low-Level Instruction Set Architecture for Hybrid Quantum--Classical Programs in a Quantum Internet},
  journal = {Quantum Science and Technology},
  volume  = {7},
  number  = {3},
  pages   = {035023},
  year    = {2022},
  doi     = {10.1088/2058-9565/ac753f},
}

@inproceedings{haner2021,
  author    = {H{\"a}ner, Thomas and Steiger, Damian S. and Hoefler, Torsten and Troyer, Matthias},
  title     = {Distributed Quantum Computing with {QMPI}},
  booktitle = {Proceedings of the International Conference for High Performance Computing, Networking, Storage and Analysis (SC '21)},
  pages     = {1--13},
  year      = {2021},
  publisher = {ACM},
  address   = {New York, NY, USA},
  doi       = {10.1145/3458817.3476172},
}

@article{mccaskey2020,
  author  = {McCaskey, Alexander J. and Lyakh, Dmitry I. and Dumitrescu, Eugene F. and Powers, Sarah S. and Humble, Travis S.},
  title   = {{XACC}: A System-Level Software Infrastructure for Heterogeneous Quantum--Classical Computing},
  journal = {Quantum Science and Technology},
  volume  = {5},
  number  = {2},
  pages   = {024002},
  year    = {2020},
  doi     = {10.1088/2058-9565/ab6bf6},
}

@inproceedings{cudaq,
  author    = {Kim, Jin-Sung and McCaskey, Alex and Heim, Bettina and Modani, Manish and Stanwyck, Sam and Costa, Timothy},
  title     = {{CUDA Quantum}: The Platform for Integrated Quantum-Classical Computing},
  booktitle = {2023 60th ACM/IEEE Design Automation Conference (DAC)},
  pages     = {1--4},
  year      = {2023},
  publisher = {IEEE},
  doi       = {10.1109/DAC56929.2023.10247886},
}

@misc{qiskitserverless,
  author       = {{IBM Quantum}},
  title        = {Qiskit Serverless},
  howpublished = {\url{https://github.com/Qiskit/qiskit-serverless}},
  year         = {2024},
  note         = {Accessed 2026-09-10},
}

@inproceedings{mantha2025,
  author    = {Mantha, Pradeep and Kiwit, Florian J. and Saurabh, Nishant and Jha, Shantenu and Luckow, Andre},
  title     = {{Pilot-Quantum}: A Middleware for Quantum-{HPC} Resource, Workload and Task Management},
  booktitle = {2025 IEEE 25th International Symposium on Cluster, Cloud and Internet Computing (CCGrid)},
  pages     = {1--10},
  year      = {2025},
  publisher = {IEEE},
  doi       = {10.1109/CCGRID64434.2025.00070},
}

@article{barral2025,
  author  = {Barral, David and Cardama, F. Javier and D{\'\i}az-Camacho, Guillermo and Fa{\'\i}lde, Daniel and Llovo, Iago F. and Mussa-Juane, Mariamo and V{\'a}zquez-P{\'e}rez, Jorge and Villasuso, Juan and Pi{\~n}eiro, C{\'e}sar and Costas, Natalia and Pichel, Juan C. and Pena, Tom{\'a}s F. and G{\'o}mez, Andr{\'e}s},
  title   = {Review of Distributed Quantum Computing: From Single {QPU} to High Performance Quantum Computing},
  journal = {Computer Science Review},
  volume  = {57},
  pages   = {100747},
  year    = {2025},
  doi     = {10.1016/j.cosrev.2025.100747},
}

@article{eisert2000,
  author  = {Eisert, J. and Jacobs, K. and Papadopoulos, P. and Plenio, M. B.},
  title   = {Optimal Local Implementation of Nonlocal Quantum Gates},
  journal = {Physical Review A},
  volume  = {62},
  number  = {5},
  pages   = {052317},
  year    = {2000},
  doi     = {10.1103/PhysRevA.62.052317},
}

@article{bravyi2005,
  author  = {Bravyi, Sergey and Kitaev, Alexei},
  title   = {Universal Quantum Computation with Ideal {Clifford} Gates and Noisy Ancillas},
  journal = {Physical Review A},
  volume  = {71},
  number  = {2},
  pages   = {022316},
  year    = {2005},
  doi     = {10.1103/PhysRevA.71.022316},
}

@book{nielsen2010,
  author    = {Nielsen, Michael A. and Chuang, Isaac L.},
  title     = {Quantum Computation and Quantum Information},
  edition   = {10th Anniversary},
  publisher = {Cambridge University Press},
  address   = {Cambridge, UK},
  year      = {2010},
  doi       = {10.1017/CBO9780511976667},
}

@article{claudino2024,
  author  = {Claudino, Daniel and Lyakh, Dmitry I. and McCaskey, Alexander J.},
  title   = {Parallel Quantum Computing Simulations via Quantum Accelerator Platform Virtualization},
  journal = {Future Generation Computer Systems},
  volume  = {160},
  pages   = {264--273},
  year    = {2024},
  doi     = {10.1016/j.future.2024.06.007},
}

@article{liu2024nonlocal,
  author  = {Liu, Xiao and Hu, Xiao-Min and Zhu, Tian-Xiang and Zhang, Chao and Xiao, Yi-Xin and Miao, Jia-Le and Ou, Zhong-Wen and Li, Pei-Yun and Liu, Bi-Heng and Zhou, Zong-Quan and Li, Chuan-Feng and Guo, Guang-Can},
  title   = {Nonlocal Photonic Quantum Gates over 7.0 km},
  journal = {Nature Communications},
  volume  = {15},
  pages   = {8529},
  year    = {2024},
  doi     = {10.1038/s41467-024-52912-3},
}

@article{zhou2023bv,
  author  = {Zhou, Xu and Qiu, Daowen and Luo, Le},
  title   = {Distributed {Bernstein--Vazirani} Algorithm},
  journal = {Physica A: Statistical Mechanics and its Applications},
  volume  = {629},
  pages   = {129209},
  year    = {2023},
  doi     = {10.1016/j.physa.2023.129209},
}

@inproceedings{chundury2025,
  title     = {Scaling Hybrid Quantum-{HPC} Applications with the {Quantum Framework}},
  author    = {Chundury, Srikar and Shehata, Amir and Kim, Seongmin and Gopalakrishnan Meena, Muralikrishnan and Lu, Chao and Gottiparthi, Kalyana and Coello Perez, Eduardo Antonio and Mueller, Frank and Suh, In-Saeng},
  booktitle = {Proceedings of the SC '25 Workshops of the International Conference for High Performance Computing, Networking, Storage and Analysis},
  publisher = {ACM},
  pages     = {1888--1897},
  year      = {2025},
  doi       = {10.1145/3731599.3767553},
  note      = {arXiv:2509.14470}
}

@misc{shehata2026,
  title        = {Quantum-{HPC} Software Stacks and the {openQSE} Reference Architecture: A Survey},
  author       = {Shehata, Amir and Austin, Brian and Beck, Tom and Burgholzer, Lukas and Chernoguzov, Alex and Churchill, Spencer and Delgado, Andrea and Eckert, Yasuko and Heckey, Jeffery and Kissell, Kevin and Klymko, Katherine and Moles, Josh and Naughton, Thomas and O'Riordan, Lee James and Ortiz Pauyac, Christian and Prawiroatmodjo, Guen and Rrapaj, Ermal and Schindler, Jiri and Schulz, Laura and Stern, Sebastian and Takeshita, Tyler and Tsuji, Miwako and Wennersteen, Aleksander and Humble, Travis and Schulz, Martin},
  year         = {2026},
  eprint       = {2604.20912},
  archivePrefix = {arXiv},
  primaryClass = {quant-ph},
  howpublished = {arXiv:2604.20912}
}

@inproceedings{burgholzer2026,
  title     = {The {Munich Quantum Software Stack}: Connecting End Users, Integrating Diverse Quantum Technologies, Accelerating {HPC}},
  author    = {Burgholzer, Lukas and Echavarria, Jorge and Hopf, Patrick and Stade, Yannick and Rovara, Damian and Schmid, Ludwig and Kaya, Erc{\"u}ment and Mete, Burak and Farooqi, Muhammad Nufail and Chung, Minh and De Pascale, Marco and Schulz, Laura and Schulz, Martin and Wille, Robert},
  booktitle = {Proceedings of the Supercomputing Asia and International Conference on High Performance Computing in Asia Pacific Region (SCA/HPCAsia '26)},
  publisher = {ACM},
  pages     = {55--67},
  year      = {2026},
  doi       = {10.1145/3773656.3773669},
  note      = {arXiv:2509.02674}
}

@misc{zhou2025cloudqc,
  title        = {{CloudQC}: A Network-aware Framework for Multi-tenant Distributed Quantum Computing},
  author       = {Zhou, Ruilin and Gan, Yuhang and Liu, Yi and Qian, Chen},
  year         = {2025},
  eprint       = {2504.20389},
  archivePrefix = {arXiv},
  primaryClass = {quant-ph},
  howpublished = {arXiv:2504.20389}
}

@article{delledonne2025,
  title   = {An operating system for executing applications on quantum network nodes},
  author  = {Delle Donne, C. and Iuliano, M. and van der Vecht, B. and Ferreira, G. M. and Jirovsk{\'a}, H. and van der Steenhoven, T. J. W. and Dahlberg, A. and Skrzypczyk, M. and Fioretto, D. and Teller, M. and Filippov, P. and Montblanch, A. R.-P. and Fischer, J. and van Ommen, H. B. and Demetriou, N. and Leichtle, D. and Music, L. and Ollivier, H. and te Raa, I. and Kozlowski, W. and Taminiau, T. H. and Pawe{\l}czak, P. and Northup, T. E. and Hanson, R. and Wehner, S.},
  journal = {Nature},
  volume  = {639},
  pages   = {321--328},
  year    = {2025},
  doi     = {10.1038/s41586-025-08704-w}
}

@article{cardama2026,
  title   = {{NetQIR}: An extension of {QIR} for distributed quantum computing},
  author  = {Cardama, F. Javier and V{\'a}zquez-P{\'e}rez, Jorge and Pi{\~n}eiro, C{\'e}sar and Pena, Tom{\'a}s F. and Pichel, Juan C. and G{\'o}mez, Andr{\'e}s},
  journal = {Future Generation Computer Systems},
  volume  = {174},
  pages   = {107989},
  year    = {2026},
  doi     = {10.1016/j.future.2025.107989}
}

@misc{zhang2025simdisq,
  title        = {An End-to-End Distributed Quantum Circuit Simulator},
  author       = {Zhang, Sen and Xiong, Lingjun and Liu, Yipie and Mark, Brian L. and Yang, Lei and Yang, Zebo and Jiang, Weiwen},
  year         = {2025},
  eprint       = {2511.19791},
  archivePrefix = {arXiv},
  primaryClass = {quant-ph},
  howpublished = {arXiv:2511.19791}
}

@article{gupta2026,
  title   = {Gate Teleportation vs.\ Circuit Cutting in Distributed Quantum Computing},
  author  = {Gupta, Shobhit and Dilley, Daniel J. and Sheshko, Nikolay and Gonzales, Alvin and Sullivan, Sean E. and Singh, Manish K. and Saleem, Zain H.},
  journal = {Advanced Quantum Technologies},
  volume  = {9},
  pages   = {e70424},
  year    = {2026},
  doi     = {10.1002/qute.70424},
  note    = {arXiv:2510.08894}
}

@article{chen2026adaptive,
  author  = {Chen, Kuan-Cheng and Burt, Felix and Panigrahy, Nitish K. and Leung, Kin K.},
  title   = {Adaptive Resource Orchestration for Distributed Quantum Computing Systems},
  journal = {IEEE Internet Computing},
  year    = {2026},
  volume  = {30},
  number  = {1},
  pages   = {16--25},
  doi     = {10.1109/MIC.2026.3652322}
}

@inproceedings{pham2026qcoem,
  author    = {Pham, Tam N. and Nguyen, Hoa T. and Le-Trung, Quan},
  title     = {{QCOEM}: Quantum Cloud Orchestration with Evolutionary Multi-Objective Optimization},
  booktitle = {2026 IEEE 19th International Conference on Cloud Computing (CLOUD)},
  year      = {2026},
  pages     = {460--466},
  publisher = {IEEE},
  doi       = {10.1109/CLOUD72782.2026.00059}
}

@inproceedings{weder2021hybrid,
  author    = {Weder, Benjamin and Barzen, Johanna and Leymann, Frank and Zimmermann, Michael},
  title     = {Hybrid Quantum Applications Need Two Orchestrations in Superposition: A Software Architecture Perspective},
  booktitle = {2021 IEEE International Conference on Web Services (ICWS)},
  year      = {2021},
  publisher = {IEEE},
  doi       = {10.1109/ICWS53863.2021.00015}
}

@inproceedings{beisel2025operations,
  author    = {Beisel, Martin and Barzen, Johanna and Leymann, Frank and Weder, Benjamin},
  title     = {Operations Patterns for Hybrid Quantum Applications},
  booktitle = {Proceedings of the 15th International Conference on Cloud Computing and Services Science (CLOSER)},
  publisher = {SciTePress},
  year      = {2025},
  pages     = {26--36},
  doi       = {10.5220/0013246900003950}
}

@inproceedings{weder2025qunicorn,
  author    = {Weder, Benjamin and Barzen, Johanna and Beisel, Martin and B{\"u}hler, Fabian and Georg, Daniel and Leymann, Frank and Stiliadou, Lavinia},
  title     = {{Qunicorn}: A Middleware for the Unified Execution Across Heterogeneous Quantum Cloud Offerings},
  booktitle = {2025 IEEE/ACM International Workshop on Quantum Software Engineering (Q-SE)},
  year      = {2025},
  pages     = {17--24},
  publisher = {IEEE},
  doi       = {10.1109/Q-SE66736.2025.00009}
}

@inproceedings{vazquezperez2026emulating,
  author    = {V{\'a}zquez-P{\'e}rez, Jorge and Cardama, F. Javier and Pena, Tom{\'a}s F. and G{\'o}mez, Andr{\'e}s},
  title     = {Emulating {NetQMPI} Applications with {CUNQA}: A Decoupled Architecture for {HPC} Environments},
  booktitle = {2026 IEEE 46th International Conference on Distributed Computing Systems Workshops (ICDCSW)},
  year      = {2026},
  pages     = {178--183},
  publisher = {IEEE},
  doi       = {10.1109/ICDCSW72724.2026.00045}
}

@article{naito2026network,
  author  = {Naito, Soshun and Suzuki, Yasunari and Tokunaga, Yuuki},
  title   = {Network-Based Quantum Computing: An Efficient Design Framework for Many-Small-Node Distributed Fault-Tolerant Quantum Computing},
  journal = {IEEE Transactions on Quantum Engineering},
  year    = {2026},
  volume  = {7},
  pages   = {3104632},
  doi     = {10.1109/TQE.2026.3722428}
}

@article{andres2024distributing,
  author  = {Andres-Martinez, Pablo and Forrer, Tim and Mills, Daniel and Wu, Jun-Yi and Henaut, Luciana and Yamamoto, Kentaro and Murao, Mio and Duncan, Ross},
  title   = {Distributing circuits over heterogeneous, modular quantum computing network architectures},
  journal = {Quantum Science and Technology},
  volume  = {9},
  number  = {4},
  pages   = {045021},
  year    = {2024},
  doi     = {10.1088/2058-9565/ad6734}
}

@misc{vandervecht2025qoala,
  author        = {van der Vecht, Bart and Y{\"u}cel, Atak Talay and Jirovsk{\'a}, Hana and Wehner, Stephanie},
  title         = {Qoala: an Application Execution Environment for Quantum Internet Nodes},
  year          = {2025},
  eprint        = {2502.17296},
  archivePrefix = {arXiv},
  primaryClass  = {quant-ph},
  doi           = {10.48550/arXiv.2502.17296}
}

\appendix
\section{Supplementary tables}
\label{app:tables}

Table~\ref{tab:hardware-fez} lists the first-session hardware results on the
second device, and Tables~\ref{tab:hardware-sessions}
and~\ref{tab:hardware-sessions-fez} both sessions on each device with every
repeat and all four controls (Section~\ref{sec:hardware}). Table~\ref{tab:planning} lists the execution
plans of Section~\ref{sec:overhead}, of which
Fig.~\ref{fig:resources} plots the teleportation rows, and Table~\ref{tab:profile} the
instruction-level counts of Section~\ref{sec:cost}, of which Fig.~\ref{fig:scaling} plots
the distributed fraction and the instruction density.

\begin{table}[h]
\centering
\caption{Execution plans produced by Algorithm~\ref{alg:plan}. Cross edges
join units of different circuits (footprint or measurement edges); intra
edges join units of one circuit; $L$ is the critical-path length in units.
Plans are byte-identical on every invocation (P1).}
\label{tab:planning}
\begin{tabular}{@{}lcccccc@{}}
\toprule
Workflow & Circuits & Qubits & Units & Cross edges & Intra edges & $L$ \\
\midrule
Dense coding ($k{=}1,2,4$) & 2 & $2k$ & $k{+}1$ & $k$ & 0 & 2 \\
Teleportation ($n{=}1,3,5,7$) & 2 & $3n$ & $2n{+}1$ & $2n$ & $n$ & 3 \\
Entanglement swapping & 3 & 4 & 5 & 3 & 2 & 5 \\
Grover (per marked state) & 2 & 4 & 18 & 7 & 17 & 11 \\
Bernstein--Vazirani ($s{=}1111$) & 2 & 7 & 19 & 7 & 18 & 11 \\
Toffoli (per input) & 3 & 6 & 51--52 & 23 & 52--53 & 30 \\
Repetition code (per case) & 3 & 9 & 41--42 & 18 & 43--44 & 22--23 \\
Accept/reject loop (per round) & 2 & 3 & 3 & 0 & 1 & 2 \\
\bottomrule
\end{tabular}
\end{table}

\begin{table}[h]
\centering
\caption{Instruction-level resource consumption per authored workflow. Local
gates exclude gates inside conditionals; the distributed fraction counts
Bell-pair requests, cross-circuit CNOTs, and conditional instructions.
Grover, Bernstein--Vazirani, Toffoli, and repetition-code counts are for
marked state $01$, hidden string $1111$, input $110$, and logical
$\lvert 1\rangle$ with a flip on B; other inputs differ by at most two
local gates or, for Bernstein--Vazirani, by the number of teleported terms.
The accept/reject row is for one round with the parity check; the
verification check of Script~\ref{lst:magic} has one more local gate.}
\label{tab:profile}
\small
\begin{tabular}{@{}lcccccccc@{}}
\toprule
Workflow & Qubits & Gates & Meas. & Bell & xCNOT & Cond. & Dist.\ frac. & Instr./qubit \\
\midrule
Dense ($k$) & $2k$ & $2k$ & $2k$ & $k$ & $k$ & 0 & 0.333 & 3.00 \\
Teleport ($n$) & $3n$ & $3n$ & $2n$ & $n$ & 0 & $2n$ & 0.375 & 2.67 \\
Swapping & 4 & 2 & 4 & 2 & 0 & 2 & 0.400 & 2.50 \\
Grover & 4 & 22 & 6 & 2 & 0 & 8 & 0.263 & 9.50 \\
Bernstein--Vazirani & 7 & 18 & 8 & 2 & 0 & 8 & 0.278 & 5.14 \\
Toffoli & 6 & 29 & 15 & 6 & 0 & 24 & 0.405 & 12.33 \\
Repetition code & 9 & 16 & 14 & 4 & 0 & 22 & 0.464 & 6.22 \\
Accept/reject loop & 3 & 3 & 1 & 0 & 0 & 1 & 0.200 & 1.67 \\
\bottomrule
\end{tabular}
\end{table}

\begin{table*}[h]
\centering
\caption{Hardware results on the second device, \texttt{ibm\_fez} (Heron r2),
2026-10-09, 17:32--17:40~UTC, in the format of Table~\ref{tab:hardware-summary}.}
\label{tab:hardware-fez}
\scriptsize
\setlength{\tabcolsep}{3pt}
\begin{tabular}{@{}p{1.8cm}p{2.6cm}p{3.2cm}p{1.0cm}p{2.7cm}p{1.5cm}@{}}
\toprule
Workflow & Configuration & Physical qubits per node & 2q / depth & Success (\%) & Control (\%) \\
\midrule
Dense coding & $k{=}1$; message 00 / 01 / 10 / 11 & A: 142; B: 143 & 2 / 15 & 96.9 / 97.0 / 96.8 / 96.3 & -- \\
Dense coding & $k{=}2$; messages (00,01) / (11,10) & A: 141, 142; B: 143, 144 & 7 / 24 & 93.7 / 94.3 & -- \\
Teleportation & $n{=}1$ & A: 142, 143; B: 144 & 2 / 18 & 97.0 [96.7, 97.3] & 52 \\
Teleportation & $n{=}2$ & A: 142, 143, 144, 145; B: 146, 147 & 10 / 32 & 91.0 [90.5, 91.5] & 27 \\
Teleportation & $n{=}3$ & A: 127, 137, 144, 145, 146, 147; B: 141, 142, 143 & 24 / 44 & 84.0 [83.4, 84.7] & 14 \\
Entanglement swapping & $Z$ / $X$ basis & A: 141; C: 142, 143; B: 144 & 3 / 15--18 & 94.5 / 92.9 & 50 / 50 \\
Grover & marked 00 / 01 / 10 / 11 & A: 141, 142; B: 143, 144 & 6 / 43 & 64.6 / 63.9 / 64.8 / 63.7 & 25 / 26 / 26 / 25 \\
Bernstein--Vazirani & $s{=}1001$ / $0110$ (one teleported term) & A: 142, 143, 144; B: 145, 146, 147, 148 & 10 / 35 & 74.7 / 74.5 & 48 / 48 \\
Bernstein--Vazirani & $s{=}1111$ (two teleported terms) & A: 142, 143, 144; B: 145, 146, 147, 148 & 20 / 57 & 18.3 [17.6, 19.0] & 22 \\
Toffoli & input 000 / 001 / 010 / 011 & A: 142, 143; B: 144, 145; C: 146, 147 & 54 / 174 & 51.5 / 43.8 / 43.9 / 36.3 & 32 / 29 / 32 / 29 \\
Toffoli & input 100 / 101 / 110 / 111 & A: 142, 143; B: 144, 145; C: 146, 147 & 54 / 174 & 38.1 / 30.6 / 32.2 / 38.5 & 30 / 28 / 17 / 18 \\
Repetition code & $\lvert 0_L\rangle$; flip none / A / B / C & A: 127, 137; B: 143, 144, 145, 146, 147; C: 141, 142 & 78 / 137--139 & 74.0 / 39.8 / 44.3 / 72.1 & 26 / 11 / 25 / 27 \\
Repetition code & $\lvert 1_L\rangle$; flip none / A / B / C & A: 127, 137; B: 143, 144, 145, 146, 147; C: 141, 142 & 78 / 137--139 & 52.6 / 33.5 / 34.7 / 51.3 & 19 / 9 / 19 / 21 \\
Repetition code & $\lvert +_L\rangle$; flip none / A / B / C & A: 127, 137; B: 143, 144, 145, 146, 147; C: 141, 142 & 78 / 137--139 & 70.6 / 67.3 / 67.0 / 70.2 & 27 / 25 / 26 / 26 \\
\bottomrule
\end{tabular}
\end{table*}

\begin{table*}[p]
\centering
\caption{Both hardware sessions on \texttt{ibm\_kingston}, 2026-10-09 and 2026-10-16,
4096 shots per repeat. Repeats: success (\%) of each of the three repeats.
$\Delta$: pooled second-session minus first-session success, with a 95\%
Newcombe interval (points). Controls: C1 / C2 / C1$'$ / C3 of the second
session (\%; C1 corrections and resets removed, C2 conditional blocks
applied at random, C1$'$ corrections removed and resets kept, C3 corrections
conditioned on an independent coin). Margin: lower bound of the 95\%
Newcombe interval of success minus each control of the same session, the
smallest over the controls, for each session (points). The physical qubits of the first session are those of Table~\ref{tab:hardware-summary}; the sector search placed most workflows on other qubits in the second.}
\label{tab:hardware-sessions}
\scriptsize
\setlength{\tabcolsep}{3pt}
\begin{tabular}{@{}lccccc@{}}
\toprule
Configuration & Repeats, 10-09 & Repeats, 10-16 & $\Delta$ & Controls, 10-16 & Margin, 10-09 / 10-16 \\
\midrule
Dense k1 00 & 97.1 / 98.0 / 97.4 & 97.4 / 97.8 / 97.5 & $+0.1$ [$-0.3$, $+0.4$] & -- & -- \\
Dense k1 01 & 97.6 / 97.1 / 97.6 & 97.1 / 97.0 / 97.1 & $-0.3$ [$-0.7$, $+0.1$] & -- & -- \\
Dense k1 10 & 97.4 / 97.8 / 97.6 & 96.3 / 96.4 / 96.8 & $-1.1$ [$-1.5$, $-0.7$] & -- & -- \\
Dense k1 11 & 97.3 / 97.2 / 97.0 & 96.7 / 96.8 / 96.3 & $-0.6$ [$-1.1$, $-0.2$] & -- & -- \\
Dense k2 00 01 & 92.5 / 92.2 / 92.0 & 92.3 / 91.6 / 91.7 & $-0.4$ [$-1.1$, $+0.3$] & -- & -- \\
Dense k2 11 10 & 93.2 / 91.8 / 92.2 & 91.4 / 92.2 / 91.8 & $-0.6$ [$-1.3$, $+0.0$] & -- & -- \\
Teleport n1 & 95.1 / 94.9 / 95.0 & 97.5 / 98.0 / 97.6 & $+2.7$ [$+2.2$, $+3.1$] & 52 / 50 / 51 / 51 & $+42.3$ / $+44.5$ \\
Teleport n2 & 89.8 / 89.8 / 90.5 & 84.8 / 82.8 / 84.0 & $-6.1$ [$-7.0$, $-5.3$] & 25 / 26 / 26 / 25 & $+63.3$ / $+56.5$ \\
Teleport n3 & 82.8 / 83.2 / 82.1 & 82.4 / 83.0 / 83.2 & $+0.2$ [$-0.8$, $+1.1$] & 11 / 14 / 11 / 13 & $+67.2$ / $+67.7$ \\
Swap Z & 91.9 / 92.5 / 91.9 & 92.4 / 92.9 / 91.8 & $+0.3$ [$-0.4$, $+0.9$] & 50 / 50 / 51 / 50 & $+40.1$ / $+40.1$ \\
Swap X & 92.5 / 91.6 / 92.5 & 93.6 / 94.2 / 94.2 & $+1.8$ [$+1.2$, $+2.4$] & 50 / 50 / 50 / 51 & $+41.1$ / $+41.9$ \\
Grover 00 & 73.3 / 72.9 / 74.6 & 75.0 / 75.3 / 75.5 & $+1.7$ [$+0.6$, $+2.7$] & 27 / 25 / 27 / 26 & $+46.1$ / $+46.8$ \\
Grover 01 & 72.3 / 72.7 / 74.3 & 73.9 / 75.2 / 74.2 & $+1.3$ [$+0.2$, $+2.4$] & 26 / 25 / 24 / 25 & $+45.9$ / $+47.2$ \\
Grover 10 & 74.0 / 72.9 / 73.6 & 75.5 / 74.1 / 74.0 & $+1.0$ [$-0.0$, $+2.1$] & 24 / 24 / 25 / 26 & $+47.1$ / $+46.9$ \\
Grover 11 & 72.4 / 71.7 / 73.0 & 74.4 / 75.5 / 74.5 & $+2.4$ [$+1.3$, $+3.5$] & 24 / 24 / 24 / 26 & $+44.6$ / $+47.4$ \\
BV 1001 & 83.8 / 82.0 / 83.4 & 64.4 / 65.5 / 64.0 & $-18.4$ [$-19.5$, $-17.3$] & 49 / 48 / 49 / 47 & $+35.1$ / $+13.8$ \\
BV 0110 & 83.8 / 83.7 / 83.3 & 63.6 / 64.6 / 64.7 & $-19.3$ [$-20.4$, $-18.3$] & 50 / 47 / 48 / 48 & $+34.3$ / $+12.9$ \\
BV 1111 & 45.8 / 43.8 / 44.7 & 26.2 / 25.9 / 24.7 & $-19.2$ [$-20.3$, $-18.0$] & 20 / 20 / 18 / 20 & $+20.9$ / $+3.9$ \\
Toffoli 000 & 73.5 / 73.2 / 73.1 & 42.0 / 41.9 / 43.3 & $-30.8$ [$-32.0$, $-29.6$] & 30 / 32 / 32 / 31 & $+40.0$ / $+9.0$ \\
Toffoli 001 & 63.8 / 63.1 / 63.9 & 35.0 / 35.5 / 34.0 & $-28.8$ [$-30.0$, $-27.6$] & 30 / 29 / 30 / 28 & $+32.1$ / $+3.3$ \\
Toffoli 010 & 57.3 / 59.8 / 58.3 & 39.1 / 38.1 / 39.7 & $-19.5$ [$-20.7$, $-18.3$] & 31 / 31 / 33 / 32 & $+25.0$ / $+4.4$ \\
Toffoli 011 & 53.8 / 50.9 / 50.6 & 34.1 / 34.1 / 32.9 & $-18.1$ [$-19.3$, $-16.9$] & 29 / 29 / 30 / 28 & $+20.9$ / $+1.7$ \\
Toffoli 100 & 54.5 / 55.9 / 55.1 & 30.9 / 31.4 / 30.1 & $-24.3$ [$-25.5$, $-23.1$] & 24 / 24 / 23 / 21 & $+24.2$ / $+5.3$ \\
Toffoli 101 & 50.6 / 49.9 / 49.4 & 26.6 / 26.2 / 25.8 & $-23.8$ [$-24.9$, $-22.6$] & 24 / 22 / 21 / 20 & $+20.6$ / $+0.7$ \\
Toffoli 110 & 43.8 / 44.8 / 45.0 & 20.7 / 22.2 / 20.6 & $-23.4$ [$-24.5$, $-22.2$] & 14 / 13 / 11 / 13 & $+26.4$ / $+5.9$ \\
Toffoli 111 & 47.6 / 46.7 / 46.3 & 24.3 / 23.2 / 24.2 & $-22.9$ [$-24.1$, $-21.8$] & 15 / 15 / 15 / 16 & $+27.0$ / $+6.9$ \\
Rep. 0 -- & 65.5 / 67.8 / 67.8 & 68.8 / 68.1 / 67.5 & $+1.1$ [$-0.1$, $+2.2$] & 28 / 25 / 27 / 13 & $+38.4$ / $+38.7$ \\
Rep. 0 A & 44.0 / 44.5 / 44.4 & 44.8 / 45.1 / 43.2 & $+0.1$ [$-1.2$, $+1.3$] & 1 / 10 / 1 / 13 & $+33.5$ / $+30.0$ \\
Rep. 0 B & 57.5 / 56.9 / 57.8 & 49.1 / 49.1 / 46.6 & $-9.1$ [$-10.3$, $-7.9$] & 26 / 9 / 25 / 12 & $+32.2$ / $+20.9$ \\
Rep. 0 C & 54.9 / 56.4 / 56.0 & 67.7 / 65.6 / 64.5 & $+10.2$ [$+8.9$, $+11.4$] & 25 / 9 / 25 / 12 & $+29.9$ / $+38.8$ \\
Rep. 1 -- & 51.8 / 53.2 / 51.5 & 50.2 / 50.5 / 48.2 & $-2.5$ [$-3.7$, $-1.2$] & 22 / 15 / 23 / 12 & $+30.8$ / $+25.5$ \\
Rep. 1 A & 38.9 / 37.4 / 39.0 & 37.2 / 37.2 / 34.7 & $-2.1$ [$-3.3$, $-0.9$] & 1 / 9 / 2 / 11 & $+28.7$ / $+23.9$ \\
Rep. 1 B & 44.8 / 43.8 / 42.4 & 41.3 / 40.1 / 37.6 & $-4.0$ [$-5.2$, $-2.7$] & 21 / 16 / 20 / 12 & $+19.5$ / $+17.0$ \\
Rep. 1 C & 44.0 / 43.8 / 45.5 & 46.3 / 48.2 / 46.1 & $+2.4$ [$+1.2$, $+3.7$] & 22 / 9 / 22 / 13 & $+19.8$ / $+23.3$ \\
Rep. $+$ -- & 67.3 / 67.7 / 67.2 & 64.6 / 65.3 / 63.5 & $-3.0$ [$-4.1$, $-1.8$] & 25 / 26 / 25 / 25 & $+39.1$ / $+37.3$ \\
Rep. $+$ A & 64.1 / 63.9 / 66.7 & 63.9 / 63.7 / 64.7 & $-0.8$ [$-2.0$, $+0.4$] & 25 / 24 / 24 / 25 & $+38.2$ / $+37.5$ \\
Rep. $+$ B & 65.4 / 67.7 / 64.7 & 64.0 / 62.7 / 64.0 & $-2.4$ [$-3.6$, $-1.2$] & 24 / 25 / 25 / 24 & $+39.0$ / $+37.2$ \\
Rep. $+$ C & 65.2 / 67.3 / 66.0 & 64.3 / 63.7 / 63.6 & $-2.3$ [$-3.4$, $-1.1$] & 24 / 25 / 26 / 25 & $+38.5$ / $+36.3$ \\
\bottomrule
\end{tabular}
\end{table*}

\begin{table*}[p]
\centering
\caption{Both hardware sessions on \texttt{ibm\_fez}, 2026-10-09 and 2026-10-16,
4096 shots per repeat. Repeats: success (\%) of each of the three repeats.
$\Delta$: pooled second-session minus first-session success, with a 95\%
Newcombe interval (points). Controls: C1 / C2 / C1$'$ / C3 of the second
session (\%; C1 corrections and resets removed, C2 conditional blocks
applied at random, C1$'$ corrections removed and resets kept, C3 corrections
conditioned on an independent coin). Margin: lower bound of the 95\%
Newcombe interval of success minus each control of the same session, the
smallest over the controls, for each session (points). On 2026-10-16 the controls were run for Bernstein--Vazirani and Toffoli only, and only C1$'$ and C3, in a separate job on a different sector (qubits 108--112, 118, 129, 130) from the nominal circuits; the differences between nominal and control on that day therefore do not isolate the effect of feedforward.}
\label{tab:hardware-sessions-fez}
\scriptsize
\setlength{\tabcolsep}{3pt}
\begin{tabular}{@{}lccccc@{}}
\toprule
Configuration & Repeats, 10-09 & Repeats, 10-16 & $\Delta$ & Controls, 10-16 & Margin, 10-09 / 10-16 \\
\midrule
Dense k1 00 & 96.6 / 97.4 / 96.7 & 96.0 / 95.5 / 95.6 & $-1.2$ [$-1.7$, $-0.7$] & -- & -- \\
Dense k1 01 & 96.7 / 97.0 / 97.2 & 95.8 / 96.6 / 96.7 & $-0.6$ [$-1.1$, $-0.2$] & -- & -- \\
Dense k1 10 & 96.5 / 97.0 / 97.0 & 95.9 / 95.5 / 95.2 & $-1.3$ [$-1.8$, $-0.8$] & -- & -- \\
Dense k1 11 & 96.0 / 96.0 / 96.9 & 96.3 / 96.0 / 95.6 & $-0.3$ [$-0.8$, $+0.1$] & -- & -- \\
Dense k2 00 01 & 93.8 / 94.1 / 93.1 & 90.4 / 90.7 / 90.0 & $-3.3$ [$-4.0$, $-2.6$] & -- & -- \\
Dense k2 11 10 & 93.9 / 94.4 / 94.7 & 89.1 / 89.4 / 89.6 & $-4.9$ [$-5.6$, $-4.3$] & -- & -- \\
Teleport n1 & 96.7 / 97.2 / 97.3 & 97.4 / 97.0 / 97.5 & $+0.3$ [$-0.1$, $+0.7$] & -- / -- / -- / -- & $+43.7$ / -- \\
Teleport n2 & 90.4 / 90.9 / 91.7 & 86.5 / 85.9 / 87.2 & $-4.5$ [$-5.2$, $-3.7$] & -- / -- / -- / -- & $+63.4$ / -- \\
Teleport n3 & 83.6 / 84.3 / 84.2 & 80.0 / 80.1 / 79.4 & $-4.2$ [$-5.1$, $-3.2$] & -- / -- / -- / -- & $+68.6$ / -- \\
Swap Z & 94.5 / 94.8 / 94.3 & 90.6 / 90.5 / 89.5 & $-4.3$ [$-5.0$, $-3.7$] & -- / -- / -- / -- & $+44.0$ / -- \\
Swap X & 92.7 / 92.9 / 93.0 & 89.1 / 89.2 / 89.0 & $-3.8$ [$-4.5$, $-3.1$] & -- / -- / -- / -- & $+41.9$ / -- \\
Grover 00 & 64.8 / 64.3 / 64.8 & 67.2 / 67.4 / 68.1 & $+2.9$ [$+1.8$, $+4.1$] & -- / -- / -- / -- & $+37.8$ / -- \\
Grover 01 & 63.9 / 63.8 / 64.0 & 66.7 / 67.3 / 67.8 & $+3.4$ [$+2.2$, $+4.6$] & -- / -- / -- / -- & $+37.2$ / -- \\
Grover 10 & 64.6 / 65.0 / 64.7 & 66.9 / 66.8 / 67.2 & $+2.2$ [$+1.0$, $+3.4$] & -- / -- / -- / -- & $+37.1$ / -- \\
Grover 11 & 63.7 / 62.5 / 64.9 & 65.7 / 66.6 / 66.1 & $+2.4$ [$+1.2$, $+3.6$] & -- / -- / -- / -- & $+37.3$ / -- \\
BV 1001 & 75.4 / 74.9 / 74.0 & 67.6 / 69.3 / 68.7 & $-6.2$ [$-7.4$, $-5.1$] & -- / -- / 50 / 47 & $+24.9$ / $+17.0$ \\
BV 0110 & 74.0 / 74.6 / 75.0 & 67.7 / 69.0 / 67.4 & $-6.5$ [$-7.6$, $-5.3$] & -- / -- / 47 / 46 & $+24.5$ / $+19.0$ \\
BV 1111 & 18.3 / 18.7 / 17.8 & 30.5 / 30.9 / 30.5 & $+12.3$ [$+11.3$, $+13.4$] & -- / -- / 21 / 22 & $-5.1$ / $+7.1$ \\
Toffoli 000 & 53.7 / 50.3 / 50.6 & 42.0 / 43.4 / 41.9 & $-9.1$ [$-10.3$, $-7.8$] & -- / -- / 32 / 35 & $+18.0$ / $+6.0$ \\
Toffoli 001 & 45.5 / 42.5 / 43.4 & 35.8 / 33.5 / 34.9 & $-9.1$ [$-10.3$, $-7.9$] & -- / -- / 28 / 24 & $+12.8$ / $+5.6$ \\
Toffoli 010 & 45.6 / 43.0 / 43.2 & 41.5 / 42.5 / 42.2 & $-1.8$ [$-3.1$, $-0.6$] & -- / -- / 31 / 35 & $+10.5$ / $+5.2$ \\
Toffoli 011 & 37.5 / 34.7 / 36.6 & 35.6 / 36.4 / 35.7 & $-0.4$ [$-1.6$, $+0.8$] & -- / -- / 27 / 22 & $+5.8$ / $+7.6$ \\
Toffoli 100 & 39.9 / 37.4 / 36.9 & 22.0 / 24.1 / 24.7 & $-14.5$ [$-15.6$, $-13.3$] & -- / -- / 26 / 20 & $+6.2$ / $-4.3$ \\
Toffoli 101 & 32.0 / 29.0 / 30.7 & 18.0 / 19.1 / 20.4 & $-11.4$ [$-12.4$, $-10.3$] & -- / -- / 22 / 14 & $+0.5$ / $-4.3$ \\
Toffoli 110 & 32.3 / 30.7 / 33.7 & 21.8 / 22.2 / 27.1 & $-8.6$ [$-9.7$, $-7.4$] & -- / -- / 14 / 10 & $+14.7$ / $+8.7$ \\
Toffoli 111 & 37.7 / 39.1 / 38.7 & 24.1 / 24.5 / 27.9 & $-13.0$ [$-14.1$, $-11.8$] & -- / -- / 16 / 18 & $+19.0$ / $+6.4$ \\
Rep. 0 -- & 74.0 / 74.7 / 73.4 & 67.7 / 67.8 / 70.1 & $-5.5$ [$-6.6$, $-4.4$] & -- / -- / -- / -- & $+46.0$ / -- \\
Rep. 0 A & 40.4 / 39.6 / 39.5 & 49.8 / 46.5 / 51.2 & $+9.4$ [$+8.1$, $+10.6$] & -- / -- / -- / -- & $+28.3$ / -- \\
Rep. 0 B & 43.7 / 45.2 / 44.1 & 51.3 / 50.5 / 53.1 & $+7.3$ [$+6.1$, $+8.6$] & -- / -- / -- / -- & $+18.0$ / -- \\
Rep. 0 C & 71.9 / 72.5 / 72.0 & 60.8 / 60.3 / 62.4 & $-10.9$ [$-12.1$, $-9.7$] & -- / -- / -- / -- & $+44.0$ / -- \\
Rep. 1 -- & 50.7 / 54.6 / 52.4 & 44.1 / 42.6 / 46.0 & $-8.3$ [$-9.6$, $-7.1$] & -- / -- / -- / -- & $+32.5$ / -- \\
Rep. 1 A & 33.4 / 33.4 / 33.7 & 34.2 / 32.5 / 34.8 & $+0.3$ [$-0.9$, $+1.5$] & -- / -- / -- / -- & $+23.9$ / -- \\
Rep. 1 B & 32.4 / 36.7 / 34.8 & 44.4 / 43.0 / 44.5 & $+9.3$ [$+8.1$, $+10.5$] & -- / -- / -- / -- & $+14.0$ / -- \\
Rep. 1 C & 49.8 / 53.1 / 51.0 & 38.5 / 37.0 / 38.7 & $-13.2$ [$-14.4$, $-12.0$] & -- / -- / -- / -- & $+28.9$ / -- \\
Rep. $+$ -- & 70.0 / 72.0 / 69.9 & 63.1 / 61.7 / 64.7 & $-7.5$ [$-8.6$, $-6.3$] & -- / -- / -- / -- & $+42.3$ / -- \\
Rep. $+$ A & 66.0 / 68.0 / 67.9 & 58.9 / 59.4 / 61.7 & $-7.3$ [$-8.5$, $-6.1$] & -- / -- / -- / -- & $+41.2$ / -- \\
Rep. $+$ B & 67.3 / 67.5 / 66.2 & 57.1 / 58.1 / 63.4 & $-7.5$ [$-8.7$, $-6.3$] & -- / -- / -- / -- & $+39.8$ / -- \\
Rep. $+$ C & 70.1 / 70.6 / 70.0 & 60.6 / 59.1 / 62.0 & $-9.7$ [$-10.9$, $-8.5$] & -- / -- / -- / -- & $+42.7$ / -- \\
\bottomrule
\end{tabular}
\end{table*}

\clearpage
\section{Workflow scripts}
\label{app:scripts}

The scripts below complete Section~\ref{sec:demos} with dense coding, entanglement
swapping, Grover search, Bernstein--Vazirani, the Toffoli gate, the repetition
code, and the accept/reject control loop. Each is the QQLink program that the
evaluation plans, simulates, and lowers to hardware.

\begin{lstlisting}[style=qqpy,caption={Multi-channel superdense coding ($k$ parallel Bell pairs).},label={lst:dense}]
ENCODING = {(0,0): None, (0,1): "X",
            (1,0): "Z",  (1,1): "Y"}

def build_dense_coding(messages, k):
    wp = WorkflowProgram("dense_coding")
    alice = wp.create_circuit("alice", num_qubits=k)
    bob = wp.create_circuit("bob", num_qubits=k)
    for i in range(k):
        bell = wp.declare_entanglement(
            "alice", i, "bob", i, link_id=f"bell_{i}")
        alice.request_bell_pair(i, bell)
        gate = ENCODING[tuple(messages[i])]
        if gate is not None:
            alice.gate(gate, [i])
        alice.cross_cnot(i, "bob", i)
        alice.gate("H", [i])
        alice.measure(i, label=f"a_{i}")
        bob.measure(i, label=f"b_{i}")
    return wp
\end{lstlisting}

\begin{lstlisting}[style=qqpy,caption={Three-node entanglement swapping with a same-basis correlation check.},label={lst:swap}]
def build_entanglement_swapping(basis="Z"):
    wp = WorkflowProgram("entanglement_swapping")
    alice = wp.create_circuit("alice", num_qubits=1)
    charlie = wp.create_circuit("charlie", num_qubits=2)
    bob = wp.create_circuit("bob", num_qubits=1)
    l1 = wp.declare_entanglement("alice", 0, "charlie", 0)
    l2 = wp.declare_entanglement("charlie", 1, "bob", 0)
    alice.request_bell_pair(0, l1)      # |Phi+> on (a0, c0)
    charlie.request_bell_pair(1, l2)    # |Phi+> on (c1, b0)
    charlie.gate("CNOT", [0, 1])        # Bell measurement
    charlie.gate("H", [0])
    m0 = charlie.measure(0, label="m0")
    m1 = charlie.measure(1, label="m1")
    bob.if_then(m1 == 1, [("X", [0])])  # swap corrections
    bob.if_then(m0 == 1, [("Z", [0])])
    if basis == "X":
        alice.gate("H", [0]); bob.gate("H", [0])
    alice.measure(0, label="check_a")
    bob.measure(0, label="check_b")
    return wp
\end{lstlisting}

\Needspace{8\baselineskip}
\begin{lstlisting}[style=qqpy,caption={Two-node Grover iteration with non-local CZ gates.},label={lst:grover}]
def add_teleported_cz(wp, alice, bob, prefix, lc):
    bob.gate("H", [0])
    lc, _, _ = add_teleported_cnot(
        wp, alice, "alice", bob, "bob", prefix, lc)
    bob.gate("H", [0])
    return lc

def build_grover(marked):
    wp = WorkflowProgram("grover")
    alice = wp.create_circuit("alice", num_qubits=2)
    bob = wp.create_circuit("bob", num_qubits=2)
    lc = 0
    alice.gate("H", [0]); bob.gate("H", [0])
    if marked[0] == 0: alice.gate("X", [0])
    if marked[1] == 0: bob.gate("X", [0])
    lc = add_teleported_cz(wp, alice, bob, "ora", lc)
    if marked[0] == 0: alice.gate("X", [0])
    if marked[1] == 0: bob.gate("X", [0])
    for g in ["H", "X"]:
        alice.gate(g, [0]); bob.gate(g, [0])
    lc = add_teleported_cz(wp, alice, bob, "dif", lc)
    for g in ["X", "H"]:
        alice.gate(g, [0]); bob.gate(g, [0])
    alice.measure(0, label="a")
    bob.measure(0, label="b")
    return wp
\end{lstlisting}

\begin{lstlisting}[style=qqpy,caption={Bernstein--Vazirani with the hidden string split across two nodes. \texttt{tcnot} is the teleported CNOT of Script~\ref{lst:gtcnot} with explicit control, target, and ancilla indices.},label={lst:bv}]
def build_bernstein_vazirani(s_alice, s_bob):
    nA, nB = len(s_alice), len(s_bob)
    wp = WorkflowProgram("bernstein_vazirani")
    alice = wp.create_circuit("alice", num_qubits=nA + 1)
    bob = wp.create_circuit("bob", num_qubits=nB + 2)
    t, anc_b, anc_a = nB, nB + 1, nA
    for i in range(nA): alice.gate("H", [i])
    for i in range(nB): bob.gate("H", [i])
    bob.gate("X", [t]); bob.gate("H", [t])     # target |->
    lc = 0
    for i, bit in enumerate(s_bob):
        if bit: bob.gate("CNOT", [i, t])       # local terms
    for i, bit in enumerate(s_alice):
        if bit:                                # teleported terms
            lc = tcnot(wp, alice, "alice", i, anc_a,
                       bob, "bob", t, anc_b, f"bv{i}", lc)
    for i in range(nA):
        alice.gate("H", [i]); alice.measure(i, label=f"s_a{i}")
    for i in range(nB):
        bob.gate("H", [i]); bob.measure(i, label=f"s_b{i}")
    return wp
\end{lstlisting}

\begin{lstlisting}[style=qqpy,caption={Three-node Toffoli via six teleported CNOTs and $T$/$T^\dagger$ rotations. \texttt{tcnot} wraps \texttt{add\_teleported\_cnot} and returns the link counter.},label={lst:toffoli}]
def build_toffoli(a_val, b_val, c_val):
    wp = WorkflowProgram("toffoli")
    nA = wp.create_circuit("nodeA", num_qubits=2)
    nB = wp.create_circuit("nodeB", num_qubits=2)
    nC = wp.create_circuit("nodeC", num_qubits=2)
    def tcnot(c, cn, t, tn, p, lc):
        return add_teleported_cnot(wp, c, cn, t, tn, p, lc)[0]
    T, Tdg = pi/4, -pi/4
    if a_val: nA.gate("X", [0])
    if b_val: nB.gate("X", [0])
    if c_val: nC.gate("X", [0])
    nC.gate("H", [0]); lc = 0
    lc = tcnot(nB, "nodeB", nC, "nodeC", "bc1", lc)
    nC.gate("RZ", [0], params=[Tdg])
    lc = tcnot(nA, "nodeA", nC, "nodeC", "ac1", lc)
    nC.gate("RZ", [0], params=[T])
    lc = tcnot(nB, "nodeB", nC, "nodeC", "bc2", lc)
    nC.gate("RZ", [0], params=[Tdg])
    lc = tcnot(nA, "nodeA", nC, "nodeC", "ac2", lc)
    nB.gate("RZ", [0], params=[T]); nC.gate("RZ", [0], params=[T])
    lc = tcnot(nA, "nodeA", nB, "nodeB", "ab1", lc)
    nA.gate("RZ", [0], params=[T]); nB.gate("RZ", [0], params=[Tdg])
    lc = tcnot(nA, "nodeA", nB, "nodeB", "ab2", lc)
    nC.gate("H", [0])
    nA.measure(0, label="a"); nB.measure(0, label="b")
    nC.measure(0, label="c")
    return wp
\end{lstlisting}

\begin{lstlisting}[style=qqpy,caption={Three-node repetition code. \texttt{tcnot} is \texttt{add\_teleported\_cnot\_q} with explicit control and target indices, each followed by the index of its node's network qubit. Qubits: A = (data, network), B = (data, $s_1$, $s_2$, decoder, network), C = (data, network).},label={lst:repetition}]
def build_repetition_code(logical=0, error=None):
    wp = WorkflowProgram("repetition_code")
    A = wp.create_circuit("nodeA", num_qubits=2)
    B = wp.create_circuit("nodeB", num_qubits=5)
    C = wp.create_circuit("nodeC", num_qubits=2)
    DA, NA = 0, 1; DC, NC = 0, 1
    DB, S1, S2, DEC, NB = 0, 1, 2, 3, 4
    if logical == 1: A.gate("X", [DA])
    elif logical == "+": A.gate("H", [DA])
    lc = 0                                    # encode |psi>|00>
    lc = tcnot(A, "nodeA", DA, NA, B, "nodeB", DB, NB, "encB", lc)
    lc = tcnot(A, "nodeA", DA, NA, C, "nodeC", DC, NC, "encC", lc)
    if error == "A": A.gate("X", [DA])        # injected bit flip
    elif error == "B": B.gate("X", [DB])
    elif error == "C": C.gate("X", [DC])
    lc = tcnot(A, "nodeA", DA, NA, B, "nodeB", S1, NB, "synA", lc)
    B.gate("CNOT", [DB, S1]); B.gate("CNOT", [DB, S2])
    lc = tcnot(C, "nodeC", DC, NC, B, "nodeB", S2, NB, "synC", lc)
    s1 = B.measure(S1, label="s1")
    B.if_then(s1 == 1, [("CNOT", [S2, DEC])])   # dec = s1 AND s2
    s2 = B.measure(S2, label="s2")
    tb = B.measure(DEC, label="tb")
    B.if_then(tb == 1, [("X", [DB])])
    A.if_then(s1 == 1, [("X", [DA])])
    A.if_then(tb == 1, [("X", [DA])])
    C.if_then(s2 == 1, [("X", [DC])])
    C.if_then(tb == 1, [("X", [DC])])
    A.measure(DA, label="out_a"); B.measure(DB, label="out_b")
    C.measure(DC, label="out_c")
    return wp
\end{lstlisting}

\begin{lstlisting}[style=qqpy,caption={Accept/reject round (verification check) and the accept/reject retry loop driven by the syndrome. The function names are those of the QQLink code, which predates the renaming of the example.},label={lst:magic}]
def build_distillation_round(noise_angle=0.0):
    wp = WorkflowProgram("distillation_round")
    factory = wp.create_circuit("factory", num_qubits=2)
    consumer = wp.create_circuit("consumer", num_qubits=1)
    factory.gate("H", [0])                       # noisy T-state
    factory.gate("RZ", [0], params=[pi/4 + noise_angle])
    factory.gate("RZ", [0], params=[-pi/4])      # T^dagger, H:
    factory.gate("H", [0])                       # verification
    syn = factory.measure(0, label="syndrome")
    factory.if_then(syn == 1, [("X", [0])])      # reset
    return wp

def run_distillation(n_rounds, max_retries, noise, rng):
    orch = Orchestrator(available_backends=BACKENDS)
    rounds, total = [], 0
    for r in range(n_rounds):
        accepted, attempts = False, 0
        for _ in range(1 + max_retries):
            attempts += 1; total += 1
            wp = build_distillation_round(
                rng.uniform(-noise, noise))
            plan = orch.plan(wp)
            ex = WorkflowExecutor(wp, backend_name="qulacs")
            orch.dispatch_plan(plan, ex.dispatch)
            if ex.get_measurement("factory", "syndrome") == 0:
                accepted = True; break
        rounds.append({"round": r, "accepted": accepted,
                       "attempts": attempts})
    return {"rounds": rounds, "total": total}
\end{lstlisting}

\end{document}